\documentclass[aps,pra,amsmath,amssymb,floatfix,twocolumn,amsmath,superscriptaddress,twocolumn,nofootinbib,tighten,letterpaper]{revtex4-1}
\usepackage{multirow}
\usepackage{bbold}
\usepackage{subfigure}
\usepackage{color}
\usepackage{mathrsfs}
\usepackage{hyperref}
\usepackage[normalem]{ulem}
\usepackage{bm}
\usepackage{comment}

\usepackage{amsfonts, relsize, color}
\usepackage{graphics}
\usepackage{graphicx}
\usepackage{subfigure}
\usepackage{hyperref}
\usepackage{color}

\graphicspath{{FinalFigures/}}

\begin{document}
\title{Topolectric circuits: Theory and construction}

\author{Junkai Dong}
\affiliation{Max-Planck-Institut f$\ddot{\mbox{u}}$r Physik komplexer Systeme, N$\ddot{\mbox{o}}$thnitzer Str. 38, 01187 Dresden, Germany}
\affiliation{Laboratory of Atomic and Solid State Physics, Cornell University, Ithaca, New York 14853, USA}

\author{Vladimir Juri\v ci\' c}
\affiliation{Nordita, KTH Royal Institute of Technology and Stockholm University, Roslagstullsbacken 23, 10691 Stockholm, Sweden}
\affiliation{Departamento de F\'isica, Universidad T\'ecnica Federico Santa Mar\'ia, Casilla 110, Valpara\'iso, Chile.}

\author{Bitan Roy}\thanks{Corresponding author: bitan.roy@lehigh.edu}
\affiliation{Max-Planck-Institut f$\ddot{\mbox{u}}$r Physik komplexer Systeme, N$\ddot{\mbox{o}}$thnitzer Str. 38, 01187 Dresden, Germany}
\affiliation{Department of Physics, Lehigh University, Bethlehem, Pennsylvania, 18015, USA}

\date{\today}

\begin{abstract}
We highlight a general theory to engineer arbitrary Hermitian tight-binding lattice models in electrical LC circuits, where the lattice sites are replaced by the electrical nodes, connected to its neighbors and to the ground by capacitors and inductors. In particular, by supplementing each node with $n$ subnodes, where the phases of the current and voltage are the $n$ distinct roots of \emph{unity}, one can in principle realize arbitrary hopping amplitude between the sites or nodes via the \emph{shift capacitor coupling} between them. This general principle is then implemented to construct a plethora of topological models in electrical circuits, \emph{topolectric circuits}, where the robust zero-energy topological boundary modes manifest through a large boundary impedance, when the circuit is tuned to the resonance frequency. The simplicity of our circuit constructions is based on the fact that the existence of the boundary modes relies only on the Clifford algebra of the corresponding Hermitian matrices entering the Hamiltonian and not on their particular representation. This in turn enables us to implement a wide class of topological models through rather simple topolectric circuits with nodes consisting of only two subnodes. We anchor these outcomes from the numerical computation of the on-resonance impedance in circuit realizations of first-order ($m=1$), such as Chern and quantum spin Hall insulators, and second- ($m=2$) and third- ($m=3$) order topological insulators in different dimensions, featuring sharp localization on boundaries of codimensionality $d_c=m$. Finally, we subscribe to the \emph{stacked topolectric circuit} construction to engineer three-dimensional Weyl, nodal-loop, quadrupolar Dirac and Weyl semimetals, respectively displaying surface and hinge localized impedance.
\end{abstract}

\maketitle

\section{Introduction}

Simple topological models, such as the Su-Schrieffer-Heeger (SSH)~\cite{SSH-original, SSH-2, SSH-review} and the Bernevig-Hughes-Zhang (BHZ) model~\cite{BHZ-2D,BHZ-3D}, have played a pivotal role in the development of topological condensed matter physics. In particular, they provide an effective description of various topologically and symmetry protected phases~\cite{Hasan-Kane-RMP, Qi-Zhang-RMP, Shen-book, Bernevig-book, Schnyder-RMP, Armitage-RMP}, which captures the topological invariant, boundary modes, and responses to external perturbations. Recently, a new frontier opened up with the advent of \emph{metamaterials} where these simple models can be directly engineered in various platforms, manifestly showing the topological features. Notable examples include photonic~\cite{Raghu-Haldane-PRL, Topo-photonics-RMP, Topo-photonics-NatPhotonics, Peleg-PRL, Wang-PRL, Wang-Nature, Hafezi-NatPhotonics, Hafezi-PRL, Bandres-PRX}, phononic~\cite{Susstrunk-Science, Yang-PRL2015, He-NaturePhysics, Chen-PRApplied, Peterson-Nature, Serra-Nature, Xue-NatMat2019, Liu-AFM2020, Zhang-NatComm2019} and topolectric~\cite{ninguyan-prx2015, albert-prl2015, imhof-natphys2018, lee-commphys2018, ezawa-prb2018, hadad-natelectronics2018, yli-natcomm2018, goren-prb2018, zhao-ananphys2018} settings where local manipulation of the lattice elements allows one to control the hopping amplitude and the phase.

In this respect, topolectric circuits, made of rather simple capacitance and inductance elements, yield a readily available route
for the realization of a plethora of topological phases~\cite{ninguyan-prx2015, albert-prl2015, imhof-natphys2018, lee-commphys2018, ezawa-prb2018, hadad-natelectronics2018, yli-natcomm2018,goren-prb2018, zhao-ananphys2018, hofmann-prl2019, helbig-natphys2020, ezawa-prb2019, wang-natcomm2019, helbig-prb2019, hofmann-prr2020, liu-research2019, chlee-arxiv2019, bao-prb2019, wzhu-prb2019, haenel-prb2019, kotwal-arxiv2019, xxzhang-prl2020, lli-commphys2019, yoshida-prr2020, rafi-njp2020, ywang-natcomm2020, liu-pra2020, yang-prr2020, olekhno-natcomm2020, wzhang-arxiv2020, liang-tao-arxiv2020, r-li-arxiv2019, rafi-commphys2020, ylu-prb2019, zqzhang-prb2019, yang-PRL-2019, luo-research-2018, yu-NatSciRev-2020, RuiChen-PRL, yyang-antichiral:arxiv}. The mapping between the response of a circuit to a locally applied voltage and a tight-binding Hamiltonian is facilitated by the frequency-dependent admittance matrix ${\hat J}(\omega)$. Its form depends on the connectivity of the circuit elements through the nodes, which is used to engineer  the parameters of a hopping model. In this platform the impedance between the two nodes, related to the admittance matrix, can be used to locally detect the boundary modes~\cite{lee-commphys2018}, and thus serves as an electric circuit analog of a tunneling probe for topological crystals.

\subsection{Summary of results}

\begin{figure}[t!]
\includegraphics[width=0.95\linewidth]{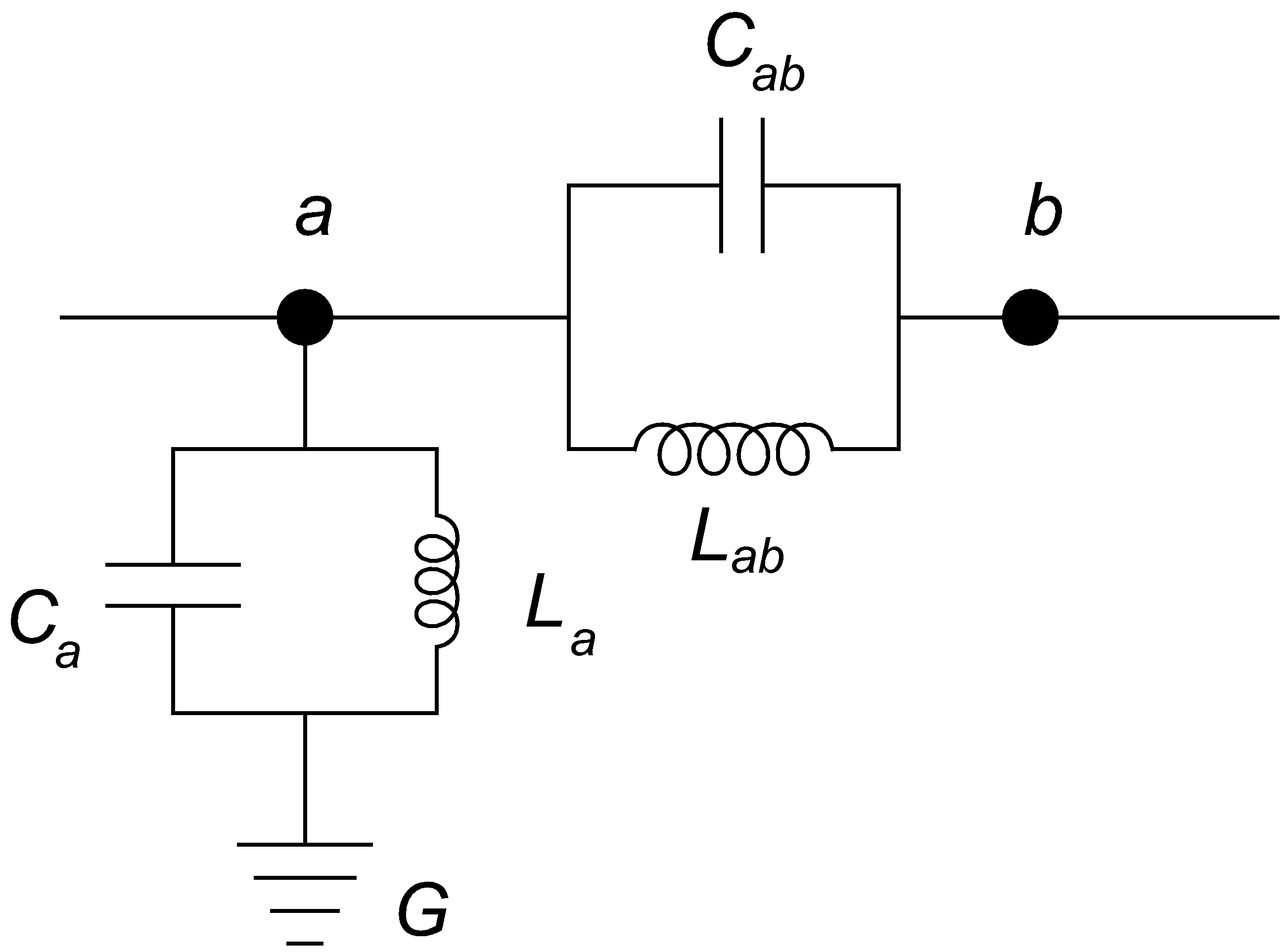}
\caption{A schematic LC circuit showing only two nodes $a$ and $b$. Here $C_{ab}$ and $L_{ab}$ are the capacitance and inductance between these two nodes, respectively. The capacitance and inductance between the node $a$ and the ground ($G$) are $C_a$ and $L_a$, respectively.
}~\label{Fig:schematiccircuit}
\end{figure}

\begin{figure}[t!]
\includegraphics[width=0.95\linewidth]{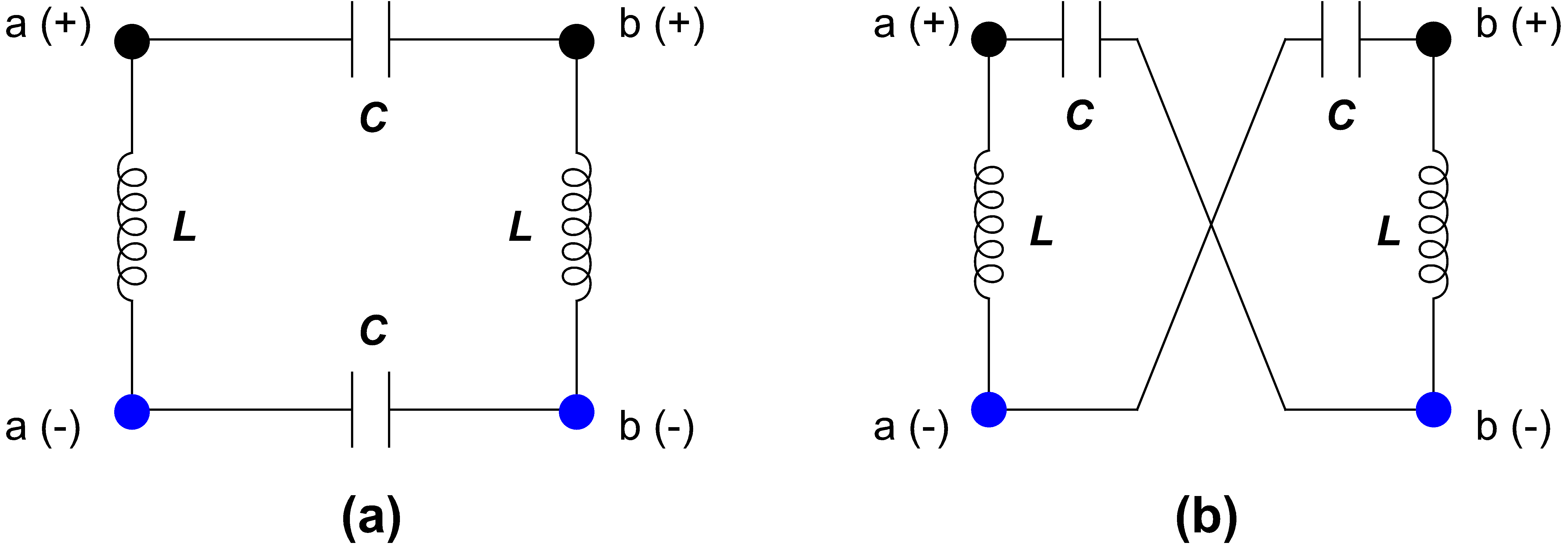}
\caption{Two possible circuit configurations with a two subnode structure. (a) The normal connection, and (b) the crossed or one-shift connection. Each subnode ($+$ or $-$) at each node ($a$ or $b$) is grounded (not shown here explicitly) via capacitance and inductance $C_g$ and $L_g$, respectively, see Fig.~\ref{Fig:schematiccircuit}.
}~\label{Fig:2node}
\end{figure}

Motivated by these developments, in this paper we use a general framework for constructing arbitrary tight-binding models in topolectric circuits to realize various gapped and gapless topological phases in one, two and three spatial dimensions. We first rederive in a rather transparent and independent way the known result~\cite{ninguyan-prx2015, albert-prl2015, zhao-ananphys2018} that a tight-binding model with arbitrary hopping amplitudes and phases can be constructed by extending a node in an LC circuit (see Fig.~\ref{Fig:schematiccircuit}) to include $n$ subnodes with the same amplitude of the input voltage but the phase factors representing $n$ different $n$th roots of unity. This method relies on the fact that each of the $n$ inequivalent connectivity configurations between the subnodes, realized with identical capacitors, directly maps into the phase factor equal to one of the $n$th roots of unity, see also Figs.~\ref{Fig:2node} and~\ref{Fig:4node} where, respectively, the cases $n=2$ and $n=4$ are displayed.

We show that a wide range of topological models can be realized in rather simple two-subnode topolectric circuits. To this end, we use the fact that the existence of the topological modes relies only on the Clifford algebra of the Hermitian matrices entering the Hamiltonian, but not on their representation. This  enables us to implement various topological models so that the hopping elements are purely \emph{real}. In turn, the corresponding topolectric circuit can be constructed  by supplementing each node (representing a lattice site) by only two subnodes, between which the phases of current and voltage differ by a factor of $\exp(i\pi)=-1$.

This method is exemplified by constructing circuit realizations of several toy models for topological phases.
Some of them have already been discussed in the literature, as detailed below, and even though our explicit circuit constructions are often different, we arrive at qualitatively similar results. Furthermore, our framework allows for the topolectric implementation of the models whose circuit realizations have not been achieved so far to the best of our knowledge.

We start the discussion with circuit implementation of the paradigmatic one-dimensional (1D) SSH model (Fig.~\ref{fig:SSH-construction}), experimentally realized in this platform in Ref.~\cite{lee-commphys2018}, and two-dimensional (2D) BHZ model (Fig.~\ref{Fig:Chern-lat}), featuring, respectively, the localized end-point and edge topological modes. We then compute the site resolved on-resonance impedance to show the hallmark boundary modes, as displayed in Figs.~\ref{Fig:SSHResults} and~\ref{fig:Chernspatial}, and to infer the global phase diagram of the models, shown in Fig.~\ref{fig:Chernphasediagram} for the square lattice topolectric Chern insulator, previously studied on brickwall or honeycomb circuits~\cite{hofmann-prl2019}. The realization of the quantum spin Hall insulator (QSHI) is displayed in Fig.~\ref{fig:QSHEsummary}. The outlined general method also can be used to construct the phases beyond the 10-fold way~\cite{schnyder-ryu-furusaki-ludwig}, such as a second-order topological insulator in two dimensions (Fig.~\ref{fig:BBHSummary2D})~\cite{imhof-natphys2018} and third-order topological insulator in three dimensions (Fig.~\ref{fig:3DHOTI-summary})~\cite{bao-prb2019}, both featuring sharp corner-localized on-resonance impedance. Besides demonstrating sharp corner localization of on-resonance impedance, we also numerically demonstrate its sublatice polarization stemming from the representation of the generator of the particle-hole symmetry, which still remains to be observed in experiments.

We also show how the hierarchy of the topological insulators can be realized in the topolectric platform in the case of a 2D second-order topological insulator by breaking the four-fold ($C_4$) rotational symmetry of the parent first-order phase (Fig.~\ref{fig:2DQSHI-HOTI}) and explore a related antiunitary symmetry protected generalized second-order topological insulator on electrical circuit (Fig.~\ref{fig:2DGHOTI}). We also construct the hierarchy of topological insulators in three dimensions, and show realizations of second- and third-order topological insulators, respectively supporting hinge and corner impedance, by breaking discrete rotational symmetries of a first-order topological insulator, yielding surface impedance (Fig.~\ref{fig:3Dhierarchy}). Finally, various 2D and three-dimensional (3D) topological nodal semimetals can be implemented  through stacking of the SSH circuits, as shown in Fig.~\ref{fig:stackedSSH}(a) for 2D Dirac semimetal (see also Ref.~\cite{lee-commphys2018}), with its hallmark boundary impedance displayed in Fig.~\ref{fig:stackedSSH}(b). The surface impedance corresponding to the drumhead states for the unknotted nodal-line semimetal is shown in Fig.~\ref{fig:stackedSSH}(c), while the Fermi arc surface states for the 3D Weyl and Dirac semimetals, respectively obtained by stacking 2D Chern and quantum spin Hall isulators, are displayed in Fig.~\ref{fig:Weylsemimetal} and~\ref{fig:Diracsemimetal}. Finally, the 1D hinge mode impedance for the quadrupolar (or second-order) Dirac and Weyl semimetals is shown in Figs.~\ref{fig:HOTDSM} and ~\ref{fig:HOTWSM}, respectively.

\subsection{Organization}

The rest of the paper is organized as follows. In Sec.~\ref{Sec:General} we present the general framework for constructing tight-binding models with arbitrary parameters. In Sec.~\ref{Sec:FOTI}  we apply this method to the realization of the first-order topological phases in SSH and BHZ models, as well as to the topological Chern insulator. Section~\ref{Sec:HOTI} is devoted to higher order topological insulators in two and three spatial dimensions. In Sec.~\ref{Sec:Antiunitary}, we present the construction of the $C_4$-symmetry breaking second-order topological insulator from the quantum spin Hall insulator and a related antiunitary symmetry protected generalized higher-order topological insulator. Section~\ref{Sec:3Dhierarchy} is devoted to the construction of the hierarchy of 3D higher-order topological insulators. In Sec.~\ref{Sec:NSM}, we implement various topological nodal semimetals in the topolectric platform, such as two- and 3D Dirac semimetals, and 3D Weyl, nodal-line and quadrupolar Dirac semimetals. In Sec.~\ref{Sec:Conclusions}, we summarize and discuss our results.

\section{General setup}~\label{Sec:General}

In this section, we highlight a general procedure for implementing an arbitrary tight binding model in LC electrical circuits. For completeness, we first discuss the relation between the admittance and the experimentally measurable \emph{impedance}~\cite{lee-commphys2018}. In particular, we lay out the connection between the \emph{divergence} of this observable and the existence of the admittance zero modes, which in turn captures the hallmark of topological phases in the circuit setup. Second, we rederive a general rule for devising a tight-binding model with \emph{an arbitrary phase of the hopping} in an LC circuit network~\cite{ninguyan-prx2015, albert-prl2015, zhao-ananphys2018} in an independent and transparent approach, which will be subsequently used to demonstrate the realization of different topological phases in topolectric circuits.

\subsection{Admittance and impedance matrices}

To make the connection between the electric circuit and a tight-binding lattice model, we start by considering a representative LC circuit, shown in Fig.~\ref{Fig:schematiccircuit}. The electric current that flows into the ground and the voltage at a particular node $a$ of an LC circuit are related by Kirchhoff's law according to
\begin{equation}\label{eq:Kirchhoff}
\dot{I}_a=\sum_b C_{ab}(\ddot{V}_a-\ddot{V}_b)+\frac{1}{L_{ab}}(V_a-V_b)+C_a\ddot{V}_a+\frac{1}{L_a}V_a.
\end{equation}
Here, $\dot{X}\equiv dX/dt$, the roman letters $a,b,\dots$ label the nodes in the circuit, and $I_a$ and $V_a$ denote the current and voltage at the node $a$, respectively. The capacitance and inductance between nodes $a$ and $b$ are, respectively, $C_{ab}$ and $L_{ab}$, while $C_{a}$ and $L_{a}$  represent these parameters between the node $a$ and the ground.

To characterize the response of an LC circuit to an applied voltage, we Fourier transform Eq.~(\ref{eq:Kirchhoff}) to obtain a  \emph{nonlocal} relation between the voltage and the current at frequency $\omega$
\begin{equation}
I_a(\omega)=\sum_b J_{ab}(\omega)V_b(\omega).
\end{equation}
The admittance matrix $J_{ab}(\omega)$ reads as
\begin{equation}\label{eq:J-matrix}
J_{ab}(\omega)=i\omega\left[N_{ab}(\omega)+\delta_{ab}W_a(\omega)\right],
\end{equation}
where
\begin{align}\label{eq:N-matrix}
N_{ab}(\omega)&=-C_{ab}+\frac{1}{\omega^2 L_{ab}},\\
W_a(\omega)&=C_a-\frac{1}{\omega^2 L_a}-\sum_c N_{ac},
\end{align}
and $N_{ab}$ and $W_a$ depend on the network structure of the circuit and on the grounding, respectively.

The experimentally measurable quantity in this context is the impedance, defined through the response of the circuit to an applied current.~\footnote{We emphasize that the measurement or computation of the impedance is performed here in the probe or linear response regime, in which the externally applied voltage and current at a single node of a topolectric circuit (containing macroscopic number of nodes), probing its impedance, do not perturb its state before their application~\cite{Wu-2004, Cervanova-2014, Cserty-2011}.}
In particular, we are interested in the voltage response when a current $I_a(\omega)\equiv I(\omega)$ is injected into a node $a$, yielding an outgoing current from node $b$, given by $I_b (\omega)=-I_a (\omega)$ (following the Kirchhoff's sign convention), and with no current inflow or outflow at any other nodes. The voltage difference between nodes $a$ and $b$ then defines the two-point impedance between these two nodes
\begin{equation}
Z_{ab}(\omega)=\frac{V_a(\omega)-V_b(\omega)}{I(\omega)}.
\end{equation}
Taking the spectral form of the admittance matrix
\begin{equation}
J_{ab}=\sum_n j_n \psi_{n,a}^* \psi_{n,b},
\end{equation}
and introducing its regularized inverse matrix
\begin{equation}
G_{ab}=\sum_{n, j_{n}\neq0} j_n^{-1}\psi^*_{n,a}\psi_{n,b},
\end{equation}
defined to exclude the zero eigenmodes of $J_{ab}$, the impedance $Z_{ab}$ reads as~\cite{lee-commphys2018, Wu-2004, Cervanova-2014, Cserty-2011}
\begin{equation}~\label{imp}
Z_{ab}=G_{aa}-G_{ab}-G_{ba}+G_{bb}=\sum_n\frac{|\psi_{n,a}-\psi_{n,b}|^2}{j_n}.
\end{equation}
Therefore when at least a single eigenvalue $j_n$ of the admittance matrix is very small, we expect a large response in the impedance $Z_{ab}$. Particularly, in the circuit realizations of topologically nontrivial phases, such a large signal corresponds to the gapless modes at the boundary of the system. Furthermore, in spatially extended circuits with translational symmetry in the bulk, the zero modes of the admittance are localized at the boundary, as we will show later in concrete examples, and directly identify the topological nature of the system.

\begin{figure}[t!]
\includegraphics[width=0.99\linewidth]{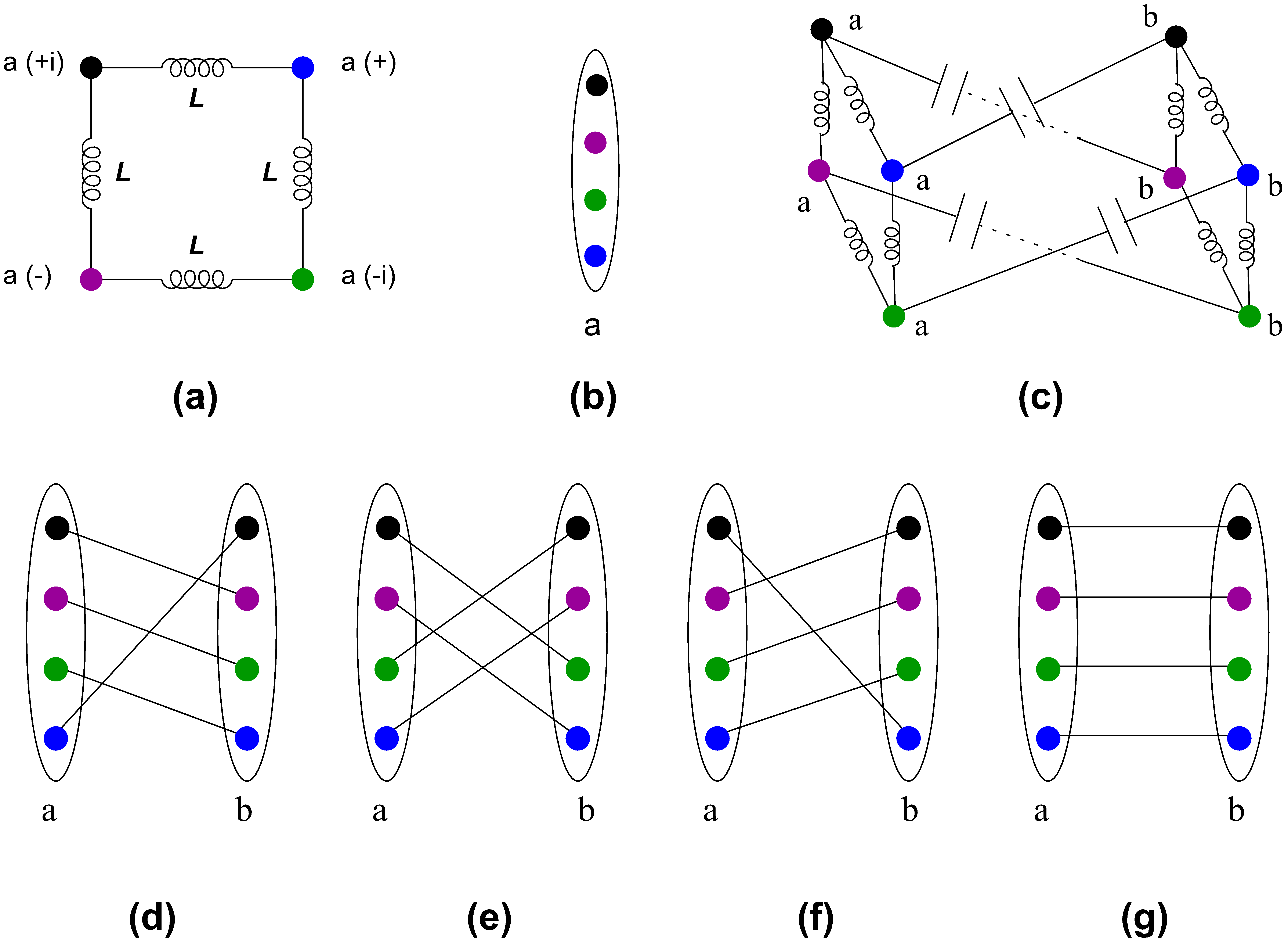}
\caption{(a) Circuit construction of a node with four subnodes $a(\alpha)$, where $\alpha=\pm, \pm i$ denote the relative phase of the current or voltage between the subnodes. (b) Schematic representation of a node with four subnodes. (c) One-shift capacitor connection with capacitance $C$ between two nodes $a$ and $b$. (d) Schematic representation of the one-shift capacitor connection from (c). Two-shift, three-shift, and no-shift or direct capacitor connections between nodes $a$ and $b$ are schematically shown in (e), (f), and (g), respectively. The effective hopping elements between these two nodes or sites are then $t_{(d)}=i C$, $t_{(e)}=-C$, $t_{(f)}=-i C$ and $t_{(g)}=C$.
}~\label{Fig:4node}
\end{figure}

\subsection{Arbitrary hopping phases}

To establish the correspondence with tight-binding models, we first notice that the admittance matrix in Eq.~\eqref{eq:J-matrix} up to the factor of $i\omega$ is Hermitian, namely $J_{ab}(\omega)=i\omega H_{ab}(\omega)$, with $H_{ab}(\omega)=H_{ba}^*(\omega)$.
Furthermore, if the nodes are thought of as the lattice sites, the matrix $\hat{H}(\omega)$ can be associated with a Hamiltonian of a tight-binding model on this lattice, with the off-diagonal elements that correspond to hoppings, while the diagonal ones represent the on-site chemical potential. Hereafter, the matrix $\hat{H}(\omega)$ is referred to as the Hamiltonian. As can be seen from the form of the admittance matrix in Eq.~\eqref{eq:J-matrix}, the corresponding Hamiltonian is completely real. Nevertheless, a hopping with an arbitrary complex phase can be implemented in this setup, as shown in Refs.~\cite{ninguyan-prx2015, albert-prl2015, zhao-ananphys2018}, which we rederive in a rather straightforward and independent manner below.

To this end, we first notice that the voltage and current at a node are defined up to a phase factor.
This observation allows one to enrich the node structure to include more subnodes within the same node with the same magnitude of the voltage and current, but the phase factors shifted with respect to each other, see Figs.~\ref{Fig:2node} and \ref{Fig:4node}. Crucially, as explicitly shown in Refs.~\cite{ninguyan-prx2015, albert-prl2015, zhao-ananphys2018} by solving the corresponding Kirchoff's equations, an arbitrary rational phase factor can be tuned between the subnodes, which is subsequently used to construct a tight-binding model with an arbitrary hopping phase between the nodes. A similar subnode-based approach has been used to engineer tight-binding models on phononic lattices with a paradigmatic example in this context being the QSHI tight-binding model with purely real hoppings~\cite{Susstrunk-Science}. Importantly, topolectric realizations of the discussed topological models can be accomplished by supplementing each node with only two subnodes, and only on rare occasions by four subnodes (see Secs.~\ref{subsec:CI} and \ref{subsec:antiunitary}), when accompanied by suitable choices of the matrices in the orbital or sublattice space. The explicit solutions of the Kirchoff's equations in the cases of four subnode and two subnode circuits, showing the realizations of desired phase distributions, are presented in Sec.~5.1(a) and Sec.~5.1(c) of Ref.~\cite{zhao-ananphys2018}, respectively.

\subsubsection{Two subnodes}~\label{subsubsec:2nodes}

To illustrate this protocol, let us start with the simplest example of a node featuring the two subnodes ($\pm$) at which the voltages are shifted by a relative phase factor $\exp[i\pi]=-1$, see Fig.~\ref{Fig:2node}. Two possible configurations of connection between the neighboring nodes, shown in Figs.~\ref{Fig:2node}(a) and ~\ref{Fig:2node}(b), correspond to the admittance matrices with the effective Hamiltonian
\begin{equation}~\label{eq:Ham-2node}
\hat{H}_{\alpha}(\omega)=\sigma_0\otimes\hat{D}+\sigma_1\otimes \hat{C}_\alpha.
\end{equation}
Here $\hat{C}_\alpha=C\tau_\alpha$, $\alpha=0$ ($\alpha=1$) corresponds to the circuit configuration in Fig.~\ref{Fig:2node}(a) [Fig.~\ref{Fig:2node}(b)], and $\otimes$ represents direct or tensor product. The Pauli matrices $\{ \sigma_\nu \}$ and $\{ \tau_\nu \}$ act on the node and the subnode spaces, respectively, where $\nu=0,\cdots,3$. Here, the diagonal part in the node space is
\begin{equation}
\hat{D}=\left( C_g-\frac{1}{\omega^2 L_g} \right) \; \tau_0 + \frac{1}{\omega^2L} \; \tau_1,
\end{equation}
where $C_g$ and $L_g$ are the capacitance and inductance of the grounding elements (displayed in Fig.~\ref{Fig:schematiccircuit}, but not shown in Fig.~\ref{Fig:2node}), respectively, which can tune the circuit to the \emph{resonance}. Since we are interested only in the configuration with the voltages tuned out of phase on the two subnodes within the same node, to obtain the hopping and the on-site chemical potential the subnode matrix is projected onto the relevant subnode subspace spanned by the vector $v_2=1/\sqrt{2}[1,-1]^\top$. The projector onto this subspace $P_{2,ij}=v_{2,i}v_{2,j}^*$  explicitly reads
\begin{align}
\hat{P}_{2}=\frac{1}{2}(\tau_0-\tau_1).
\end{align}
Consequently, the hopping matrix element for the two configurations in Fig.~\ref{Fig:2node} respectively read
\begin{equation}
t_\alpha={\rm Tr}(\hat{P}_{2} \hat{C}_\alpha \hat{P}_{2}),
\end{equation}
for $\alpha=0$ and $1$, yielding
\begin{equation}~\label{eq:eff-hopping2node}
t_{\alpha}=(-1)^\alpha \; C,
\end{equation}
after using Eq.~\eqref{eq:Ham-2node}. On the other hand, the on site chemical potential for these two configurations is
\begin{equation}
\mu_\alpha={\rm Tr}(\hat{P}_{2} \hat{D}_\alpha \hat{P}_{2}).
\end{equation}
As a consequence of Eq.~\eqref{eq:Ham-2node}, we find for both circuit network configurations $\mu_0=\mu_1 \equiv \mu$, where
\begin{equation}
\mu=C_g-\frac{1}{\omega^2 L_g}-\frac{1}{\omega^2L}.
\label{eq:eff-cp2node}
\end{equation}
Obviously, by applying different grounding elements on the two sites, the difference of the onsite chemical potentials can be realized.
Consequently, by periodically repeating two subnode circuit configurations an arbitrary tight-binding model with pure \emph{real} hopping can be realized in topolectric circuits. In fact, as we show, a wide class of topological models  can be cast in such a form since the topological features do not depend on the representation of the Hermitian matrices entering the Hamiltonian, but only on their Clifford algebra. Finally, by generalizing the above construction to include more than two subnodes on a single node a hopping with an arbitrary phase factor can be obtained.

\begin{figure}[t]
\includegraphics[width=0.95\linewidth]{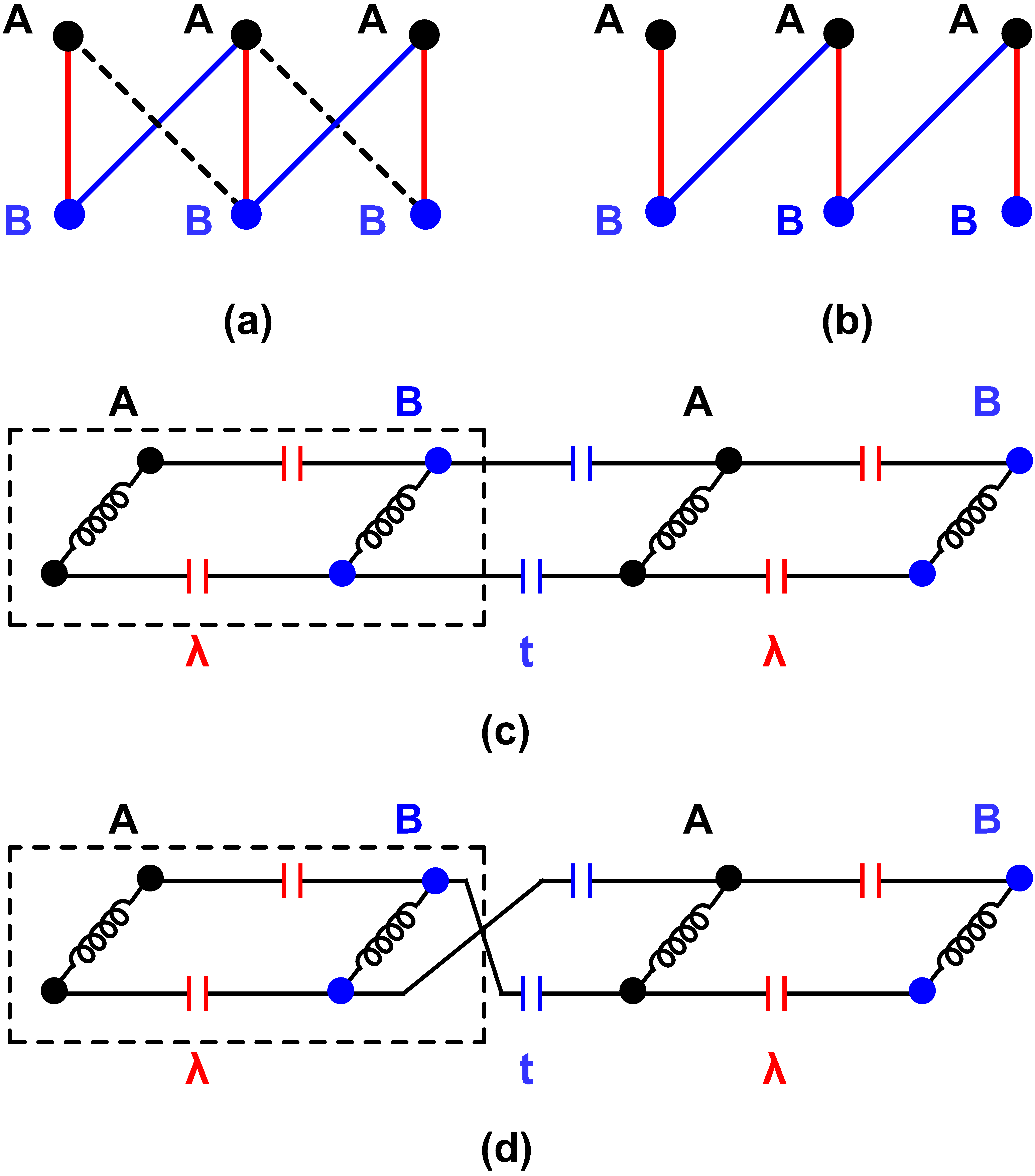}
\caption{The SSH model and its topolectric circuit realization. (a) The real space realization of the SSH model. A and B are two orbitals on each lattice site. The solid red lines correspond to hopping $\lambda$, and the hopping amplitudes ${\tilde t}_1=(t+t_1)/2$ and ${\tilde t}_2=(t-t_1)/2$ are, respectively, represented by solid blue and dashed black lines; see Eq.~\eqref{eq:SSH-real-space}. When $|\lambda/t|<1$ the system is in the topological phase. (b) The hopping pattern in the limit $t=t_1$; see Eq.~(\ref{eq:SSH-momentum-space}). (c) A circuit realization of the SSH model in the limit $t=t_1$, with two subnodes at each node or orbital. Two nodes residing within the dashed rectangle constitute the unit cell of the SSH model. The corresponding capacitances in the circuit are $\lambda$ (red) and $t$ (blue). The inductors connecting each subnode to the ground (not shown explicitly here, see Fig.~\ref{Fig:schematiccircuit}) and two subnodes at each node are all with the same inductance equal to $L$. (d) The circuit realization of the SSH model with an \emph{effective} negative nearest-neighbor hopping amplitude due to the crossed or one-shift connection (see Fig.~\ref{Fig:2node}), yielding access to the parameter regime $\lambda/t<0$, which is not possible in a one subnode structure with only capacitor connections; see also Ref.~\cite{lee-commphys2018}.
}~\label{fig:SSH-construction}
\end{figure}

\subsubsection{Four subnodes}~\label{subsubsec:4nodes}

Next we consider the four subnode configurations, as shown in Fig.~\ref{Fig:4node}(a), where the phase differences between the subnode $a(+)$ and the rest of the subnodes [including the subnode $a(+)$] are $\pm 1, \pm i$. Therefore the relevant subnode subspace is spanned by the vector $v_4=(1/2)[1,-i,-1,i]^\top$, graphically represented in Fig.~\ref{Fig:4node}(b). The corresponding projector $P_{4,ij}=v_{4,i} v_{4,j}^*$, with the explicit form given by
\begin{equation}
\hat{P}_4=\frac{1}{4}
\begin{pmatrix}
1 & i & -1 & -i \\
-i & 1 & i & -1 \\
-1 & -i & 1 & i \\
i & -1 & -i & 1 \\
\end{pmatrix}.
\end{equation}
The hopping element between the nodes $a$ and $b$ is then the projection of the connectivity matrix between the subnodes onto the subspace spanned by the vector $v_4$, namely
\begin{equation}\label{eq:4node-hopping}
t_\alpha={\rm Tr}(\hat{P}_4 \hat{C}_\alpha \hat{P}_4),
\end{equation}
with $\hat{C}_\alpha$ as the connectivity matrices for four possible configurations, schematically shown in Figs.~\ref{Fig:4node}(d)-(g). For instance, the circuit in Fig.~\ref{Fig:4node}(c) corresponds to the graph in Fig.~\ref{Fig:4node}(d), and the corresponding connectivity matrix is
\begin{equation}
\hat{C}_{(d)}=C
\begin{pmatrix}
0 & 0 & 0 & 1 \\
1 & 0 & 0 & 0 \\
0 & 1 & 0 & 0 \\
0 & 0 & 1 & 0 \\
\end{pmatrix},
\end{equation}
where $C$ is the capacitance between two neighboring nodes. Then Eq.~\eqref{eq:4node-hopping} yields $t_{(d)}=i C$. Similarly, one can readily check that the circuit with the graph shown in Fig.~\ref{Fig:4node}(e) [Fig.~\ref{Fig:4node}(g)] yields the hopping $t_{(e)}=-C$ [$t_{(g)}=C$], while for the remaining graph in Fig.~\ref{Fig:4node}(f) the hopping $t_{(f)}=-i C$. A straightforward calculation then shows that the corresponding on-site potential when the grounding elements on all four subnodes are the same, is given by $\mu=C_g-1/(\omega^2 L_g)-1/(\omega^2 L)$. Again, by choosing different grounding elements for different nodes, one can tune the difference of the on site chemical potentials. Therefore, the four subnode LC circuit can realize an arbitrary tight-binding model with hopping phases being four fourth roots of unity, namely $\exp[2\pi i m/4]$, with $m=1,2,3,4$, i.e. either purely real or purely imaginary.

\subsubsection{$n$ subnodes}~\label{subsubsec:nnodes}

This construction can now be further generalized to a hopping with phase factors $\Phi_{n,m}=\exp[2i\pi (1-m)/n]$, with $m=1,...,n$, which are the $n$th roots of unity. Now each node contains $n$ subnodes. The hopping elements are obtained by projecting the connectivity matrices onto the subspace spanned by the $n$-component vector $v_n$, with the elements $v_{n,m}=\Phi_{n,m}$. The matrix corresponding to the ``one-shift" in the connectivity $1\rightarrow n, 2\rightarrow1, 3\rightarrow2, 4\rightarrow3,..., n\rightarrow n-1$ generates the hopping equal to $t_{(1)}=\exp[2i\pi/n]$. See the Appendix~\ref{app:general-proof} for the proof. More generally, an $s$-shift results in the hopping phase equal to $t_{(s)}=\exp[2i\pi s/n]$, as also shown in the Appendix~\ref{app:general-proof}.

In the following, we apply this general protocol to construct various insulating and gapless or nodal topological phases, both more conventional first-order ones and the higher-order ones, in one, two, and three spatial dimensions.

\section{First-order topolectric insulators}~\label{Sec:FOTI}

The hallmark of any topological phase of matter is the bulk-boundary correspondence, which ensures the existence of topologically protected robust boundary modes~\cite{Kane-Mele-PRL2005,BHZ-2D, Hasan-Kane-RMP, Qi-Zhang-RMP, Shen-book, Bernevig-book, Schnyder-RMP, Armitage-RMP, Fu-Kane-PRB2006,Fu-Kane-PRB2007,BHZ-3D,CXLiu-PRB2010,Fu-PRL2011,Slager-NatPhys}. Typically, a $d$-dimensional topological phase supports such modes that reside on $(d-1)$-dimensional boundaries, which are also characterized by the co-dimension $d_c=d-(d-1)=1$, yielding first-order topological phases. Some well known examples of topological modes include the endpoint modes in 1D SSH topological insulator, 1D edge modes for 2D Chern and quantum spin Hall insulators, and the surface states of 3D topological insulators. In this section we demonstrate realizations of some of these insulating phases in topolectric circuits. In Secs.~\ref{subsec:DSM-NLSM}, ~\ref{subsec:3DWSM} and ~\ref{subsec:3DDSM} we demonstrate circuit realizations of first-order nodal or gapless topological phases.

\subsection{Su-Schrieffer-Heeger model}~\label{subsec:SSH}

The Su-Schrieffer-Heeger (SSH) model is a paradigmatic example of a topological state in one dimension~\cite{SSH-original, SSH-2, SSH-review}. Here we show its circuit realization within the general construction principle, discussed in the previous section. The Hamiltonian of the SSH model in the real space takes the following form
\begin{equation}~\label{eq:SSH-real-space}
H^{\rm real}_{\rm SSH}=\sum_i [\lambda a_i^\dagger b_i + {\tilde t}_1  a_i^\dagger b_{i+1}+{\tilde t}_2  b_i^\dagger a_{i+1}],
\end{equation}
where the annihilation operators $a_i$ and $b_i$ act on the orbitals A and B localized at the site $i$, respectively. The hopping amplitudes $\lambda$ and ${\tilde t}_{1,2}=(t\pm t_1)/2$ are purely real, see Fig.~\ref{fig:SSH-construction}(a) and~\ref{fig:SSH-construction}(b). The corresponding momentum space representation of the SSH model reads as
\begin{equation}~\label{eq:SSH-momentum-space}
H^{\rm mom}_{\rm SSH} = [\lambda+t\cos(k)] \; \tau_1 + t_1 \sin(k) \; \tau_2,
\end{equation}
where the Pauli matrices $\{ \tau_\mu\}$ with $\mu=0,\cdots,3$ act on the orbital space. For the range of parameters $-1<\lambda/t<1$, the SSH model is in the topological regime and features a localized zero mode at each end of the system. Furthermore, the matrix $\tau_3$ \emph{anticommutes} with the Hamiltonian $H^{\rm mom}_{\rm SSH}$ and ensures the spectral symmetry. Consequently, all zero-energy modes are eigenstates of the matrix $\tau_3$, and as such they are completely localized on either of the two orbitals. By contrast, the system becomes a trivial insulator when $|\lambda/t|>1$. This model belongs to class BDI in the ten-fold classification~\cite{schnyder-ryu-furusaki-ludwig}.

\begin{figure}[t!]
\includegraphics[width=0.47\linewidth]{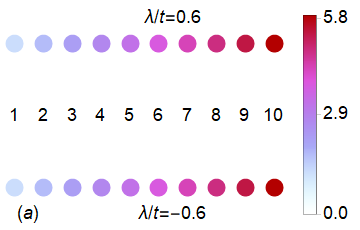}
\includegraphics[width=0.47\linewidth]{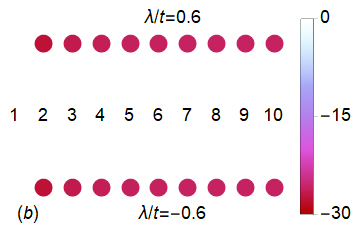}
\includegraphics[width=0.48\linewidth]{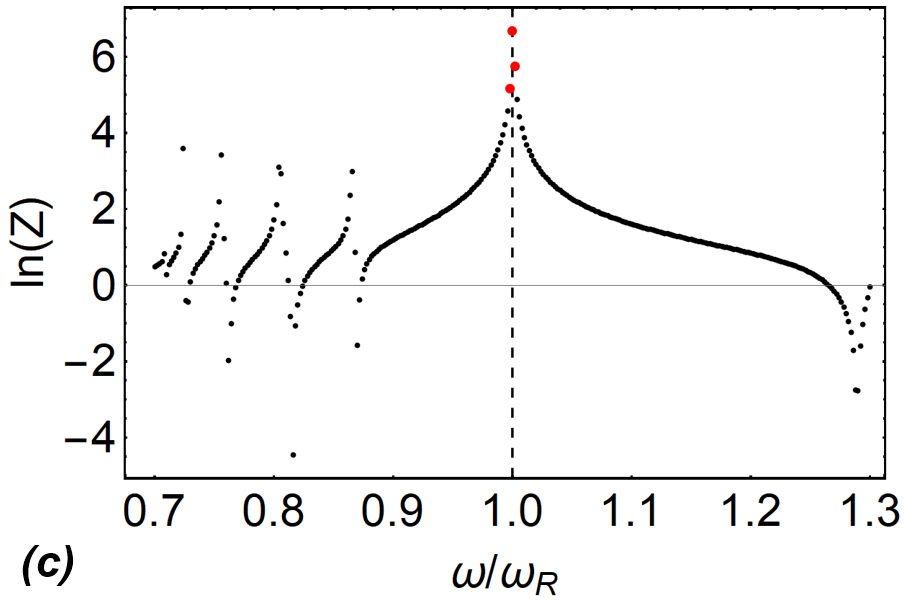}
\includegraphics[width=0.48\linewidth]{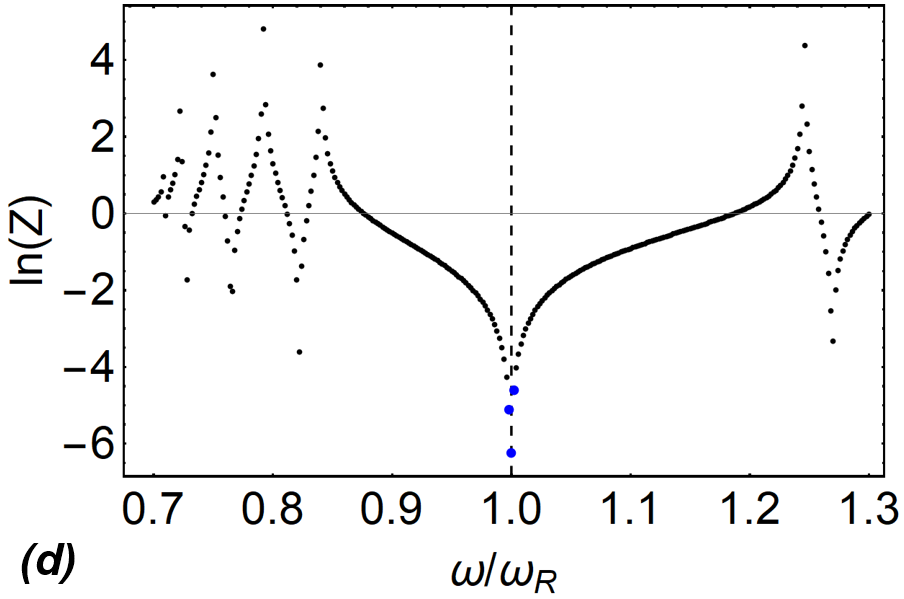}
\caption{Numerical computation of the impedance of the SSH chain realized on an electrical circuit, see Figs.~\ref{fig:SSH-construction}(a) and~\ref{fig:SSH-construction}(b). (a) On-resonance logarithm of the impedance $\ln (Z)$ in an SSH circuit of linear dimension $\ell =10$ with input point on the orbital A at site 1 and output point on the orbital B at various sites of the circuit for $\lambda/t=0.6$ (top) and $\lambda/t=-0.6$ (bottom). (b) Same as (a), but with the output point on the orbital A at various sites of the circuit. Frequency dependence of the impedance in an SSH circuit of linear dimension $\ell=10$ for (c) $\lambda/t=\pm 0.55$ (topological phase) and (d) $\lambda/t=\pm 2.00$ (trivial phase). Here we compute the impedance with the orbital A at site 1 as the input point and the orbital B at site 10 as the output point. While the peak (red dots) of impedance in (c) at the resonance [$\omega=\omega_R=1/\sqrt{L(\lambda+t)}$] indicates the topological nature of the state with endpoint zero modes, the dip (blue dots) at the resonance in (d) confirms that the phase is a trivial insulator.
}~\label{Fig:SSHResults}
\end{figure}

The form of the Hamiltonian in Eq.~(\ref{eq:SSH-real-space}) in terms of only real hopping parameters allows a realization of the SSH model through an electric circuit with two nodes corresponding to the two orbitals per lattice site. Furthermore, when each node contains two subnodes, it allows access to the parameter regimes $\lambda/t>0$ as well as $\lambda/t<0$, which is not possible in a one subnode structure with pure capacitor (or inductor) connections; see Fig.~\ref{Fig:2node}. Concrete circuit realizations to access these two parameter regimes are, respectively, shown in Figs.~\ref{fig:SSH-construction}(c) and ~\ref{fig:SSH-construction}(d). Note that the SSH model has been studied in the topolectric circuit extensively in Ref.~\cite{lee-commphys2018}, as well as in other circuit setups~\cite{hadad-natelectronics2018, goren-prb2018, wang-natcomm2019, helbig-prb2019, liu-research2019}. Here we discuss this model for the sake of completeness and to establish generic features of a topolectric circuit.

Now consider a periodic SSH circuit [see Fig.~\ref{fig:SSH-construction}(a)] with the Hamiltonian matrix, ${\hat H}_{\rm SSH}(\omega)\equiv{\hat J}(\omega)/i\omega$, which according to Eq.~\eqref{eq:J-matrix}, is given by
\begin{eqnarray}
&&{\hat H}_{\rm SSH}(\omega)= \nonumber \\
&&\left[ \begin{array}{cccc}
\lambda+t-\dfrac{1}{\omega^2 L}&-\lambda&0&\dots\\
-\lambda&\lambda+t-\dfrac{1}{\omega^2 L}&-t&0\\
0&-t&\lambda+t-\dfrac{1}{\omega^2 L}&-\lambda\\
0&0&-\lambda&\ddots
\end{array}\right]. \nonumber\\
\end{eqnarray}
It can be seen that this Hamiltonian matrix represents an SSH Hamiltonian with an overall shift in chemical potential equal to $\lambda+t-\frac{1}{\omega^2L}$. Therefore, the SSH circuit is at \emph{resonance} when the ac frequency $\omega=\omega_R=1/\sqrt{L(\lambda+t)}$. We notice that the two zero-energy modes localized on one particular orbital mix due to a finite size effect. Because of such an orbital mixing between the zero-energy modes, the measured impedance with one orbital (or sublattice) as the input point and the other one as the output point shows a large orbital selective on-resonance peak in the topological regime. This mechanism for the observed large orbital or sublattice selective on-resonance topological impedance is operative on all the topolectric circuits that we discuss in this paper. Furthermore, in a finite system zero energy states also split symmetrically about the zero admittance eigenvalue. Therefore the impedance between the same sublattice (or same orbital) at two sites is extremely small, as can be seen from Eq.~\eqref{imp}, which approaches zero with the increasing system size. On the other hand, the impedance between two distinct sublattices or orbitals at the two ends of the circuit is \emph{large}, since the zero admittance state is predominantly localized on only one of the two orbitals at a given end. These outcomes are demonstrated through a numerical simulation of the SSH circuit in the topological and trivial phases, shown in Fig.~\ref{Fig:SSHResults}(a) and~\ref{Fig:SSHResults}(b), respectively. One can also see that the impedance grows with the separation between the two sites, consistent with the topological mode being localized at the endpoint of the system.

\begin{figure}[t!]
\includegraphics[width=.95\linewidth]{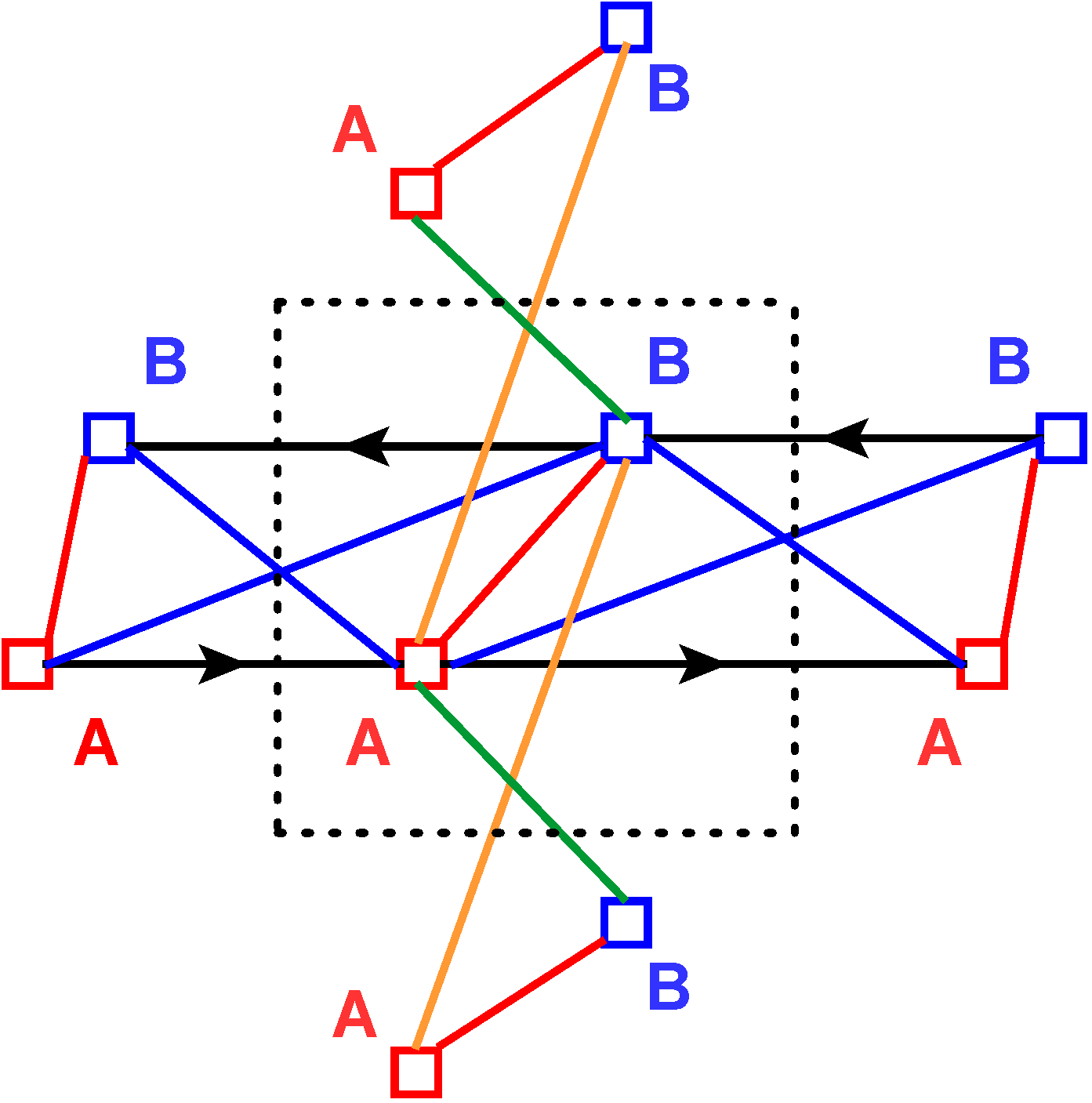}
\caption{Chern topolectric circuit. The red (blue) squares represent sublattices or nodes A (B) and the lines indicate the hopping between them. The dashed box corresponds to the two-sublattice unit cell of the Chern insulator. The green, orange, red and blue lines denote a hopping amplitude equal to $(t-t_0)/2$, $-(t+t_0)/2$, $m$ and $-t_0/2$, respectively. The arrowed lines correspond to a hopping of $it/2$ in its direction. Such hopping amplitudes are obtained by supplementing each node with four subnodes (not shown here explicitly) and connecting subnodes from nearest-neighbor nodes by capacitors of capacitance equal to modulus of the requisite hopping, following the prescription from Fig.~\ref{Fig:4node} and Sec.~\ref{subsubsec:4nodes}. Each subnode is grounded with inductor of inductance $L$. For the range of parameters $-2t_0<m<2 t_0$ the system is topological and thus supports boundary states that in turn yield enhanced impedance ($Z$) on the edge of the circuit, see Figs.~\ref{fig:Chernspatial} and ~\ref{fig:Chernphasediagram}.
}~\label{Fig:Chern-lat}
\end{figure}

To further corroborate the topological nature of the circuit, we measure impedance over a wide range of frequencies between orbital A from site $1$ and orbital B of site $10$, located at two opposite ends of the circuit, see Fig.~\ref{Fig:SSHResults}(c). The \emph{peak} of the impedance at the resonance frequency ($\omega=\omega_R$) signals that the system is topological when $|\lambda/t|<1$. On the other hand, for $|\lambda/t|>1$, the \emph{dip} in the impedance at the resonance frequency is a signature of the topologically trivial nature of this phase, as shown in Fig.~\ref{Fig:SSHResults}(d).

\subsection{Chern insulator}~\label{subsec:CI}

\begin{figure}[t!]
\includegraphics[width=0.95\linewidth]{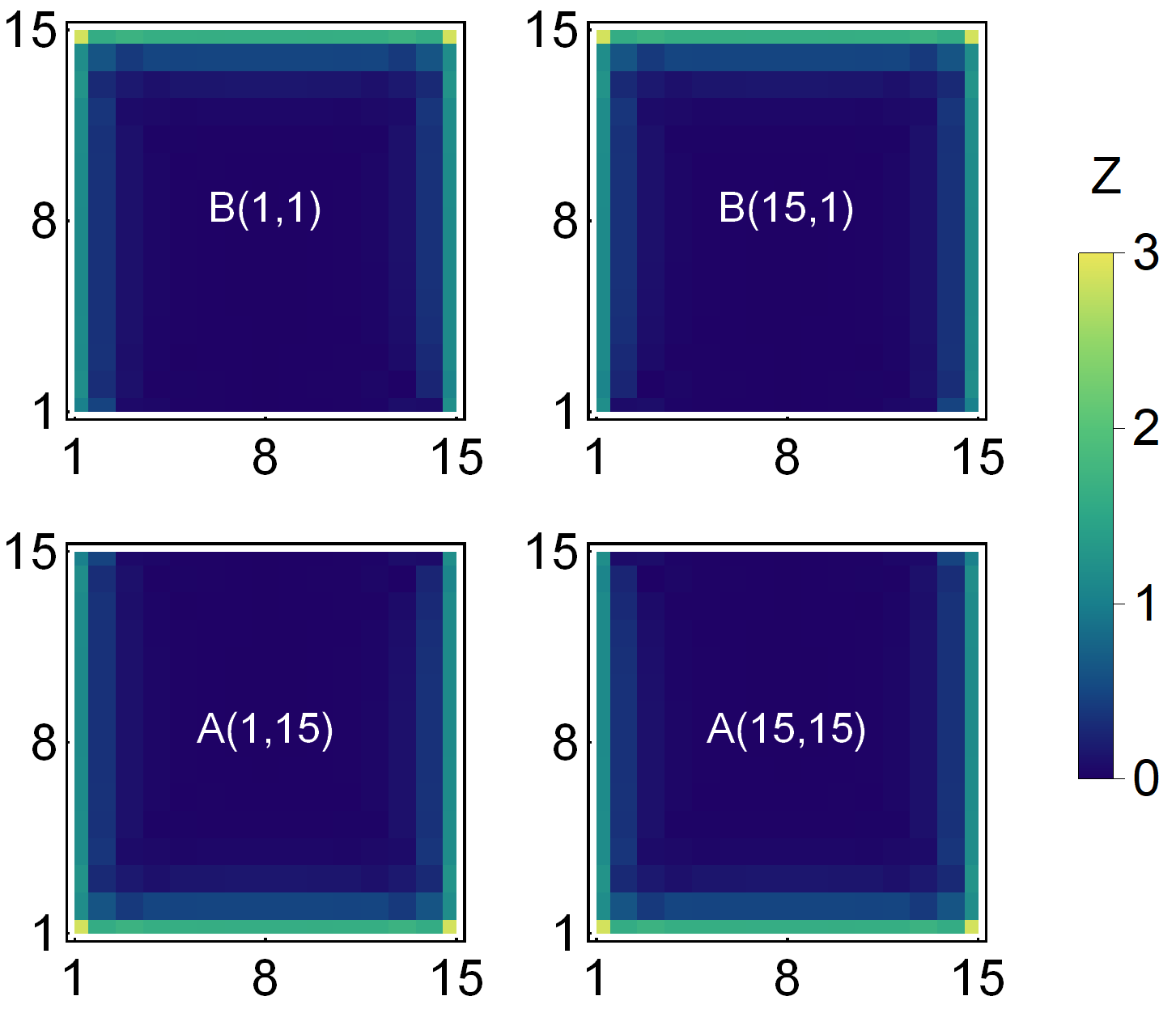}
\caption{Impedance of the Chern circuit of linear dimension (number of sites) $\ell =15$ in each direction at resonance when $m=0.7$ and $t=t_0=1$ [see Eq.~\eqref{eq:Chern-Hamiltonian}]. The quantity ${\rm X}(p,q)$ in each subfigure denotes the location $(p,q)$ of the input point, which is fixed on the ${\rm X}={\rm B}$ (top) and A (bottom) sublattices. The output point is then on the A and B sublattices, respectively, which we sweep over the entire system. The sharp peak of the impedance near the edges of the circuit (irrespective of the choice of the input point) shows the topological nature of the Chern insulator.
}~\label{fig:Chernspatial}
\end{figure}

\begin{figure}[t!]
\includegraphics[width=0.95\linewidth]{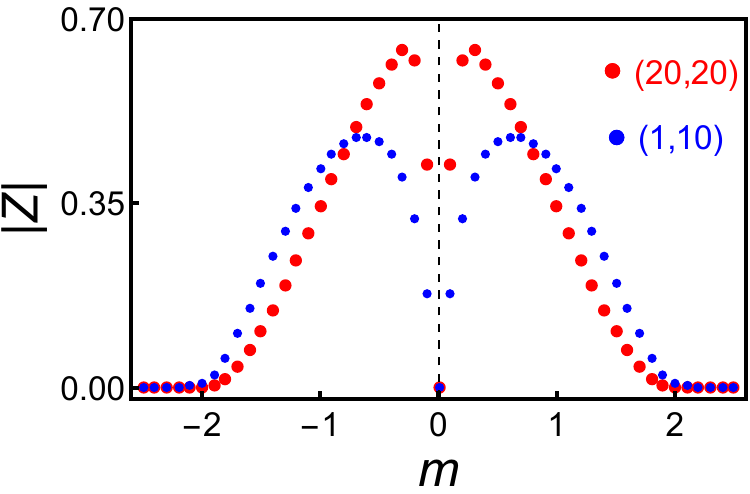}
\caption{The global phase diagram of $H_{\rm CI}$ as a function of the parameter $m$ [see Eq.~(\ref{eq:Chern-Hamiltonian})], obtained by computing the on-resonance impedance ($Z$), with a fixed input point on the A node at $(1,1)$ and for two choices (quoted in the figure) of the output point on the B nodes, residing on the edges of the circuit with linear dimension $\ell=20$ in each direction, when $t=t_0=1$. The magnitude of $Z$ shows that the system is a Chern insulator with finite $Z$ for $-2<m<0$ and $0<m<2$, with a bandgap closing at $m=0$, where $Z$ (almost) vanishes. It also shows the existence of trivial insulators for $m>2$ and $m<-2$, where as well $Z$ (almost) vanishes.
}~\label{fig:Chernphasediagram}
\end{figure}

We now consider the circuit realization of the 2D Chern insulator. The corresponding Hamiltonian in momentum space is given by
\begin{eqnarray}~\label{eq:Chern-Hamiltonian}
H_{\rm CI} &=& t \left[ \sin(k_x) \; \tau_3 + \sin(k_y) \; \tau_2 \right] \nonumber \\
&+& \left\{ m-t_0 \left[ \cos(k_x)+\cos(k_y) \right] \right\} \; \tau_1.
\end{eqnarray}
Pauli matrices $\{ \tau_\mu \}$ with $\mu=0, \cdots, 3$ act on two orbitals in a unit cell. While $t$ and $t_0$ are the inter-unit cell hopping parameters, $m$ denotes the intra-unit cell hopping. This model breaks the time-reversal symmetry. The system is in the topological regime when $-2<m/t_0<2$, describing the Chern insulator with 1D edge mode. By contrast, the system describes trivial insulators when $|m/t_0|>2$. The band gap closing between the Chern and trivial insulators occurs at $m/t_0=\pm 2$. On the other hand, the transition between two distinct Chern insulators takes place at $m/t_0=0$, also through a band gap closing. We note that the system possesses an \emph{antiunitary} particle-hole symmetry represented by the operator $A=\tau_3 {\mathcal K}$ which anticommutes with the Hamiltonian of the system $H_{\rm CI}$, where ${\mathcal K}$ is the complex conjugation. Consequently, the zero-energy edge mode of the Chern insulator is an eigenstate of the antiunitary operator $A$. Since the unitary part of $A$ is $\tau_3$, the edge modes reside on either one of two orbitals. Previously a Chern insulator has been constructed in a brickwall or honeycomb topolectric circuit~\cite{hofmann-prl2019}. Here we realize the Chern insulator in a square topolectric circuit [see Eq.~\eqref{eq:Chern-Hamiltonian} and Fig.~\ref{Fig:Chern-lat}] and in addition to probing its topological edge mode [see Fig.~\ref{fig:Chernspatial}], we also construct its phase diagram [see Fig.~\ref{fig:Chernphasediagram}].

The tight-binding Hamiltonian in Eq.~\eqref{eq:Chern-Hamiltonian}, according to the previously discussed correspondence between the hopping amplitudes and the circuit elements, is realized on the circuit consisting of the components shown in Fig.~\ref{Fig:Chern-lat}. These elements are then connected so that the 2D circuit network features a two-node (A and B) unit cell (the dashed box in Fig.~\ref{Fig:Chern-lat}). Since the lattice model $H_{\rm CI}$ involves both real and imaginary hopping elements in the real space, each node possesses four subnodes. Then the subnodes from the nearest-neighbor nodes are connected by capacitors with appropriate capacitance following the general prescription discussed in Sec.~\ref{subsubsec:4nodes}, see also Fig.~\ref{Fig:4node}. All four subnodes of A and B nodes are grounded with inductor of inductance $L$, so that the resonance frequency of the Chern circuit is $\omega_R =1/\sqrt{L m}$. The numerical computations of the impedance are performed for $t=t_0=1$.

\begin{figure*}[t!]
\subfigure[]{\includegraphics[width=0.38\linewidth]{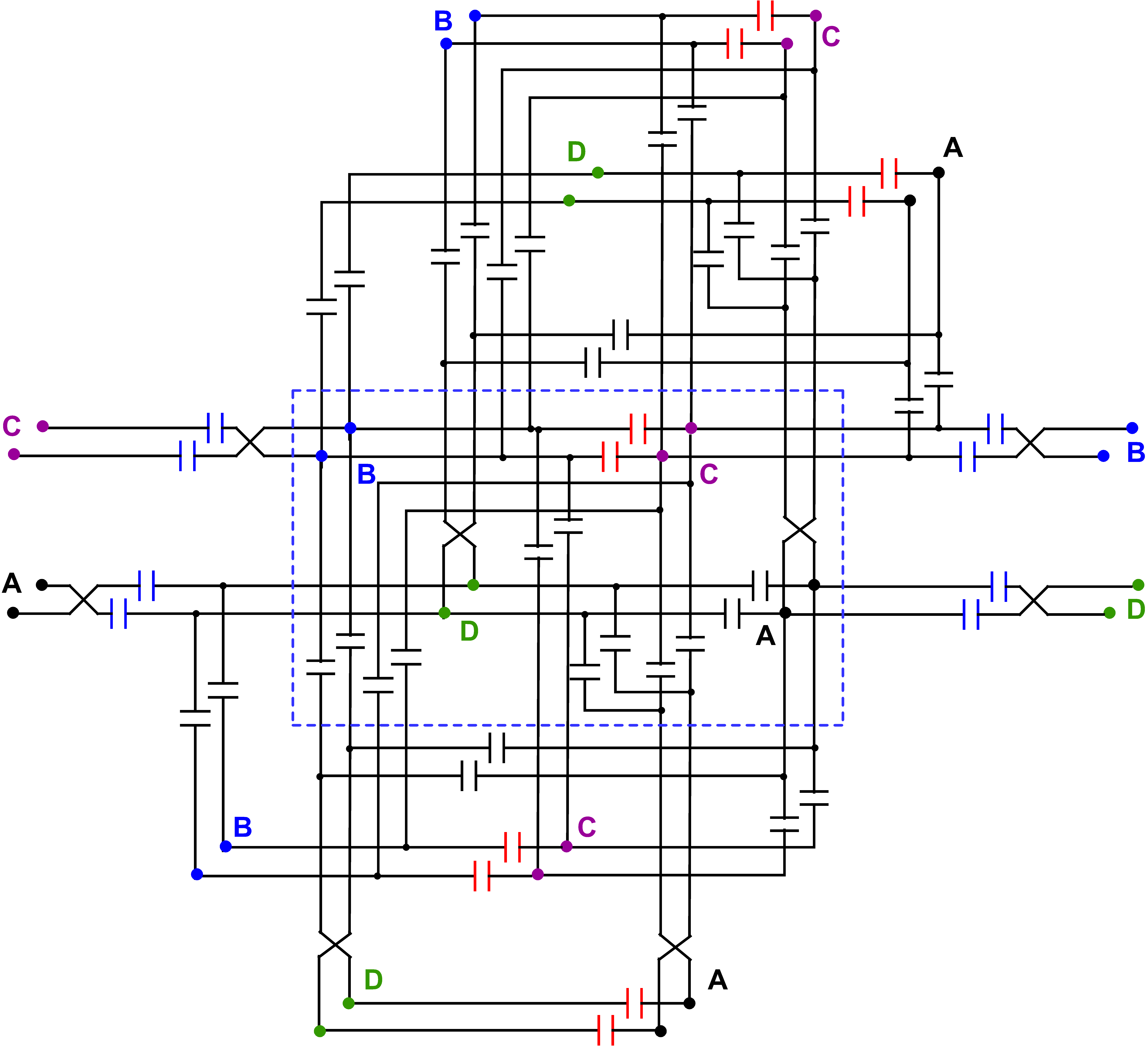}}
\subfigure[]{\includegraphics[width=0.265\linewidth]{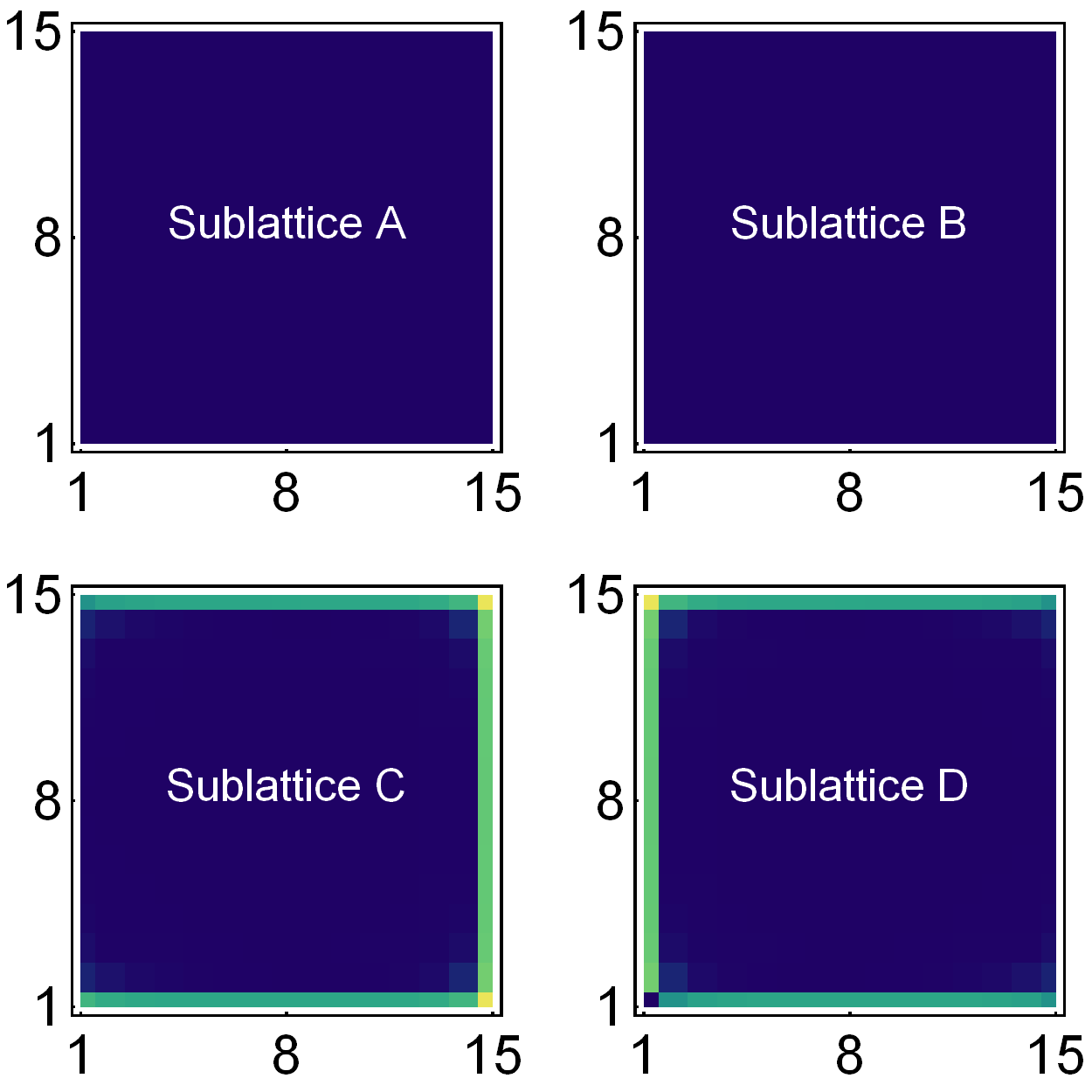}}
\subfigure[]{\includegraphics[width=0.315\linewidth]{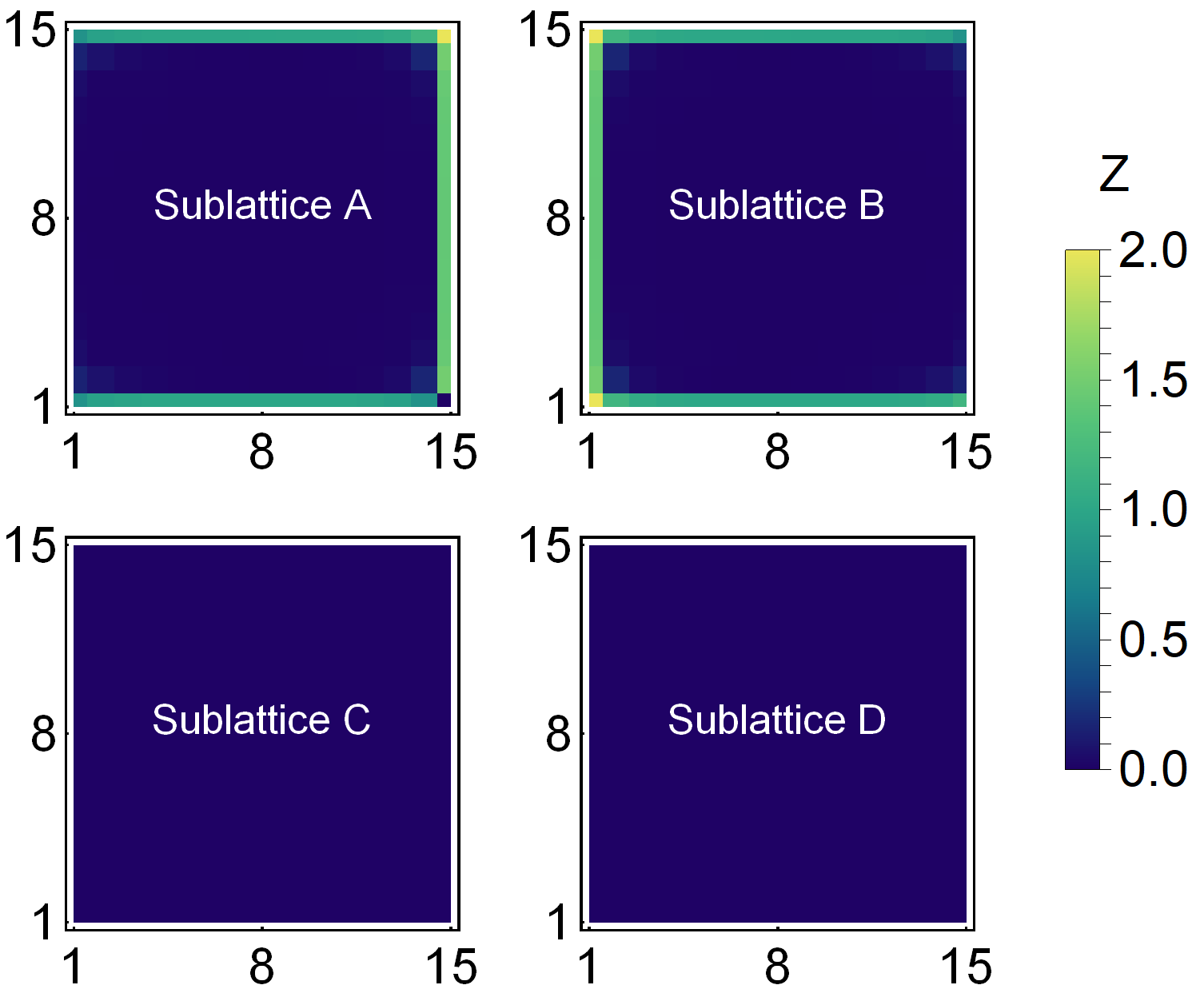}}
\caption{Quantum spin Hall effect in a topolectric circuit. (a) The circuit realization of the quantum spin Hall insulator (QSHI) model [see Eq.~(\ref{eq:QSHImodel})], when $t=t_0$. The capacitances of the black, blue and red capacitors are $t/2$, $t$ and $m$, respectively. Big dots are the subnodes of a given node (color coded), representing the orbitals A, B, C, and D of the lattice model. The small black dots denote the connection between two lines. All the subnodes are connected to the ground via an inductor of inductance $L$ (not shown explicitly). The dashed box encircles four nodes or orbitals, constituting the unit cell of the corresponding lattice model. (b) Spatial distribution of the on-resonance impedance ($Z$) on the entire circuit for $t=t_0=m=1$, with the B node at $(1,1)$ as the fixed input point. The output point scans the entire system on four individual sublattices or nodes separately (mentioned in the figures). (c) Same as (b), but with the C node at $(1,15)$ as the input point. The sublattice (or equivalently spin) selective sharp localization of impedance at the edges of the circuit manifests QSHI in the topolectric circuit.
}~\label{fig:QSHEsummary}
\end{figure*}

To capture the topological edge modes in a Chern topolectric circuit, we choose either an A or B node from one of the four corners of the circuit as the input point and numerically compute the impedance with a B or A node as the output point. For a fixed input A or B node, we scan all the B or A nodes in the system to capture the spatial resolution of the on-resonance impedance. This measurement is then repeated for different corner choices of the A or B nodes. The results are summarized in Fig.~\ref{fig:Chernspatial}. We find that only when $m<2$ the impedance increases sharply at the edges of the circuit, signaling the appearance of the boundary modes, which is consistent with the system becoming a Chern insulator. Also when we scan the impedance over a wide range of frequency, it shows an on-resonance ($\omega=\omega_R$) \emph{peak} only when the system is in the topological regime. On the other hand, in the trivial phases the on-resonance impedance shows a \emph{dip}. These features are qualitatively similar to the ones we previously reported for the SSH chain, see Figs.~\ref{Fig:SSHResults}(c) and \ref{Fig:SSHResults}(d). Hence, we do not display these results in the paper.

Finally, we construct the global phase diagram of the model $H_{\rm CI}$ from the scaling of the on-resonance impedance ($Z$) as a function of $m$, see Fig.~\ref{fig:Chernphasediagram}. While in the Chern insulator phases $Z$ is finite, it (almost) vanishes in the trivial insulating phases. Finally, at the transition point between two Chern insulators ($m=0$) the impedance $Z$ (almost) vanishes. Therefore, measurement of the on-resonance impedance in topolectric circuit can be instrumental not only to identify gapless topological modes but also to construct the global phase diagram of various topological models, and the topological phase transitions therein.

\subsection{Quantum spin Hall insulator (QSHI)}~\label{subsec:QSHI}

Next we focus on the lattice model of QSHI and demonstrate its realization on topolectric circuits. The corresponding tight-binding model takes the following form in the momentum space
\begin{eqnarray}~\label{eq:QSHImodel}
H_{\rm QSHI}&=& t \left[ \sin(k_x) \; \Gamma_1 + \sin(k_y) \; \Gamma_2 \right] \nonumber \\
&+& \left\{ m-t_0 \left[ \cos(k_x)+\cos(k_y) \right] \right\} \; \Gamma_3,
\end{eqnarray}
where ${\boldsymbol \Gamma}$ are mutually anticommuting four-component Hermitian matrices. When the $\Gamma$ matrices belong to the representation
\begin{eqnarray}
\Gamma_1 &=& \sigma_3 \otimes \tau_1, \; \Gamma_2=\sigma_0 \otimes \tau_2, \Gamma_3=\sigma_0 \otimes \tau_3, \nonumber \\
\Gamma_4 &=& \sigma_1 \otimes \tau_1, \; \Gamma_5 = \sigma_2 \otimes \tau_1,
\end{eqnarray}
one can identify two sets of Pauli matrices $\{ \sigma_\mu \}$ and $\{ \tau_\mu \}$ operating on the spin ($\uparrow$ and $\downarrow$) and sublattice or orbital (A and B) indices, respectively, with $\mu=0, \cdots, 3$. The above model is in the topological regime for $|m/t_0|<2$, where it describes a QSHI supporting counter-propagating 1D edge modes for opposite spin projections. However, topology of the above quadratic Hamiltonian is insensitive to the representation of the $\Gamma$ matrices. We judiciously commit to the following representation
\begin{eqnarray}~\label{eq:gammarepresentationQSHI}
\Gamma_1 &=& \sigma_1 \otimes \tau_2, \; \Gamma_2= \sigma_2 \otimes \tau_0, \; \Gamma_3= \sigma_1 \otimes \tau_1, \nonumber \\
\Gamma_4 &=& \sigma_1 \otimes \tau_3, \; \Gamma_5 = \sigma_3 \otimes \tau_0,
\end{eqnarray}
such that in the real space all the hopping elements associated with $H_{\rm QSHI}$ are completely \emph{real}. This is so, because in this representation purely imaginary (real) $\Gamma$ matrices multiply the sine (cosine and real constant) functions. In the following sections, we will subscribe to such representation whenever possible. In this representation, two sets of Pauli matrices $\{ \sigma_\mu\}$ and $\{ \tau_\mu \}$ respectively operate on the orbitals or sublattices (C,D) and (A,B). Also note that $H_{\rm QSHI}$ anticommutes with the unitary operator $\Gamma_5$, which in turn generates the spectral or particle-hole symmetry of the system. Consequently, the 1D edge modes at zero energy are eigenstates of $\Gamma_5$, and reside either on (A,B) or (C,D) sublattices.

\begin{figure*}[t!]
\subfigure[]{\includegraphics[width=0.24\linewidth]{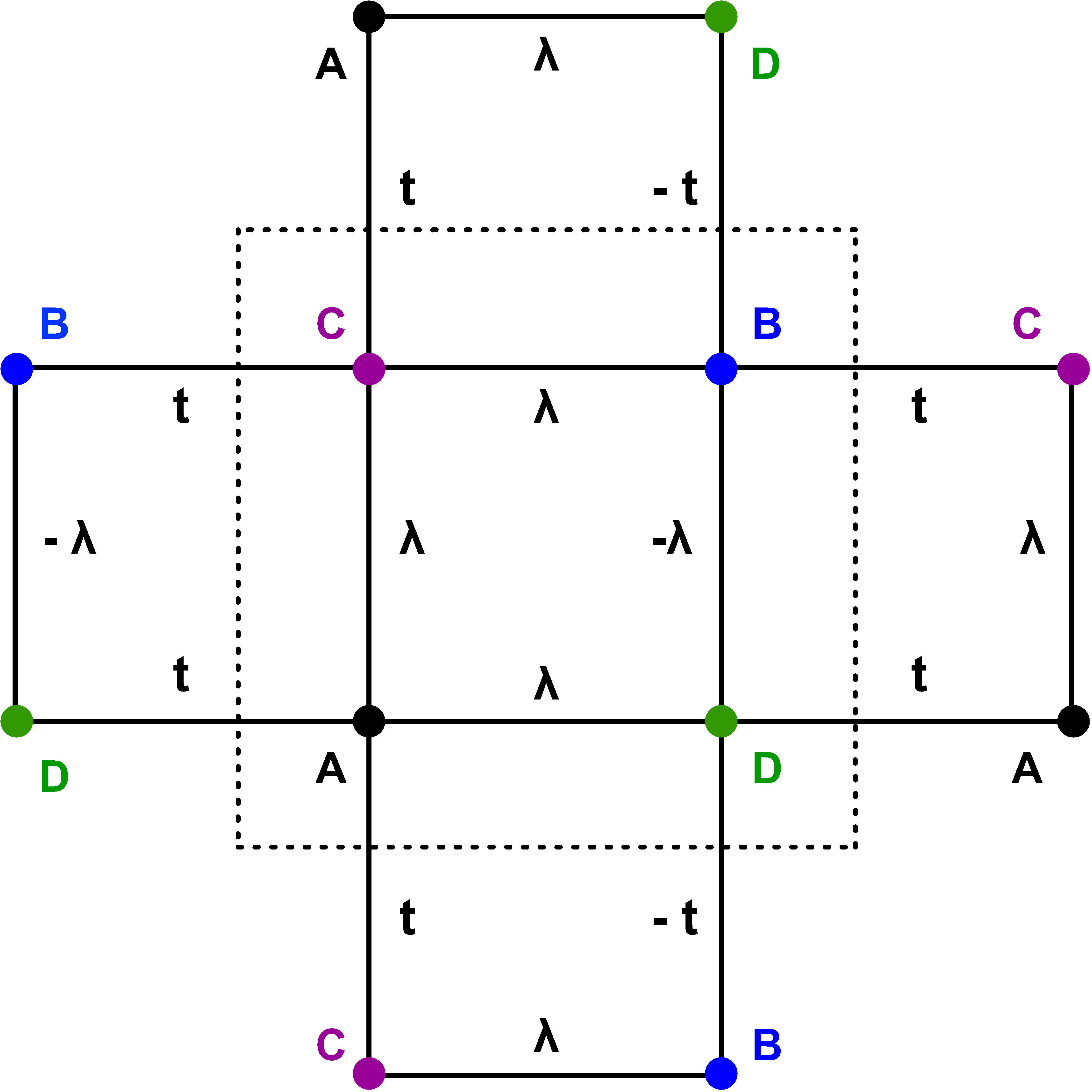}}
\subfigure[]{\includegraphics[width=0.24\linewidth]{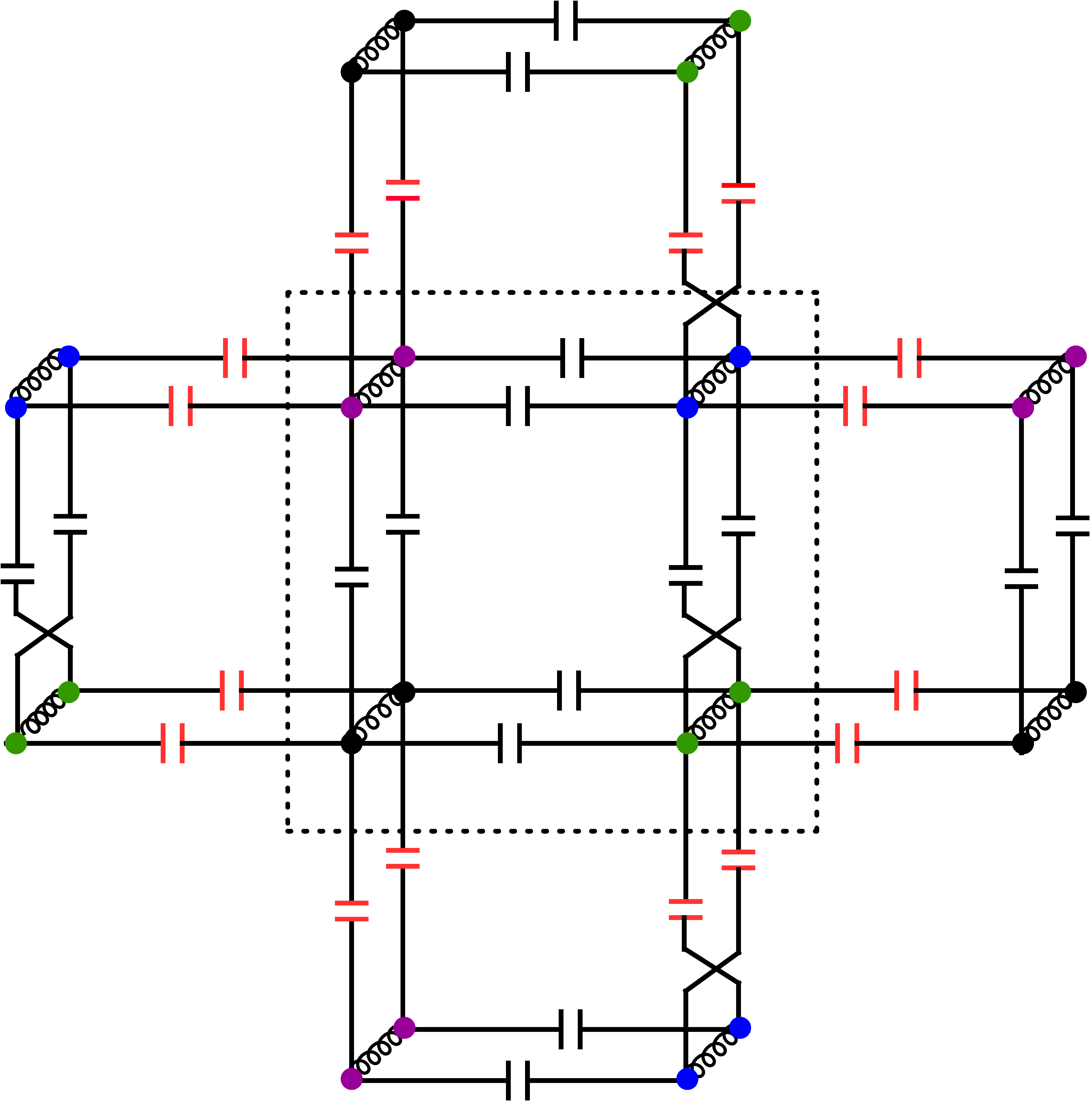}}
\subfigure[]{\includegraphics[width=0.23\linewidth]{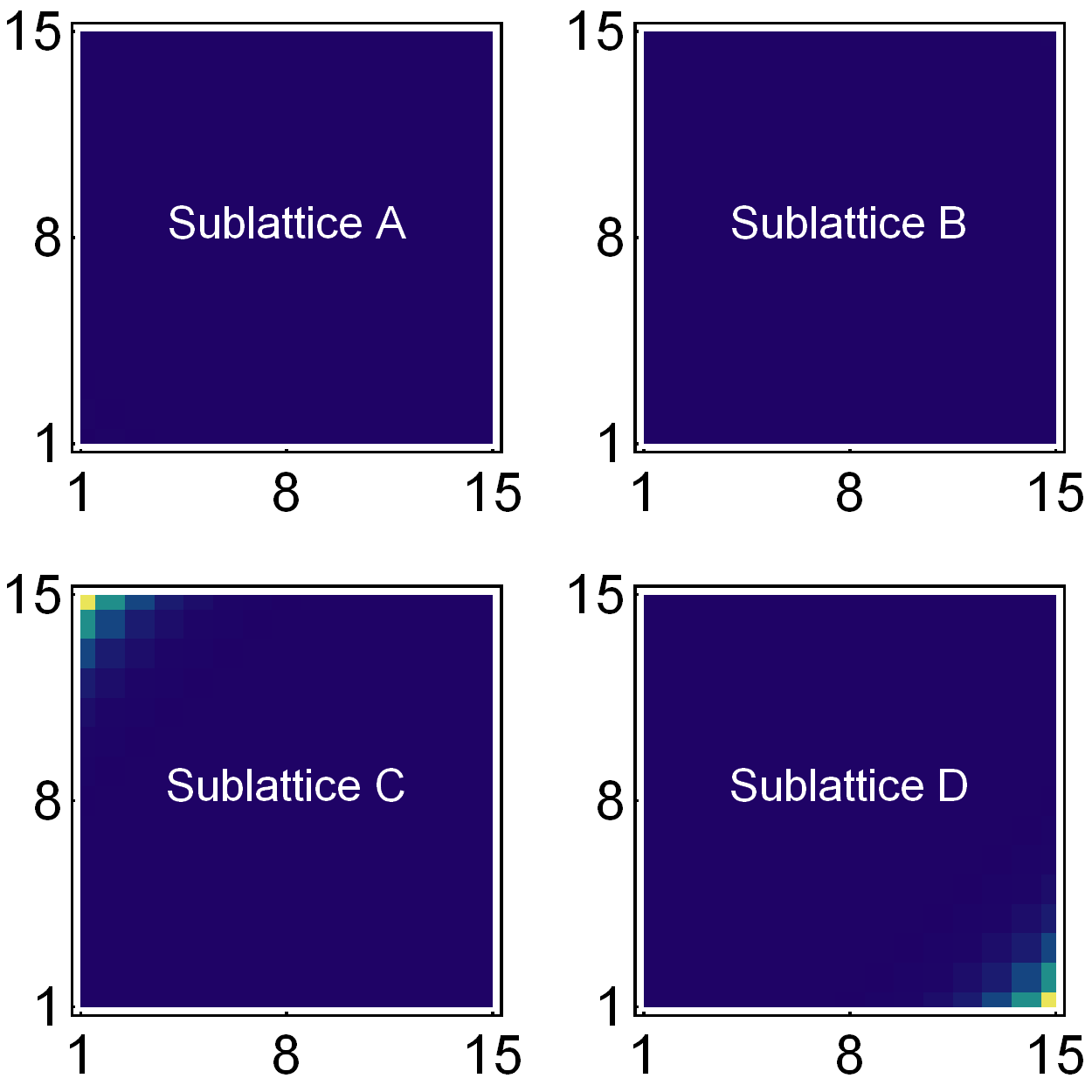}}
\subfigure[]{\includegraphics[width=0.27\linewidth]{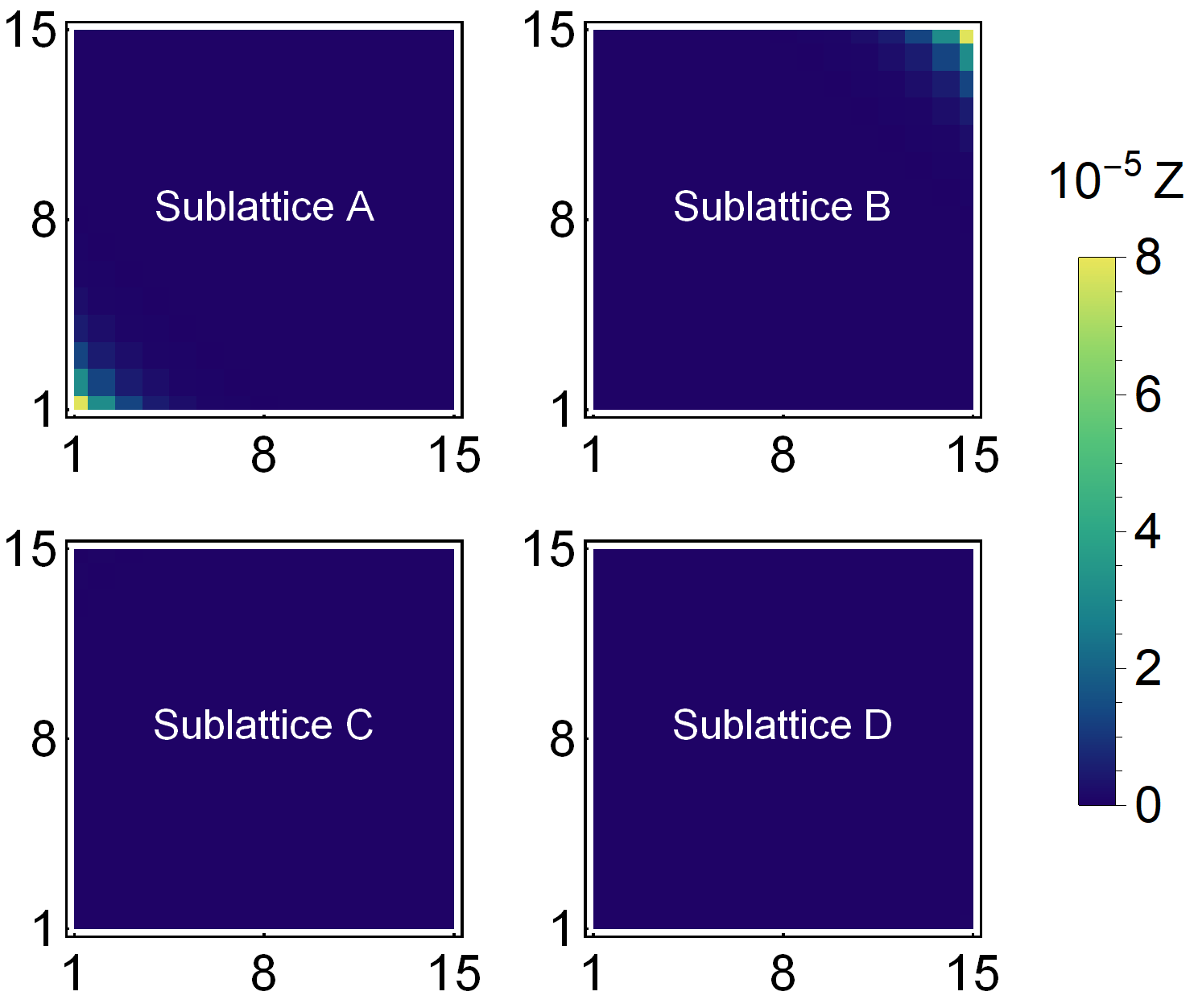}}
\caption{Higher-order topological insulator in a topolectric circuit. (a) A schematic representation of the BBH Hamiltonian. The (real) hopping parameters are $\lambda$ and $t$. This model can feature HOTI state that supports four corner localized zero energy modes. (b) Circuit realization of the BBH model. Each node (color coded) contains two subnodes, and all the subnodes are connected to the ground via an inductor of inductance $L$ (not shown explicitly). The capacitances of the black and red capacitors are respectively $\lambda$ and $t$. The negative hopping amplitudes are implemented via one-shift couplings discussed in Sec.~\ref{Sec:General}, and are denoted by the crossed lines. The dashed boxes in (a) and (b) denote the unit cell of the BBH model. (c) Spatial distribution of the on-resonance impedance ($Z$) for capacitances $t=1$ and $\lambda=0.4$ (ensuring the topological regime of the BBH model), with the A node at $(1,1)$ as the fixed input point. The output point scans the entire system on individual sublattices separately (mentioned in the figures). (d) Same as (c), but with the C node at $(1,15)$ as the fixed input point. Node selective corner localization of impedance shows the circuit realization of a 2D HOTI and the sublattice symmetry of the corner modes.
}
~\label{fig:BBHSummary2D}
\end{figure*}

By virtue of committing to such representation of the $\Gamma$ matrices, we can engineer $H_{\rm QSHI}$ in an electric circuit by supplementing each sublattice or node by two subnodes (see Sec.~\ref{subsubsec:2nodes} and  Fig.~\ref{Fig:2node}). A concrete circuit realization of $H_{\rm QSHI}$ is shown in Fig.~\ref{fig:QSHEsummary}(a). By numerically computing the on-resonance ($\omega=\omega_R=1/\sqrt{L m}$) impedance, we confirm that only in the topological regime impedance is sharply peaked around the edges of the circuit. Furthermore, by performing node or sublattice resolved computation of the on-resonance impedance, we find that the edge mode is either localized on the sublattices (C,D) [see Fig.~\ref{fig:QSHEsummary}(b)] or (A,B) [see Fig.~\ref{fig:QSHEsummary}(c)]. Later we will show (in Sec.~\ref{subsec:discretesymmetrybreaking}) how one can break the discrete four-fold ($C_4$) rotational symmetry of this circuit to realize a higher-order topological (HOT) insulator, supporting corner impedance. Prior to that we first discuss some standard models for HOT phases in two and three dimensions, and their realizations in topolectric circuits.

\section{Higher-order topolectric circuits}~\label{Sec:HOTI}

Recently, the notion of the bulk-boundary correspondence has been generalized to include low-dimensional topological modes that reside on boundaries with integer codimension $d_c>1$, giving rise to the notion of higher-order topological (HOT) phases~\cite{BBH-Science, BBH-PRB, Langbehn-PRL2017, Schindler-SciAdv2018, Khalaf-PRB2018, Miert-PRB2018, Hsu-PRL2018, Liu-Hughes-PRB2018, Trifunovic-PRX2019, calugaru-juricic-roy, Stern-PRL2019, Varjas-PRL2019, agarwala-PRR2020, ca-li-prb2020, zeng-PRB-2020, DasSarma-arxiv2019, ghorashi-arxiv2020, jiang-arxiv2020}~\cite{RuiChen-PRL}. The well studied examples of the low-dimensional boundary modes are the corner (with $d_c=d$) and hinge (with $d_c=d-1$) modes; namely, an $n$th order topological phase supports boundary modes of codimension $d_c=n$. So far, we have discussed first-order topological insulators and their realizations on topolectric circuits. Next we present realizations of HOT insulators on topolectric circuits. In Sec.~\ref{subsec:3D-QDSM}, we will discuss a circuit realization of a HOT semimetal.

\subsection{Two-dimensional HOT insulator}~\label{subsec:BBH2D}

We start by considering the example of 2D Benalcazar-Bernevig-Hughes (BBH) model~\cite{BBH-Science}, a second order topological insulator featuring four corner modes with $d_c=2$. The model is defined on a square lattice with four sublattices (A, B, C, and D) per unit cell. The lattice Hamiltonian reads as $H^{\rm 2D}_{\rm BBH}=h_x+h_y$, where
\allowdisplaybreaks[4]
\begin{align}
h_x&= \left[ \lambda + t \cos(k_x) \right] \; \Gamma_1 + t \sin(k_x) \; \Gamma_2, \nonumber \\
h_y&= \left[ \lambda + t \cos(k_y) \right] \; \Gamma_3 + t \sin(k_y) \; \Gamma_4.
\end{align}
The mutually anticommuting $\Gamma$ matrices ($\{\Gamma_i,\Gamma_j\}=2\delta_{ij}$) acting on the sublattice space can be chosen as
\allowdisplaybreaks[4]
\begin{align}
\Gamma_1&=\sigma_1\otimes\tau_1,\; \Gamma_2=\sigma_1\otimes\tau_2,\; \Gamma_3=\sigma_1\otimes\tau_3,\nonumber\\
\Gamma_4&=\sigma_2\otimes\tau_0,\; \Gamma_5=\sigma_3\otimes\tau_0.
\end{align}
Namely, the sets of Pauli matrices $\{ \tau_\mu \}$ and $\{ \sigma_\mu \}$ respectively operate on the sublattice space spanned by (A,B) and (C,D), with $\mu=0, \cdots, 3$. Notice that $h_x$ ($h_y$) assumes the form of the SSH model along the $x$ ($y$) direction [see Sec.~\ref{subsec:SSH}]. Therefore, $h_x$ and $h_y$ individually support a string of zero-energy end modes along the $y$ and $x$ directions, respectively, when $|\lambda/t|<1$. However, $\{ h_x,h_y \}=0$, i.e., $h_x$ acts as a mass for the end modes of $h_y$ and vice-versa. Consequently, only the four corner modes survive in the spectra of $H^{\rm 2D}_{\rm BBH}$ and we realize a 2D second-order topological insulator.

This specific representation of the $\Gamma$ matrices is convenient for our purposes as it features three real and two imaginary $\Gamma$ matrices. It allows us to construct the BBH model with all the hopping elements being real [see Fig.~\ref{fig:BBHSummary2D}(a)] and therefore the nodes (representing sublattices) with only two subnodes are sufficient to be implemented in the circuit realization of the BBH model, as shown in Fig.~\ref{fig:BBHSummary2D}(b). Each subnode is grounded by an inductor of inductance $L$. Consequently, the resonance frequency of the circuit is $\omega_R=1/\sqrt{L (\lambda+t)}$. Furthermore, note that the matrix $\Gamma_5=\sigma_3\otimes\tau_0$ anticommutes with the Hamiltonian $H^{2D}_{\rm BBH}$. Hence, the spectrum is particle-hole symmetric and the zero-energy corner modes are eigenstates of $\Gamma_5$. Therefore, the topological corner modes are localized either on the sublattices A and B or on the sublattices C and D, as we also explicitly demonstrate from the concrete circuit realization of the BBH model.

The spatial distributions of the impedance are shown in Figs.~\ref{fig:BBHSummary2D}(c) and ~\ref{fig:BBHSummary2D}(d) on a circuit with $\ell =15$ sites in each direction. In the topological regime ($|\lambda/t|<1$) of the BBH model, we first fix the input point at (1,1) and on node A. The spatial variation of the on-resonance impedance on four individual nodes over the entire system then shows sharp corner localization on the C and D nodes, see Fig.~\ref{fig:BBHSummary2D}(c). We arrive at similar conclusions by choosing the B node at (1,1) as the input point (results are not shown here explicitly). On the other hand, when the input point is fixed on the C node located at (1,15), the corner localized peaks of impedance appear on nodes A and B, see Fig.~\ref{fig:BBHSummary2D}(d). Similar conclusions are found with the D node at (1,15) as the input point, for which the results are not displayed here. These features of the node (or sublattice) and site resolved impedance confirm the realization of a 2D HOT insulator on topolectric circuit and the sublattice or node selection of the corresponding corner impedance. We note that the 2D topolectric BBH circuit has already been realized in Ref.~\cite{imhof-natphys2018}. Even though our explicit circuit construction is distinct from the one engineered in Ref.~\cite{imhof-natphys2018}, here we arrive at qualitatively similar results. On the other hand, the observed sublattice polarization of the on-resonance corner impedance is yet to be demonstrated in experiments.

\subsection{Three-dimensional HOT insulator}~\label{subsec:BBH3D}

\begin{figure*}[t!]
\subfigure[]{\includegraphics[width=0.20\linewidth]{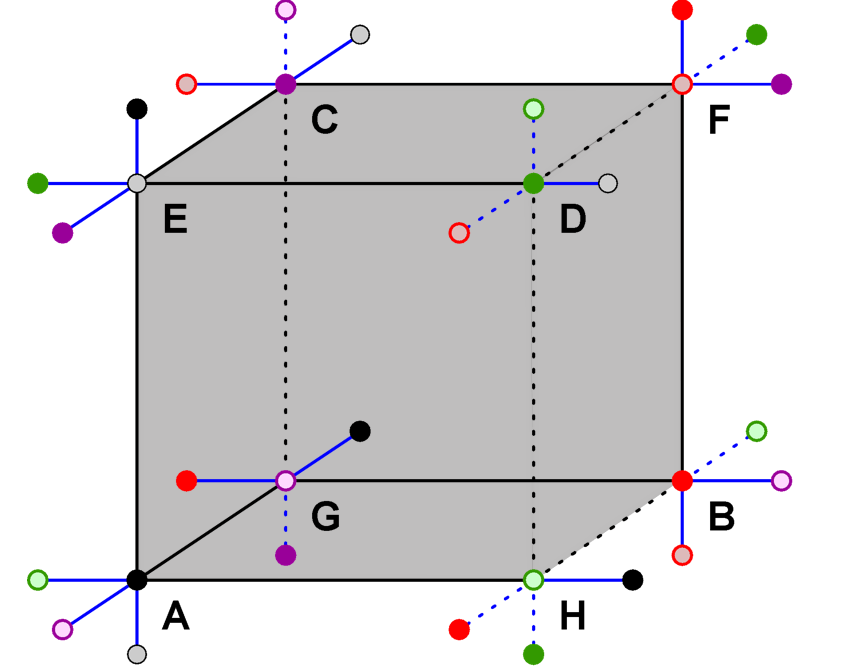}}
\subfigure[]{\includegraphics[width=0.23\linewidth]{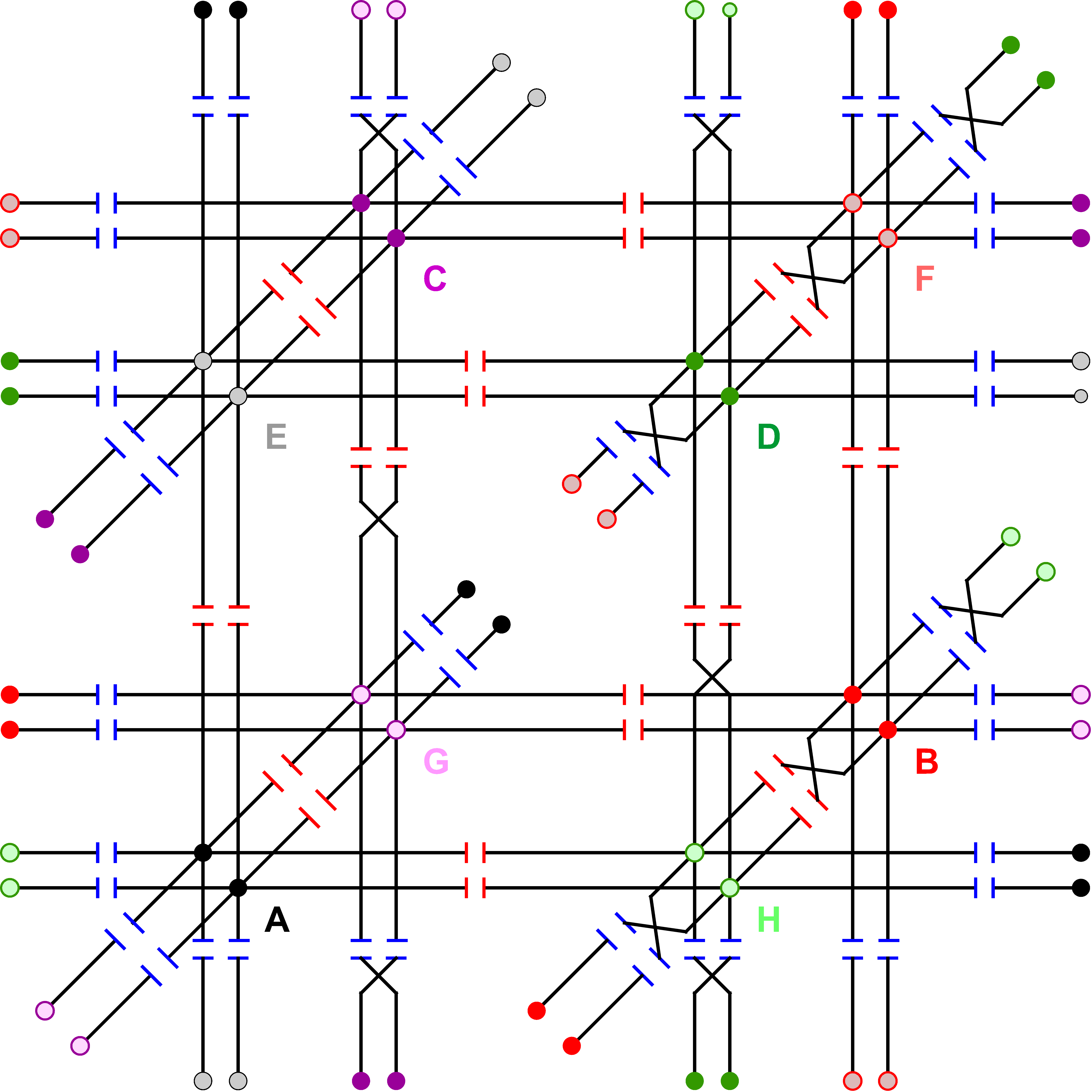}}
\subfigure[]{\includegraphics[width=0.55\linewidth]{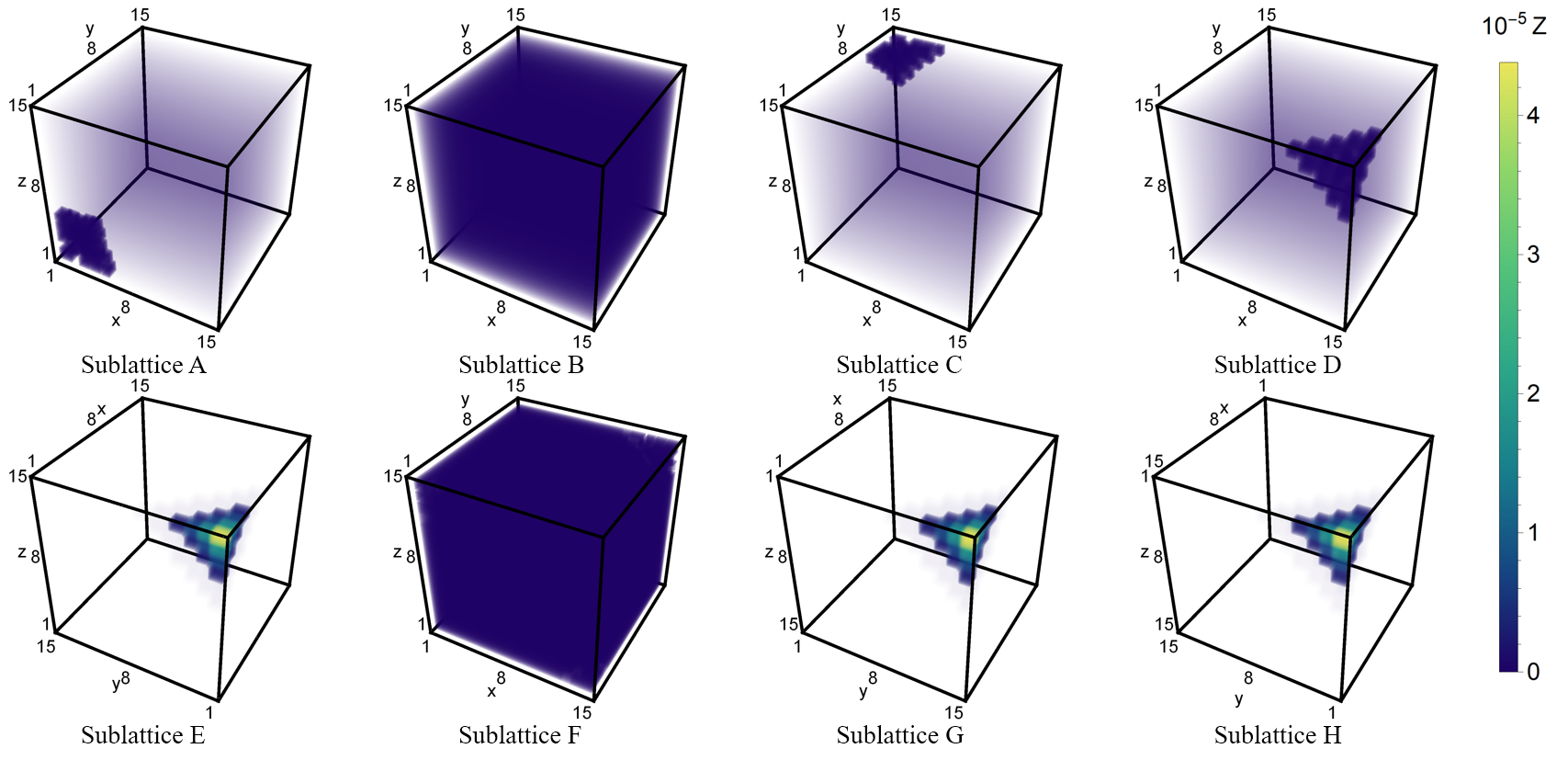}}
\caption{Circuit realization of a 3D higher-order topological insulator (HOTI). (a) A unit cell (the shaded cube) in the lattice construction of  3D HOTI [see Eq.~(\ref{eq:hamil3Dhoti})]. The black (blue) lines represent intra-unit cell (inter-unit cell) hopping with absolute value $\lambda$ ($t$). The solid lines represent positive hopping amplitudes and the dashed lines represent negative ones. The system is topological when $|\lambda/t|<1$. (b) A unit cell in the circuit construction of 3D HOTI, where each node (or sublattice) contains only two subnodes. The red capacitors have capacitance equal to $\lambda$ and the blue capacitors have capacitance equal to $t$. The negative hopping amplitudes [see (a)] are realized through one-shift capacitor coupling between the subnodes from the neareast-neighbor nodes (see Sec.~\ref{subsubsec:2nodes} and Fig.~\ref{Fig:2node}). (c) On-resonance impedances of the 3D HOTI circuit of linear dimension (number of sites) $\ell=15$ in each direction for $\lambda=0.4$ and $t=1$. Throughout the input point is fixed at $(1,1,1)$ and on the node A. The output point scans all eight nodes (or sublattices) over the entire system, as labeled in the caption. One can see a sublattice selection of the sharp corner localized impedance, which confirms realization of 3D HOTI on topolectric circuit. We rotate the cubes for E, G, and H nodes (or sublattices) for the best visualization of sharp corner impedance.
}~\label{fig:3DHOTI-summary}
\end{figure*}

Similar to the 2D HOTI, one can construct a BBH Hamiltonian for its 3D counterpart~\cite{BBH-Science}, yielding a thrid-order topological insulator that supports eight corner modes with $d_c=3$. Each unit cell then contains \emph{eight} sublattices (A, $\cdots$, H), see Fig.~\ref{fig:3DHOTI-summary}, and in the Fourier space the corresponding Hamiltonian reads as
\begin{equation}~\label{eq:hamil3Dhoti}
H^{\rm 3D}_{\rm BBH}=h_x+h_y+h_z,
\end{equation}
where
\allowdisplaybreaks[4]
\begin{eqnarray}
h_x &=& \left[ \lambda + t \cos (k_x) \right] \; \Gamma_1 + t \sin(k_x) \; \Gamma_2, \nonumber \\
h_y &=& \left[ \lambda + t \cos (k_y) \right] \; \Gamma_3 + t \sin(k_y) \; \Gamma_4, \\
h_z &=& \left[ \lambda + t \cos (k_z) \right] \; \Gamma_5 + t \sin(k_z) \; \Gamma_6. \nonumber
\end{eqnarray}
Here ${\boldsymbol \Gamma}$ are mutually anticommuting eight-dimensional Hermitian matrices that satisfy the Clifford algebra $\{ \Gamma_i, \Gamma_j \}=2\delta_{ij}$. Even though topology of $H^{\rm 3D}_{\rm BBH}$ is insensitive to the representation of the $\Gamma$ matrices, for the sake of convenience (about which more in a moment), we choose
\allowdisplaybreaks[4]
\begin{align}
\Gamma_1&=\Sigma_1\otimes\sigma_1\otimes\tau_1, \; \Gamma_2=\Sigma_1\otimes\sigma_1\otimes\tau_2, \nonumber \\
\Gamma_3&=\Sigma_1\otimes\sigma_1\otimes\tau_3, \; \Gamma_4=\Sigma_1\otimes\sigma_2\otimes\tau_0, \nonumber \\
\Gamma_5&=\Sigma_1\otimes\sigma_3\otimes\tau_0, \; \Gamma_6=\Sigma_2\otimes \sigma_0\otimes \tau_0,
\end{align}
where $\{ \Sigma_\mu \}$,$\{ \sigma_\mu \}$ and $\{ \tau_\mu \}$ are three independent sets of Pauli matrices. Since the maximal number of mutually anticommuting eight-dimensional Hermitian matrices is \emph{seven}, we can always find an Hermitian matrix, namely $\Gamma_7$, that satisfies $\{ \Gamma_7, \Gamma_j\}=2 \delta_{7j}$ for $j=1,\cdots, 7$, and therefore anticommutes with $H^{\rm 3D}_{\rm BBH}$. The $\Gamma_7$ matrix generates a unitary particle-hole symmetry of $H^{\rm 3D}_{\rm BBH}$, and in the announced representation $\Gamma_7=\Sigma_3\otimes \sigma_0\otimes \tau_0$.

The above model is in the topological regime for $|\lambda/t|<1$ for which it supports eight corner localized zero energy modes. This is so because all three components of $H^{\rm 3D}_{\rm BBH}$, namely $h_x$, $h_y$ and $h_z$ assume the form of the 1D SSH model. Consequently, $h_x$ supports a collection of end-point zero-energy modes, localized on the $yz$ planes, and similarly $h_y$ on the $zx$ planes and $h_z$ on the $xy$ planes. However, these three components of $H^{\rm 3D}_{\rm BBH}$ mutually anticommute with each other. Consequently, only the zero modes at eight corners survive, where $xy$, $yz$ and $zx$ planes meet, yielding the corner modes. Due to the particle-hole or spectral symmetry of $H^{\rm 3D}_{\rm BBH}$, the corner modes are eigenstates of $\Gamma_7$. With the above specific choices of the $\Gamma$ matrices, the zero-energy corner states are therefore localized on either the sublattices (A,B,C,D) or (E,F,G,H).

By virtue of the above representation of the $\Gamma$ matrices, all the hopping elements associated with $H^{\rm 3D}_{\rm BBH}$ are purely \emph{real}, see Fig.~\ref{fig:3DHOTI-summary}(a). Hence, this model can be implemented on a topolectric circuit by supplementing each node with only two subnodes, as shown in Fig.~\ref{fig:3DHOTI-summary}(b). Each subnode is grounded by an inductor of inductance $L$, and therefore the resonance frequency of the circuit is $\omega_R=1/\sqrt{L(\lambda+t)}$. The numerical computation of the on-resonance ($\omega=\omega_R$) impedance then shows that in the topological regime (realized on the circuit for $t=1$ and $\lambda=0.4$) it is highly localized at the corners of the cubic circuit, see Fig.~\ref{fig:3DHOTI-summary}(c). Moreover, when we fix the input point for the measurement of impedance on the A node at (1,1,1), the sharp corner localized impedance is realized on the E, G and H nodes, confirming its node (or sublattice) selection, stemming from the particle-hole symmetry of $H^{\rm 3D}_{\rm BBH}$. Experimental realization of 3D BBH model in the topolectric platform has been recently reported in Ref.~\cite{bao-prb2019}, where an enhanced impedance was detected at the corners, but no sublattice-selective measurement was carried out (despite the differences in the explicit circuit construction). Our discussion should therefore motivate future experiments to search for the predicted sublattice selection rule of corner impedance.


\section{Discrete symmetry breaking and antiunitary symmetry}~\label{Sec:Antiunitary}

So far we have shown realizations of various topological phases in one-, two-, and 3D topolectric circuits. In this section, by focusing on the specific example of 2D QSHI, we first demonstrate how one can introduce a discrete four-fold ($C_4$) symmetry breaking Wilson-Dirac mass on a circuit to convert the QSHI into a 2D HOTI, supporting corner localized impedance. Subsequently, we also show how one can exploit an underlying antiunitary spectral symmetry to generalize the HOTI model~\cite{roy-aniunitaryHOTI, Nag-prr2019} in a circuit.

\begin{figure*}[t!]
\subfigure[]{\includegraphics[width=0.380\linewidth]{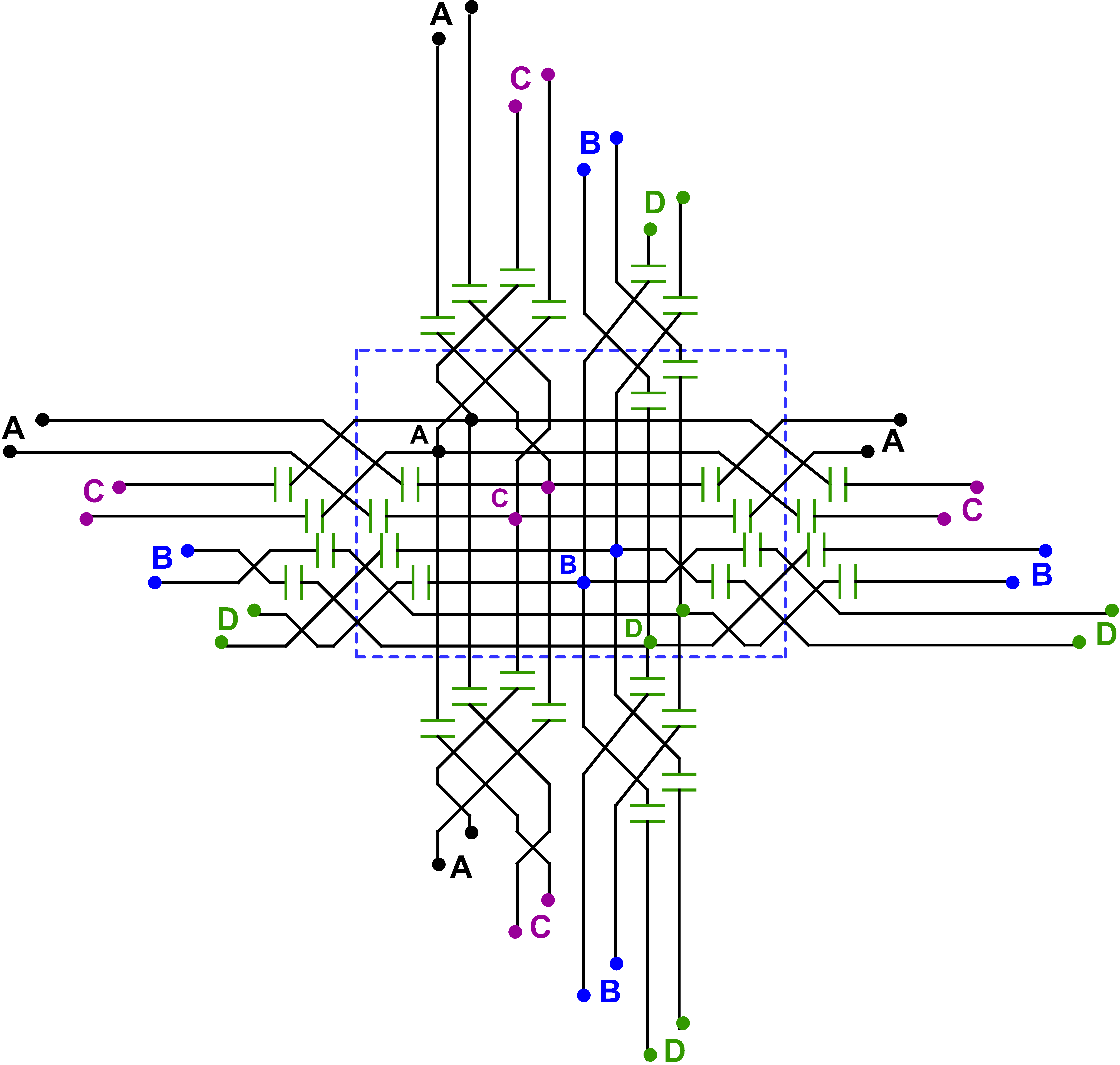}}
\subfigure[]{\includegraphics[width=0.265\linewidth]{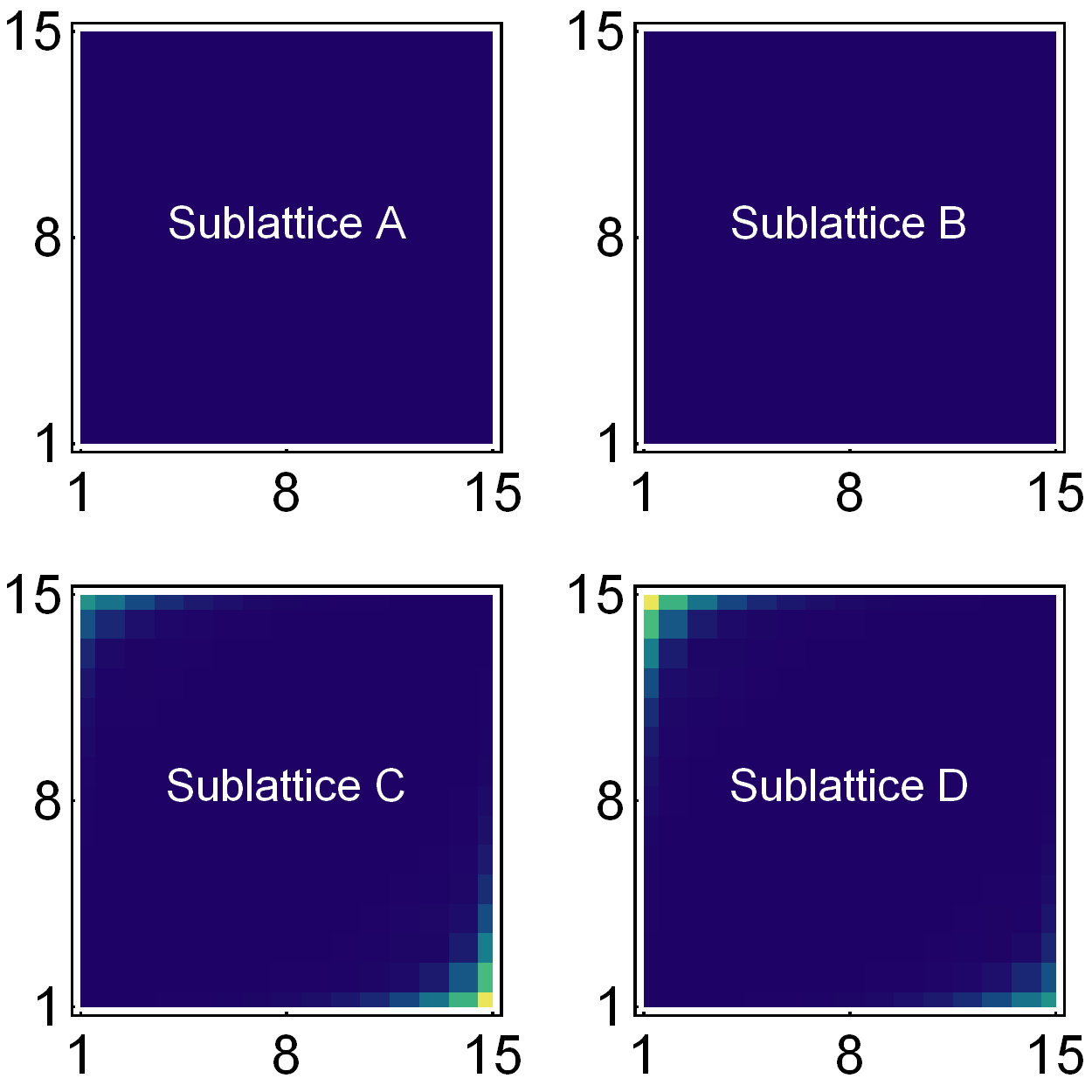}}
\subfigure[]{\includegraphics[width=0.315\linewidth]{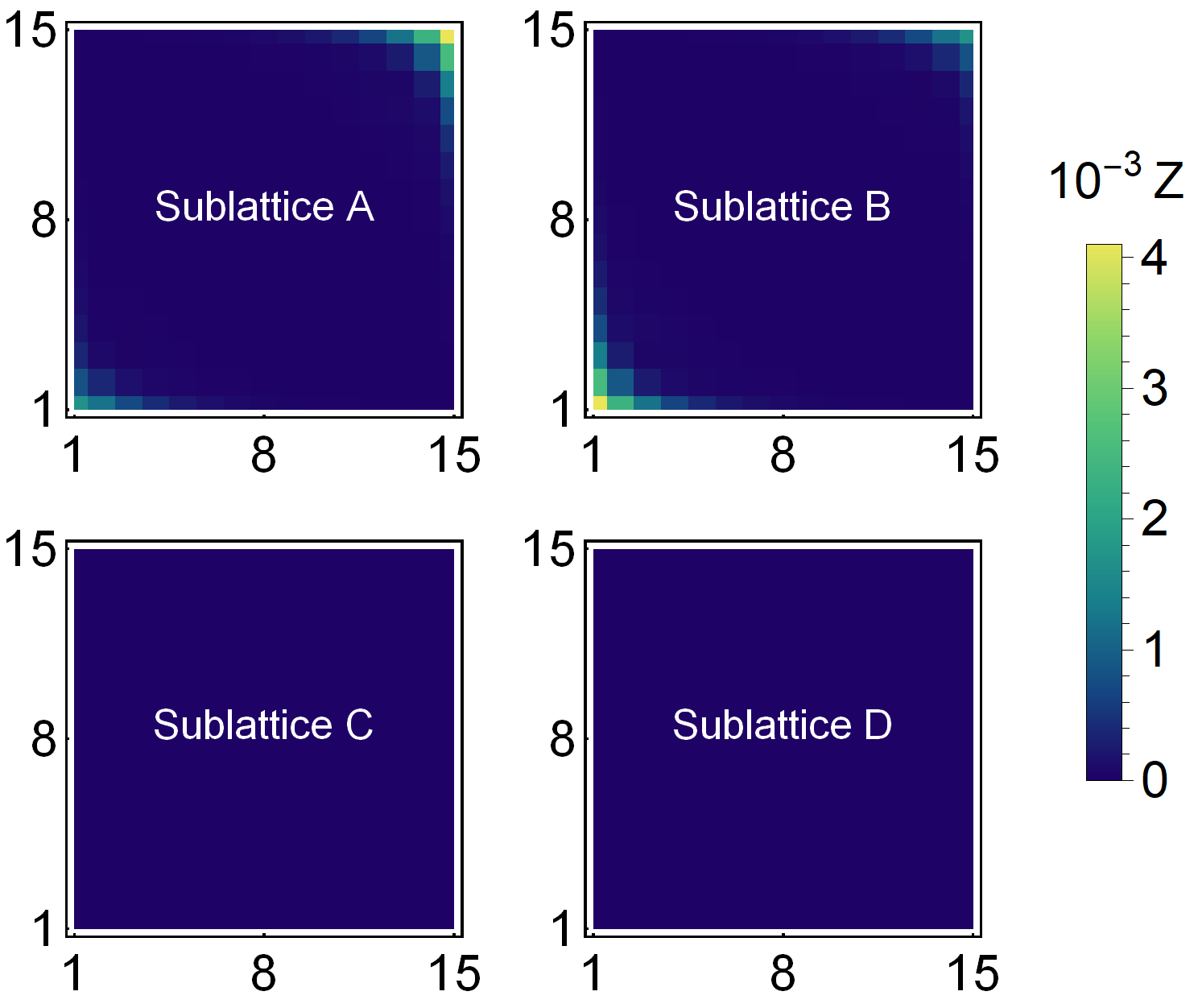}}
\caption{Discrete symmetry breaking in topolectric circuit and realization of 2D HOTI from QSHI. (a) Circuit realization of $H_\Delta$ [see Eq.~(\ref{eq:discreteC4breaking})] that breaks four-fold ($C_4$) rotational symmetry of the QSHI circuit, shown in Fig.~\ref{fig:QSHEsummary}(a). All the capacitors have capacitance $\Delta$, and all the subnodes are grounded via inductors of inductance $L$. (b) Numerical computation of on-resonance impedance with the B node at (1,1) as the input point for $t=t_0=m=1$ and $\Delta=0.5$. The output point scans the entire system and all four individual nodes (or sublattices) separately. (c) Same as (b), but with the C node at (1,15) as the input point. Results show node or sublattice selective sharp corner localization of the on-resonance impedance due to the $C_4$ symmetry breaking by a circuit analog of the Wilson-Dirac mass $H_\Delta$, ultimately yielding a 2D HOTI phase.
}~\label{fig:2DQSHI-HOTI}
\end{figure*}

\subsection{Discrete symmetry breaking: QSHI to HOTI}~\label{subsec:discretesymmetrybreaking}

Recall that a QSHI supports two counter-propagating edge modes for opposite spin projections. If we now add a term, namely
\begin{equation}~\label{eq:discreteC4breaking}
H_\Delta= \Delta \; \left[ \cos(k_x)-\cos(k_y) \right] \; \Gamma_4,
\end{equation}
to the Hamiltonian for the QSHI $H_{\rm QSHI}$ [see Eq.~(\ref{eq:QSHImodel})], it gaps out the edge modes, since $\{ H_{\rm QSHI}, H_\Delta  \}=0$. Here the $\Gamma$ matrices follow the representation from Eq.~(\ref{eq:gammarepresentationQSHI}). However, notice that $H_\Delta$ changes sign under the $C_4$ rotation, and as such breaks the four-fold rotational symmetry of $H_{\rm QSHI}$. Therefore, $H_\Delta$ acts as a mass for 1D edge modes of $H_{\rm QSHI}$, with the profile of a domain wall mass that changes sign across each corner of the system. Consequently, the edge modes are only partially gapped, and according to generalized Jackiw-Rebbi mechanism give rise to four sharply corner localized modes~\cite{jackiw-rebbi, calugaru-juricic-roy}. We then realize a 2D HOTI by lifting the $C_4$ symmetry of the system via the Wilson-Dirac mass $H_\Delta$.

We now show how to break such discrete rotational symmetry in a topolectric circuit and implement $H_\Delta$ to realize HOTI from QSHI. Notice that $\Gamma_4$ is a real Hermitian matrix [see Eq.~(\ref{eq:gammarepresentationQSHI})]. Therefore, hopping matrix elements associated with $H_{\Delta}$ in the real space are completely \emph{real}. We can then introduce $H_\Delta$ in a circuit consisting of nodes that are accompanied by only two subnodes, as shown in Fig.~\ref{fig:2DQSHI-HOTI}(a). Note that the total Hamiltonian $H_{\rm QSHI}+H_\Delta$ anticommutes with $\Gamma_5=\sigma_3 \otimes \tau_0$. Consequently, the on-resonance ($\omega=\omega_R=1/\sqrt{Lm}$) impedance shows a node or sublattice selective sharp corner localization, displayed in Figs.~\ref{fig:2DQSHI-HOTI}(b) and ~\ref{fig:2DQSHI-HOTI}(c), confirming the realization of $C_4$ symmetry breaking HOTI in a topolectric circuit.

Note that here we present an alternative realization of HOTI in two dimensions, in comparison to the previous one from the BBH model in Sec.~\ref{subsec:BBH2D}. However, these two seemingly distinct realizations of 2D HOTI are \emph{equivalent}~\cite{roy-aniunitaryHOTI}. The main purpose of the present discussion is to highlight a concrete path to explore the hierarchy of orders for topological states (such as the first and second in this case) in topolectric circuits. Next we discuss a further generalization of 2D HOTI by using its antiunitary spectral or particle-hole symmetry in a topolectric circuit.

\subsection{Antiunitary symmetry protected HOTI in 2D}~\label{subsec:antiunitary}

So far we discussed a plethora of topological phases in different spatial dimensions, in which robust zero-energy boundary modes are protected by the spectral or particle-hole symmetry, typically generated by \emph{unitary} operators. A much less explored situation is when the zero-energy modes are protected by \emph{antiunitary} operator. We have already encountered one such example, the edge mode of the Chern insulator in Sec.~\ref{subsec:CI}, that is protected by antiunitary spectral symmetry. The notion of the antiunitary spectral symmetry is also germane for 2D HOTI and its corner modes, which we discuss next.

\begin{figure*}[t!]
\subfigure[]{\includegraphics[width=0.245\linewidth]{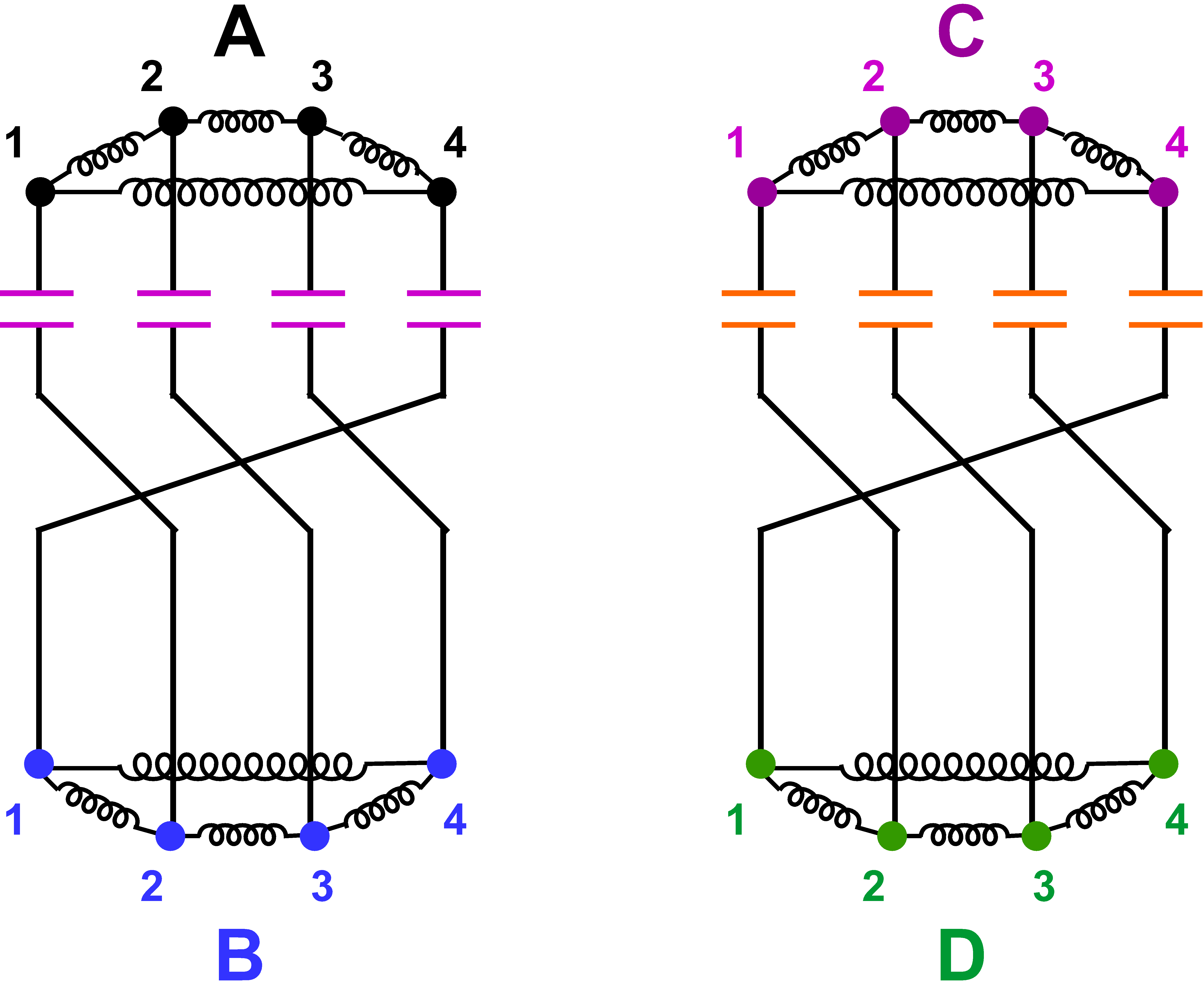}}
\subfigure[]{\includegraphics[width=0.245\linewidth]{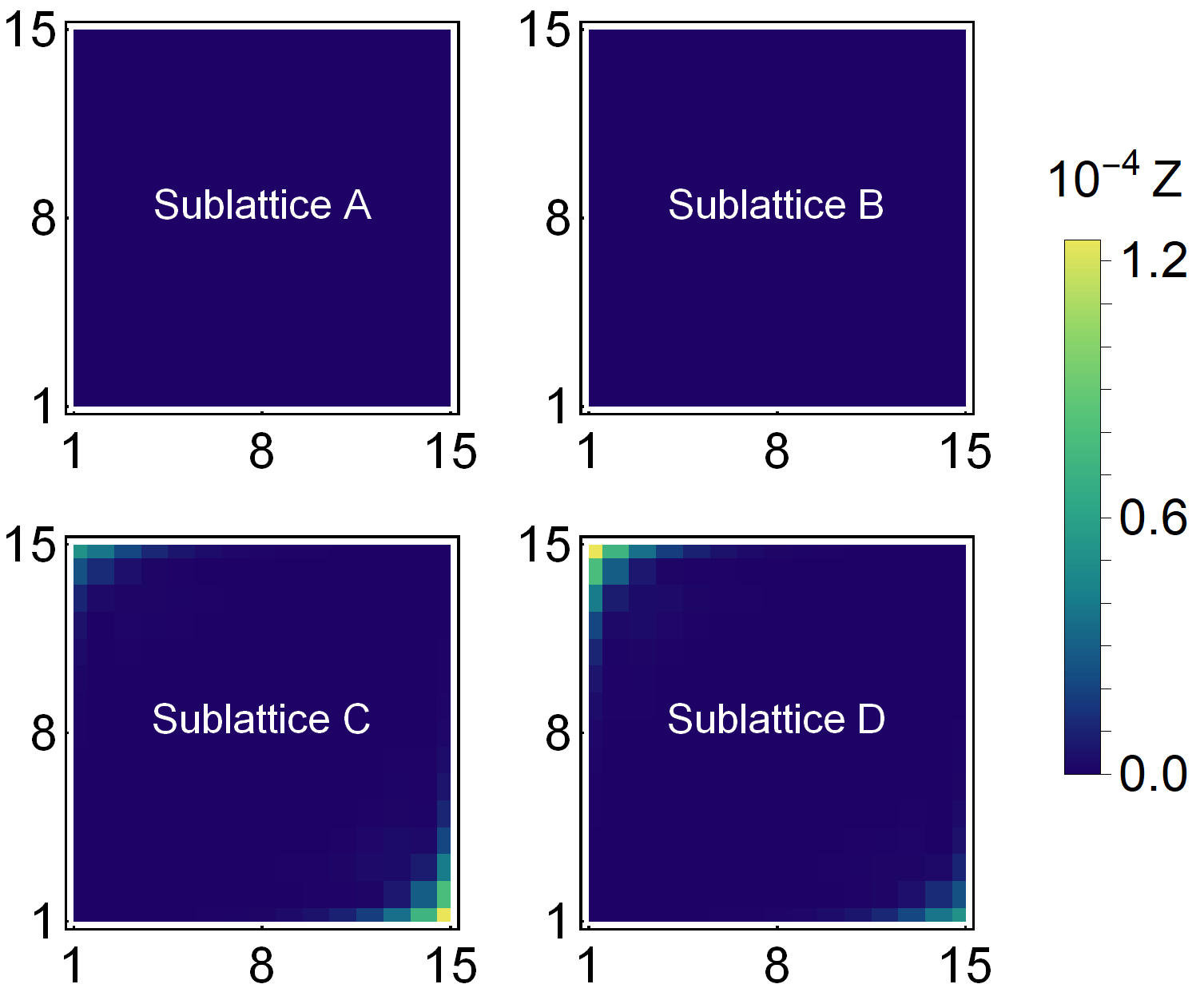}}
\subfigure[]{\includegraphics[width=0.245\linewidth]{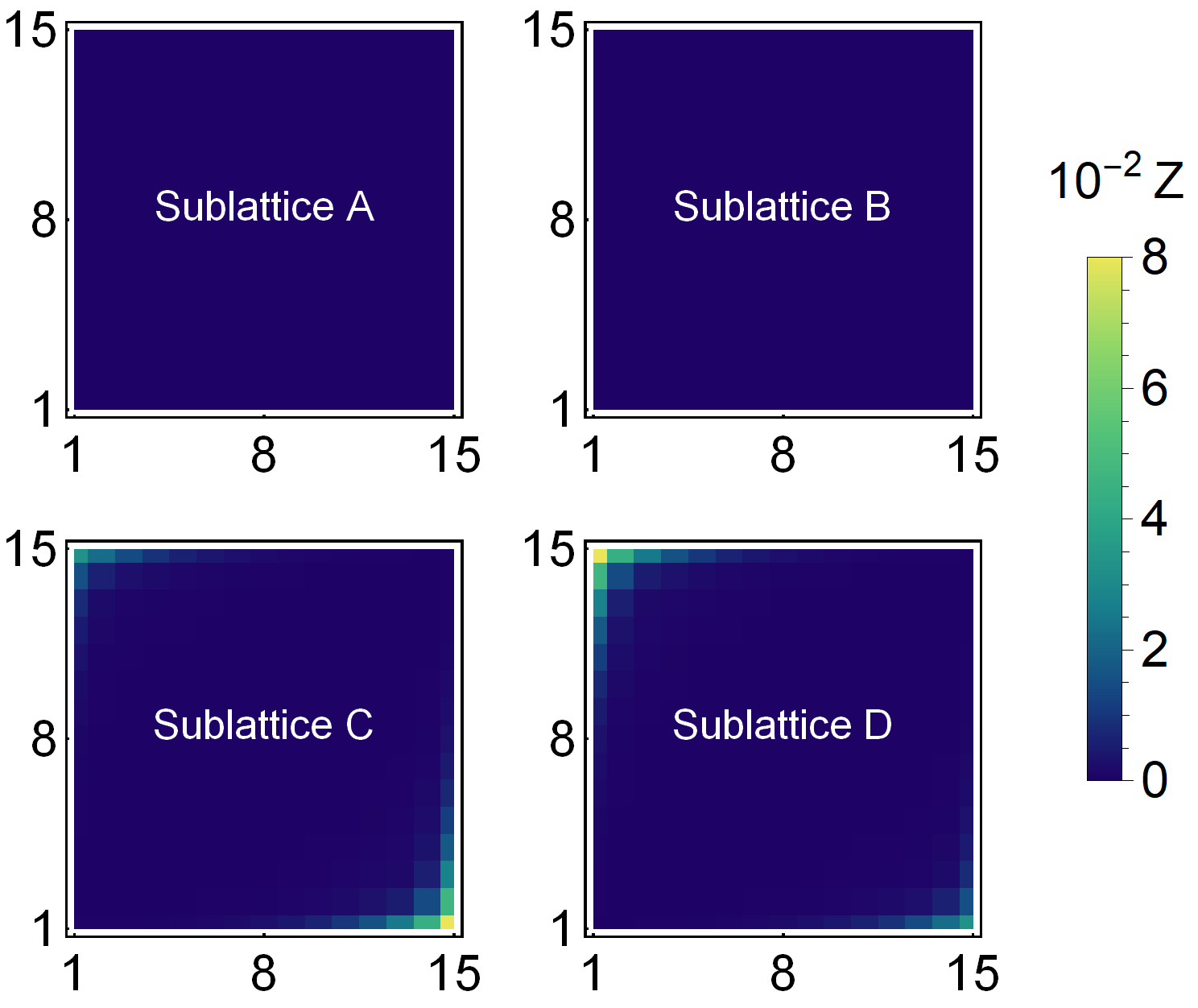}}
\subfigure[]{\includegraphics[width=0.245\linewidth]{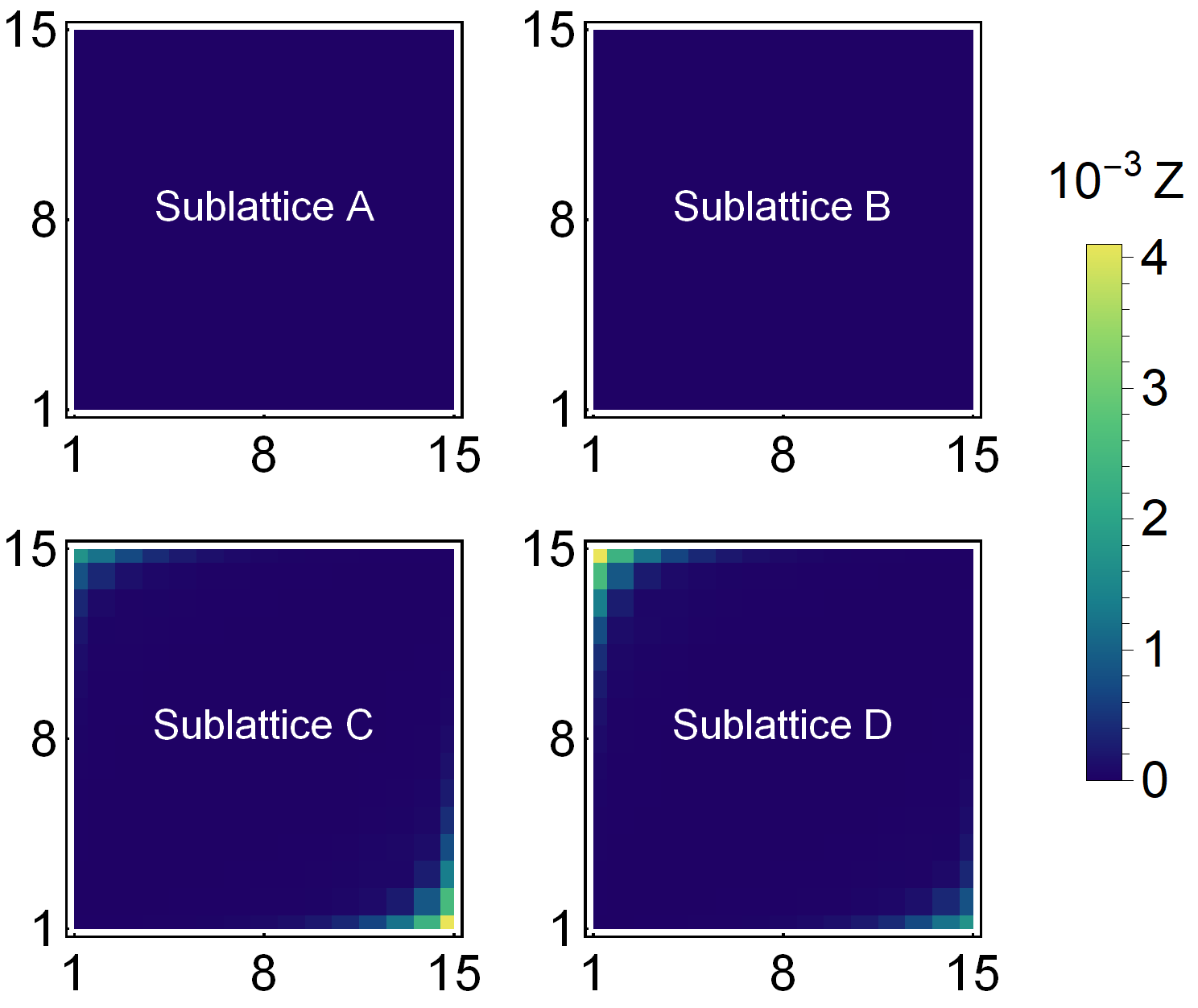}}
\caption{Circuit realization of an antiunitary symmetry protected generalized higher-order topological insulator (GHOTI). (a) Circuit realization of $H_p$ [see Eq.~(\ref{eq:HpGHOTI})] in a unit-cell of GHOTI, where each node (equivalent of sublattice) is now accompanied by four subnodes. Each subnode is grounded by an inductor of inductance $L$. The purple (orange) capacitors have capacitance equal to $\Delta_1+\Delta_2$ ($|\Delta_1-\Delta_2|$). On-resonance computation of impedance with the B node at (1,1) as the input point for $t=t_0=m=2\Delta=1$, and (b) $(\Delta_1,\Delta_2)=(0.2,0.0)$, (c) $(\Delta_1,\Delta_2)=(0.0,0.3)$ and (d) $(\Delta_1,\Delta_2)=(3.0,3.0)$. The output point scans four nodes separately over the entire system. Node (or sublattice) selective sharp corner localization of the on-resonance impedance confirms topolectric circuit realization of GHOTI for small $\Delta_1$ or $\Delta_2$, and \emph{arbitrarily} large $\Delta_1=\Delta_2$.
}~\label{fig:2DGHOTI}
\end{figure*}

Recall first that the model Hamiltonian for 2D HOTI, namely $H^\prime_{\rm QSHI}=H_{\rm QSHI}+H_\Delta$ [see Eqs.~(\ref{eq:QSHImodel}) and (\ref{eq:discreteC4breaking})] possesses unitary particle-hole symmetry generated by $\Gamma_5$. This model also enjoys an antiunitary particle-hole symmetry generated by $A=\Gamma_5 {\mathcal K}$, as $\{ H_{\rm QSHI}+H_\Delta, A \}=0$. Here ${\mathcal K}$ is the complex conjugation, and by virtue of the $\Gamma$ matrix representation in Eq.~(\ref{eq:gammarepresentationQSHI}), the unitary component of $A$ is identical to the unitary particle-hole symmetry generator ($\Gamma_5$).

The model Hamiltonian for 2D generalized higher-order topological insulator (GHOTI) reads~\cite{roy-aniunitaryHOTI}
\begin{equation}
H_{\rm GHOTI}= H_{\rm QSHI}+ H_\Delta + H_p,
\end{equation}
where
\begin{equation}~\label{eq:HpGHOTI}
H_p= \Delta_1 \left( i \Gamma_1 \Gamma_2 \right) + \Delta_2 \left( i \Gamma_3 \Gamma_4 \right),
\end{equation}
with $\Delta_1$ and $\Delta_2$ as real parameters. Notice that $\{ H_{\rm GHOTI}, A \}=0$. Hence, $H_{\rm GHOTI}$ enjoys antiunitary particle-hole symmetry, but loses the particle-hole symmetry with respect to the unitary operator $\Gamma_5$. The global phase diagram of this model has already been analyzed in Ref.~\cite{roy-aniunitaryHOTI}, which we do not discuss here in details. In brief, $H_{\rm GHOTI}$ supports four corner localized zero-energy modes for (1) small $\Delta_1$ or $\Delta_2$, and (2) arbitrarily large $\Delta_1=\Delta_2$. Next we implement $H_{\rm GHOTI}$ on topolectric circuit and test the validity of these predictions from the numerical measurement of the on-resonance corner impedance.

First note that we cannot find any representation of the $\Gamma$ matrices in which all the onsite and hopping elements of $H_{\rm GHOTI}$ are real. So we stick to the old representation of the $\Gamma$ matrices [see Eq.~(\ref{eq:gammarepresentationQSHI})], and first supplement each node associated with the circuit realization of $H_{\rm QSHI}+H_\Delta$ by four subnodes. Such a doubling of the HOTI circuit does not alter any outcome we discussed so far. Nevertheless, when each node (or sublattice) contains four subnodes one can implement $H_p$ in a topolectric circuit following the design shown in Fig.~\ref{fig:2DGHOTI}(a).

\begin{figure*}[t!]
\subfigure[]{\includegraphics[width=0.49\linewidth]{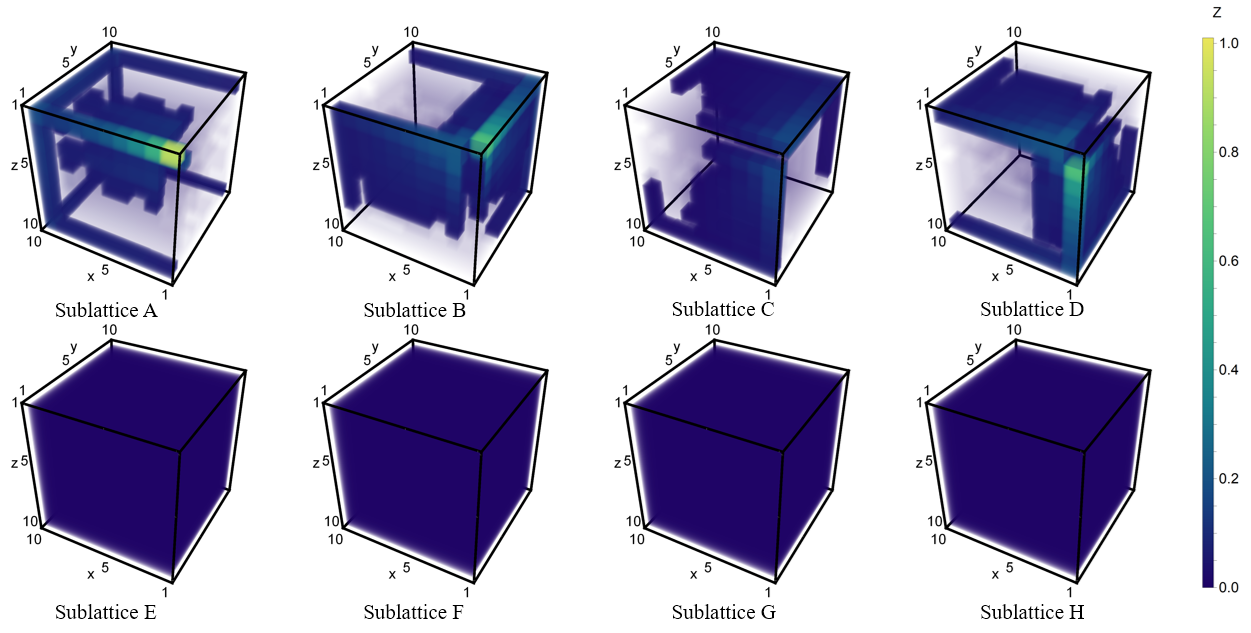}}
\subfigure[]{\includegraphics[width=0.24\linewidth]{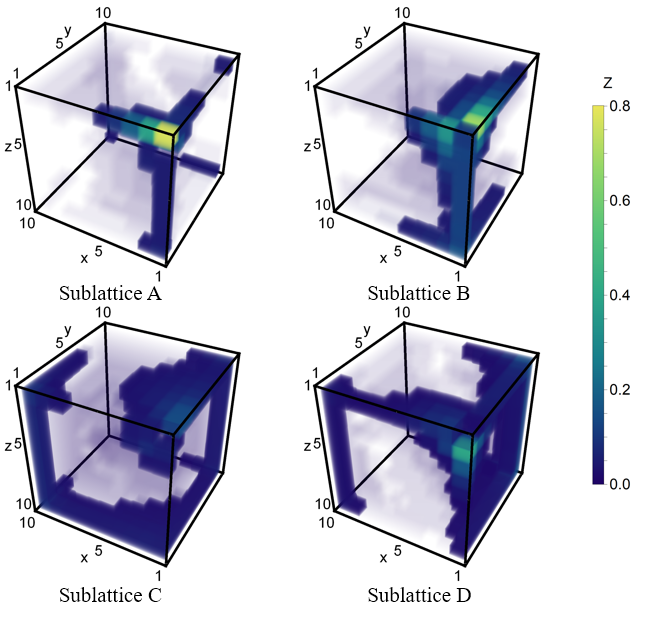}}
\subfigure[]{\includegraphics[width=0.24\linewidth]{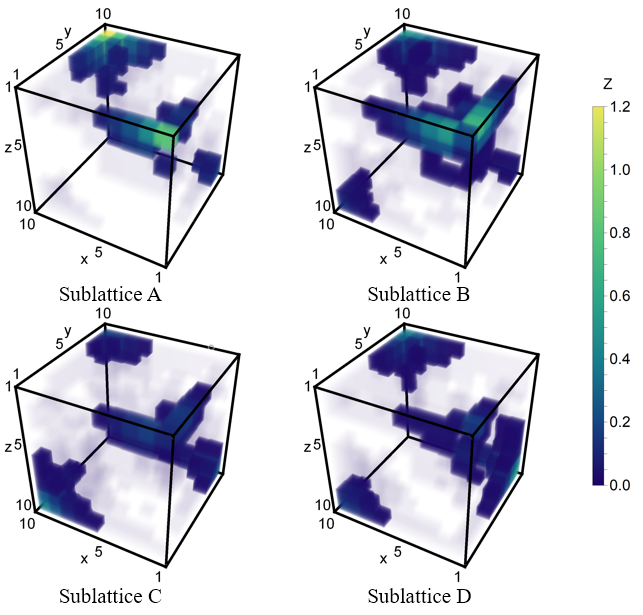}}
\caption{On-resonance impedance of (a) first-order, (b) second-order and (c) third-order topological insulators in three-dimensions, always measured with the site on the H sublattice at (1,1,1) as the input point. The output point scans all eight sublattices over the entire system. In a first-order topological insulator the surface impedance is localized on the A, B, C and D sublattices (sublattice polarization), and computed by setting $t=t_0=1$ and $m=1.5$ [Eq.~(\ref{eq:FOTI})]. By contrast, the on-resonance impedance on the E, F, G and H sublattices are (almost) zero, and we do not show them in panels (b) and (c). The on-resonance impedance for a second-order topological insulator is predominantly localized along the hinges in the $z$ direction and on the $xy$ surfaces as shown in (b) for $\Delta_1=0.5$ [Eq.~(\ref{eq:SOTI})]. Finally, in a third-order topological insulator the on-resonance impedance is localized at the corners on the A, B, C and D sublattices, as shown in (c) for $\Delta_2=0.7$ [Eq.~(\ref{eq:TOTI})].
}~\label{fig:3Dhierarchy}
\end{figure*}

As the corner modes of GHOTI are protected by $A=\Gamma_5 {\mathcal K}$, with $\Gamma_5=\sigma_3 \otimes \tau_0$, their sublattice selection for finite $\Delta_1$ and/or $\Delta_2$ remains unchanged. From the measurement of the on-resonance impedance with the B node at (1,1) as the fixed input point, we find the following. Node (or sublattice) selective sharp corner localization of impedance does not change for (1) small $\Delta_1$ [see Fig.~\ref{fig:2DGHOTI}(b)], (2) small $\Delta_2$ [see Fig.~\ref{fig:2DGHOTI}(c)], and (3) large $\Delta_1=\Delta_2$ [see Fig.~\ref{fig:2DGHOTI}(d)]. These findings are in agreement with Ref.~\cite{roy-aniunitaryHOTI}, and we find a circuit realization of GHOTI for which the corner impedance is protected by an antiunitary operator. When $\Delta_1=\Delta_2$, the system always   remains in the GHOTI phase, while for large $\Delta_1$ or $\Delta_2$ it undergoes a transition into a trivial phase, without any corner modes. In this regime the on-resonance impedance does not show any corner localization, which we do not display here.

We point out that the same generalization is also applicable for the 2D BBH model for HOTI, discussed in Sec.~\ref{subsec:BBH2D}. Furthermore the 3D BBH model from Sec.~\ref{subsec:BBH3D} also possesses an antiunitary spectral symmetry, generated by the eight-dimensional operator $\Gamma_7 {\mathcal K}$, with $\Gamma_7=\Sigma_3 \otimes \sigma_0 \otimes \tau_0$. In the future, it will be interesting to find a generalized model for 3D HOTI.

\section{Hierarchy of HOT insulators in 3D}~\label{Sec:3Dhierarchy}

Following the spirit of the previous section, we now show how one can construct the hierarchy of 3D topological insulators and capture their signatures in topolectric circuits. Specifically, in what follows we construct 3D second-order and third-order topological insulators by respectively adding one and two discrete symmetry breaking Wilson-Dirac masses to a first-order topological insulator. Recall that first-, second- and third-order topological insulators support gapless 2D surface states, 1D hinge modes and pointlike corner modes, respectively. Our starting point is the following Hamiltonian
\begin{eqnarray}~\label{eq:FOTI}
H_{\rm FOTI} &=& t\left[ \sin(k_x)\Gamma_1+\sin(k_y) \Gamma_2+\sin (k_z)\Gamma_3 \right] \nonumber \\
&+& \left[ m-t_0\sum_{i=x,y,z}\cos(k_i) \right] \Gamma_4,
\end{eqnarray}
describing a first-order topological insulator with the band inversion at the $\Gamma=(0,0,0)$ point of the cubic Brillouin zone for $1<m/t_0 <3$, that supports 2D gapless surface states on six surfaces of a cubic system. As we will show,  the realization of third-order topological insulator demands addition of two discrete symmetry breaking Wilson-Dirac masses to $H_{\rm FOTI}$, which must be accompanied by two additional mutually anticommuting $\Gamma$ matrices, which also anticommute with four mutually anticommuting $\Gamma$ matrices appearing in $H_{\rm FOTI}$. Therefore, altogether we require \emph{six} mutually anticommuting $\Gamma$ matrices. This implies that the representation of the $\Gamma$ matrices must be \emph{eight-dimensional}, which, on the other hand, accommodates maximal \emph{seven} mutually anticommuting matrices. Out of them \emph{four} can be chosen to be purely \emph{real} and the remaining \emph{three} to be purely \emph{imaginary}. Following the discussion so far, we make a judicious choices for the $\Gamma$ matrices to be
\allowdisplaybreaks[4]
\begin{eqnarray}
\Gamma_1 &=& \Sigma_1\otimes\sigma_1\otimes\tau_2, \:\: \Gamma_2=\Sigma_1\otimes\sigma_2\otimes\tau_0, \nonumber \\
\Gamma_3 &=& \Sigma_2\otimes \sigma_0\otimes \tau_0, \:\: \Gamma_4=\Sigma_1\otimes\sigma_1\otimes\tau_1, \nonumber \\
\Gamma_5 &=& \Sigma_1\otimes\sigma_1\otimes\tau_3, \:\: \Gamma_6=\Sigma_1\otimes\sigma_3\otimes\tau_0,
\end{eqnarray}
such that the tight-binding model always contains only real hopping amplitudes and we can implement it on a topolectric circuit by supplementing each node by two subnodes. Finally, the seventh mutually anticommuting matrix is given by $\Gamma_7=\Sigma_3\otimes \sigma_0\otimes \tau_0$.

\begin{figure*}
\subfigure[]{\includegraphics[width=0.31\linewidth]{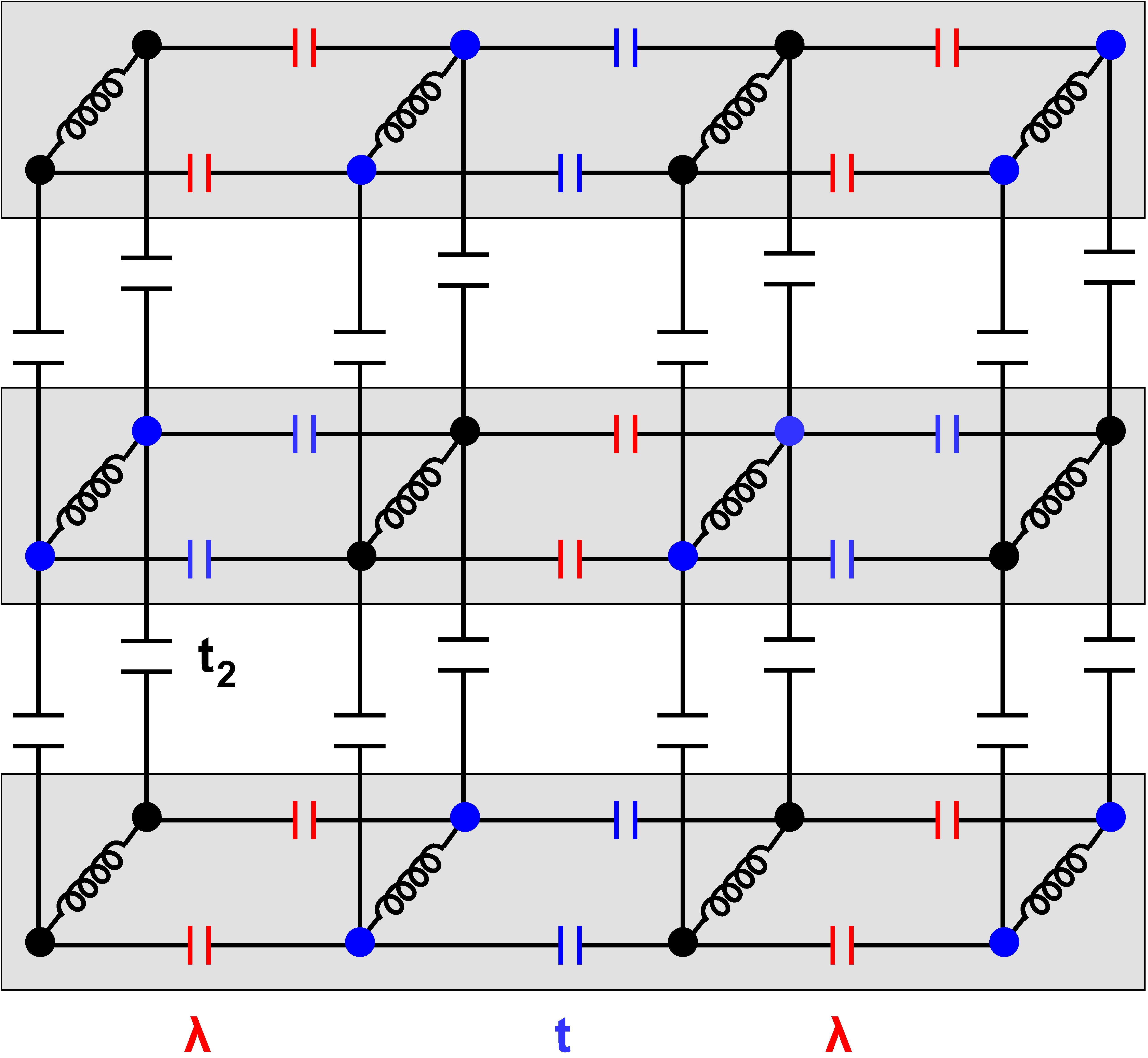}}
\subfigure[]{\includegraphics[width=0.31\linewidth]{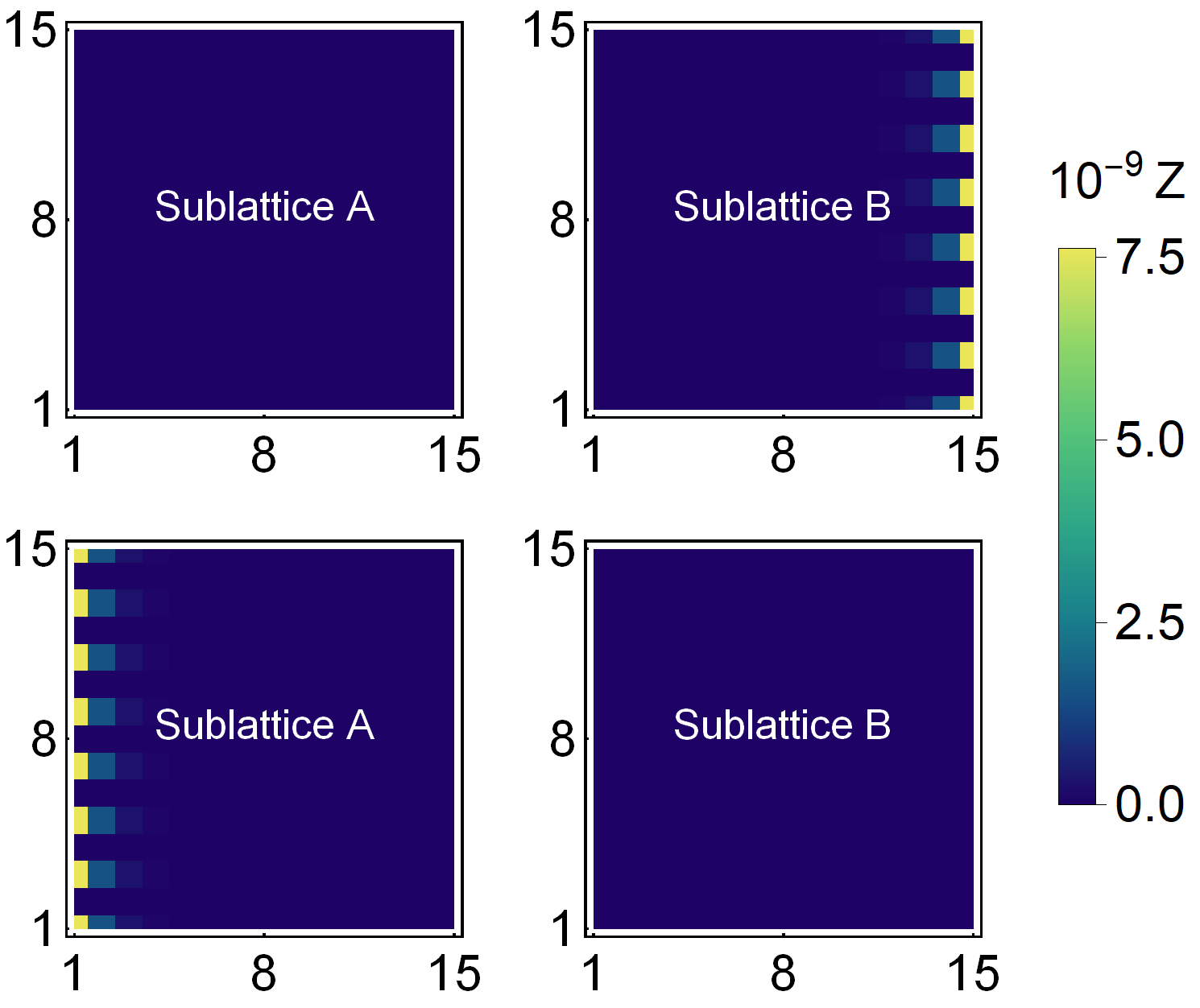}}
\subfigure[]{\includegraphics[width=0.35\linewidth,height=3.5cm]{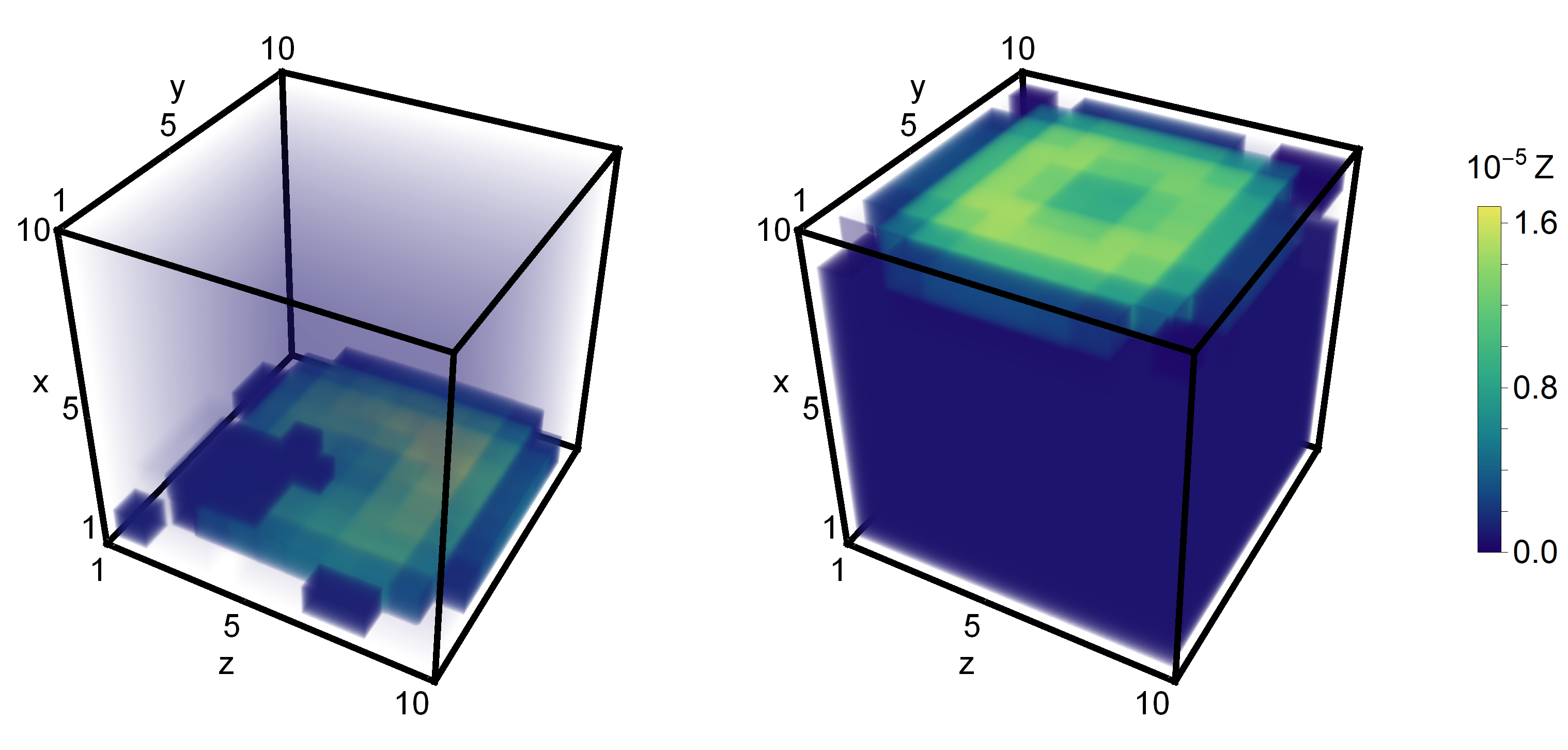}}
\caption{Realization of 2D and 3D topological semimetals from stacked 1D Su-Schrieffer-Heeger (SSH) chains, see also Fig.~\ref{fig:SSH-construction}(c), in topolectric circuits. (a) Stacking of 1D SSH circuit along the $y$-direction by capacitors of capacitance $t_2$ in a translationally invariant fashion that produces 2D Dirac semimetal (DSM) for suitable choices of $\lambda, t$ and $t_2$ (see text and Ref.~\cite{lee-commphys2018}). For the clarity of presentation we shift the location of the sublattices or nodes in the middle SSH circuit with respect to the first and the third one. Extending such stacking construction in the $z$-direction one produces 3D nodal-line semimetal (NLSM), which we do not show here explicitly. (b) Numerically computed on-resonance impedance ($Z$) with the A (top) and B (bottom) nodes at (1,1) and (15,1), respectively, as the input point for $t=1$, $\lambda=0.2$ and $t_2=3.0$ [see Eq.~(\ref{eq:stackedSSH2D})]. The impedance is then computed with both nodes or orbitals as the output point (mentioned explicitly in each subfigure), which we scan over the entire system. The edge localized impedance only along the $y$ direction captures the fingerprint of the Fermi arc associated with a 2D DSM. (c) Numerically computed impedance for a 3D NLSM with the A node at (1,1,1) as the fixed input point for $t=1$, $\lambda=1$ and $t_2=1$ [see Eq.~(\ref{eq:stackedSSH3D})]. The output point is chosen to be on the A (left) and B (right) nodes, which we scan over the entire systems. The top and bottom $yz$ surface localized impedance captures the signature of the \emph{drumhead} surface states in the real space of a topolectric circuit.
}~\label{fig:stackedSSH}
\end{figure*}

So far, we have shown realization of a variety of tight-binding models on topolectric circuits. Following the same spirit, one can implement $H_{\rm FOTI}$ on a cubic topolectric circuit. However, the explicit circuit realization of $H_{\rm FOTI}$ is somewhat involved and as such it is not very instructive. Therefore, here we do not show it explicitly, but rather focus the spatial distribution of the on-resonance impedance. The results are displayed in Fig.~\ref{fig:3Dhierarchy}(a). As $\{ H_{\rm FOTI}, \Gamma_7 \}=0$, the 2D surface states are localized either on the sublattices A, B, C and D or on the sublattices E, F, G and H. The on-resonance impedance also shows the signature of such sublattice polarization. Namely, when we choose the input point for the measurement of the on-resonance impedance ($Z$) on the H sublattice, located at (1,1,1), $Z$ is highly localized on the surfaces on A, B, C and D subalttices, whereas $Z$ on the remaining four sublattices are \emph{almost zero}. It should be noted that even though the 2D surface states of a first-order topological insulator equally populate all six surfaces of a cube (due to the cubic symmetry), any choice of the input point breaks such symmetry. Therefore, despite showing strong sublattice polarized surface localization, the on-resonance impedance is not fully cubic symmetric.

Next we add the Wilson-Dirac mass
\begin{equation}~\label{eq:SOTI}
H_{1}= \Delta_1 \left[ \cos(k_x)-\cos(k_y) \right] \Gamma_5
\end{equation}
to $H_{\rm FOTI}$. Note that $H_1$ breaks discrete $C_4$ rotational symmetry and as such it changes sign across four corners in the $xy$ plane for any $z$. Therefore, addition of this term gaps out the surface states residing on the $xz$ and $yz$ planes, leaving four intersections in the $z$ direction gapless. Consequently, $H_{\rm SOTI}= H_{\rm FOTI} + H_1$ accommodates four zero energy gapless hinge modes in the $z$ direction and we realize a second-order topological insulator. On the other hand, the Wilson-Dirac mass $H_1$ vanishes at the center of the surface Brillouin zone on the $xy$ planes, located at $(k_x,k_y)=(0,0)$, where the apex of the surface Dirac cone is placed at. Hence, the $xy$ surfaces continue to host gapless states in a second-order topological insulator. The realization of such $C_4$ symmetry breaking Wilson-Dirac mass on a 2D topolectric circuit has already been shown in Fig.~\ref{fig:2DQSHI-HOTI}(a), which can be generalized to three dimensions. Here, we discuss only the results from the spatial distribution of the on-resonance impedance, shown in Fig.~\ref{fig:3Dhierarchy}(b). As $\{H_{\rm SOTI}, \Gamma_7 \}=0$, the $z$ directional hinge and $xy$ surface modes, and the corresponding on-resonance impedance continue to show sublattice polarization. Otherwise, comparing with Fig.~\ref{fig:3Dhierarchy}(a), we find a clear dimensional reduction of the on-resonance impedance, which is localized along the four hinges in the $z$ direction. At the same time we also find remnant surface impedance on the $xy$ surfaces. These results in turn guarantee a realization of a second-order topological insulator on a topolectric circuit.

Finally, we introduce the second Wilson-Dirac mass
\begin{equation}~\label{eq:TOTI}
H_2 = \Delta_2 \left[ 2\cos(k_z)-\cos(k_x)-\cos(k_y) \right] \Gamma_6
\end{equation}
to $H_{\rm SOTI}$. Notice that $H_2$ vanishes along eight body-diagonal directions $(\pm 1,\pm 1, \pm 1)$, as well as along $(\pm \sqrt{2},0,\pm 1)$ and $(0,\pm \sqrt{2}, \pm 1)$ directions. However, the latter directions are already gapped by $H_1$. Therefore, the total Hamiltonian $H_{\rm TOTI}=H_{SOTI}+H_2$ supports only localized gapless modes at eight corners of a cubic system residing in the $(\pm 1, \pm 1, \pm 1)$ directions, and we realize a third-order topological insulator. As $\{ H_{\rm TOTI}, \Gamma_7 \}=0$, the corner modes continue to display the sublattice polarization. Concomitantly, the on-resonance impedance also displays sublattice polarization, besides being highly localized at the corners of a cubic topolectric circuit, as shown in Fig.~\ref{fig:3Dhierarchy}(c). The corner localization of on-resonance impedance in turn ensures the realization of a third-order topological insulator in a topolectric circuit. We should also note that the Hamiltonian describing a third-order topological insulator $H_{\rm TOTI}$ can be exactly mapped onto the 3D BBH model~\cite{nagjuricicroy:3D}, discussed in Sec.~\ref{subsec:BBH3D}.

\section{Topolectric nodal semimetals}~\label{Sec:NSM}

So far we have discussed realizations of first-, second- and third-order topological insulators on topolectric circuits and their identification from the boundary localized on-resonance impedance. On the other hand, there exists a whole family of topological phases of matter, known as \emph{topological semimetals}, where the bulk quasiparticle spectra are gapless, but the bulk-bounadary correspondence remains operative therein~\cite{Armitage-RMP, Bernevig-book, Schnyder-RMP, Shen-book}. Typically, topological semimetals are constructed by stacking lower-dimensional topological insulators, while preserving the translational symmetry in the stacking direction, which is the approach we use to construct their topolectric realizations. In this section, we show that some prominent nodal topological phases can be realized in topolectric circuits and identify them from the boundary localized on-resonance impedance.

\subsection{Stacked SSH chain: 2D Dirac and 3D nodal-line semimetals}~\label{subsec:DSM-NLSM}

In contrast to Weyl and Dirac semimetals where the valence and conduction bands touch at isolated points, in nodal-line semimetals (NLSMs) the band touching takes place along a closed curve in momentum space~\cite{Burkov-PRB2011, Phillips-PRB2014, mullen-PRL2015, bzdusek-nature2016, Roy-NLSM-PRB2017, Kim-PRL2017, Ramamurthy-PRB2017, Geilhufe-PRB2019}. In this section, we first construct a 2D Dirac semimetal (DSM) by stacking a collection of 1D SSH chains in the $y$-direction in a translationally invariant fashion, such that the symmetry class of the system, namely BDI, remains unchanged. The corresponding Hamiltonian in the momentum space takes the form
\begin{equation}
H^{\rm 2D}_{\rm DSM}= \left[ \lambda + t \cos(k_x) + t_2 \cos (k_y) \right] \tau_1 + t \sin(k_x) \tau_2,
\end{equation}
where $t_2$ denotes the strength of the inter-SSH chain hopping along the $y$ direction. Depending on the relative strength of various hopping parameters ($\lambda$, $t$ and $t_2$), the system supports a 2D DSM, topological and trivial insulators. Specifically for $\lambda/t>0$, the Dirac points are located at ${\bf k}=(\pi, \pm k^\star_y)$, where
\begin{equation}~\label{eq:stackedSSH2D}
k^{\star}_y=\pi-\cos^{-1}\left( \frac{t}{t_2} \left[ 1 + \frac{\lambda}{t} \right]  \right).
\end{equation}
As one can see, depending on the values of $t$, $\lambda$ and $t_2$, Eq.~\eqref{eq:stackedSSH2D} has a real solution for $k^\star_y$, which then describes a 2D DSM. Otherwise, the system is an insulator. Furthermore, each of the 1D SSH topological insulators stacked between two Dirac nodes supports endpoint zero modes. The collection of these zero modes ultimately constitutes Fermi arc states connecting two Dirac points in the momentum space. On the other hand, in the real space the Fermi arc states are localized along only two edges in the $y$ direction. This scenario is supported  from the numerical computation of the on-resonance impedance in a 2D circuit constructed by stacking a 1D SSH circuit in the $y$ direction. We note that the 2D DSM has been realized in topolectric circuits in Refs.~\cite{lee-commphys2018, yli-natcomm2018, helbig-prb2019}, and our results are in qualitative agreement. The main purpose of this discussion is to develop a concrete path for stacked topolectric circuits to realize various gapless topological phases, which we systematically explore next. As we show now, by extending this construction in the $z$ direction, we can find a 3D \emph{unknotted} nodal-line semimetal. We point out that even though various topological knots in the momentum space have been engineered in topolectric circuits~\cite{chlee-arxiv2019}, explicit demonstration of simple nodal-line semimetal has not been reported so far to the best of our knowledge.

Since the hopping parameter in the $y$ direction is completely \emph{real}, one can construct the circuit corresponding to Eq.~(\ref{eq:stackedSSH2D}) by coupling 1D SSH circuits, shown in Fig.~\ref{fig:SSH-construction}(c), by capacitors of capacitance $t_2$ in the $y$-direction between nodes A and B, as shown in Fig.~\ref{fig:stackedSSH}(a). Next we numerically compute the on-resonance impedance with either an A or B node of the circuit as an input point, and scanning all the sites as the output point. Results are shown in Fig.~\ref{fig:stackedSSH}(b), displaying a sharp $y$-edge localization of the on-resonance impedance, which in turn captures the existence of Fermi arcs states and a 2D DSM in a topolectric circuit.

Next we continue with the stacking protocol and extend it along the $z$ direction. For the sake of simplicity we consider the inter-layer hopping amplitude along the $z$ direction to be $t_2$ as well. Then the corresponding Hamiltonian in the momentum space reads as
\begin{align}~\label{eq:stackedSSH3D}
H^{\rm 3D}_{\rm NLSM}&= \left[ \lambda + t \cos(k_x) + t_2 \sum_{j=y,z} \cos (k_j) \right] \tau_1 \nonumber \\
&+ t \sin(k_x) \tau_2.
\end{align}
This model supports a NLSM, in which the valence and conduction band touch each other over a closed curve in the $k_x=\pi$ plane when $t=1$, $\lambda=2$ and $t_2=1$, for example. The implicit form of the closed curve in the $k_x=\pi$ plane is then given by  $\cos(k^\star_y)+\cos(k^\star_z)=1$. A NLSM is therefore constructed by stacking 1D SSH topological insulator in the $y$ and $z$ direction within the perimeter of the closed curved mentioned above. A collection of endpoint zero modes associated with the SSH insulator within the perimeter of such closed curve ultimately constitutes the \emph{drumhead} surface states on the $(k_y,k_z)$ planes, which are the surface projections of the bulk nodal loop. On the other hand, in the real space the surface zero-energy modes are localized on the entire $yz$ plane.

A circuit realization of NLSM can be engineered by extending the construction from Fig.~\ref{fig:stackedSSH}(a) in the $z$ direction, along which the hopping elements are also real. Therefore, inter-layer hopping in the $z$ direction is accomplished by capacitor connections with capacitance $t_2$. The explicit circuit construction for NLSM is, however, not shown here. The numerical computation of the on-resonance impedance clearly displays a strong surface localization on the top and bottom $yz$ planes, capturing the imprint of the drumhead surface states associated with a NLSM in a 3D topolectric circuit, as shown in Fig.~\ref{fig:stackedSSH}(c).

\subsection{Stacked Chern insulator: 3D Weyl semimetal}~\label{subsec:3DWSM}

Next we construct a 3D Weyl semimetal (WSM) by stacking 2D Chern insulators in the $z$-direction in a translationally invariant fashion. The resulting Hamiltonian in the momentum space then reads as
\begin{align}~\label{eq:stackedCIWSM}
H^{\rm 3D}_{\rm WSM} &= t \left[ \sin(k_x) \; \tau_3 + \sin(k_y)\; \tau_2 \right] \nonumber \\
&+ \left[t_z \cos(k_z) +m - t_0 \sum_{j=x,y} \cos(k_j) \right] \; \tau_1,
\end{align}
where $t_z$ denotes the interlayer hopping amplitude in the $z$ direction. The global phase diagram of this model has been reported in Refs.~\cite{chen-song, slager-juricic-roy:PRX}, which we do not discuss here in details. Conveniently, we set $t=t_0=1$, $m=2$ and $t_z=2.5$ for the rest of the discussion. For this set of parameters the system supports a WSM, with two Weyl nodes located at ${\bf k}^\star=(0,0,\pm \cos^{-1}(4/5))$. The two Weyl nodes represent the bandgap closing points for the collection of Chern insulators stacked in the direction connecting them, and the zero-energy states associated with each Chern insulator layer ultimately constitute the Fermi arc surface states. In the surface Brillouin zone on the $(k_j,k_z)$ plane the Fermi arc connects two Weyl points at $k^\star_z=\pm \cos^{-1}(4/5)$ and placed along the $k_j=0$ line, where $j=x$ or $y$~\cite{Armitage-RMP, krempa:review, slager-juricic-roy:Fermiarc}. In the real space, the Fermi arc occupies the $xz$ and $yz$ planes. Next we demonstrate these features in a circuit realization of the WSM. The existing literature reporting topolectric circuit realizations of WSM is concerned with systems possessing an \emph{even} number of pairs of Weyl nodes~\cite{lee-commphys2018, rafi-njp2020, r-li-arxiv2019, ylu-prb2019, luo-research-2018}. By contrast, here we demonstrate circuit realization of the minimal WSM, supporting only two Weyl points, by stacking layers of 2D Chern insulators.

\begin{figure}[t!]
\subfigure[]{\includegraphics[width=0.49\linewidth]{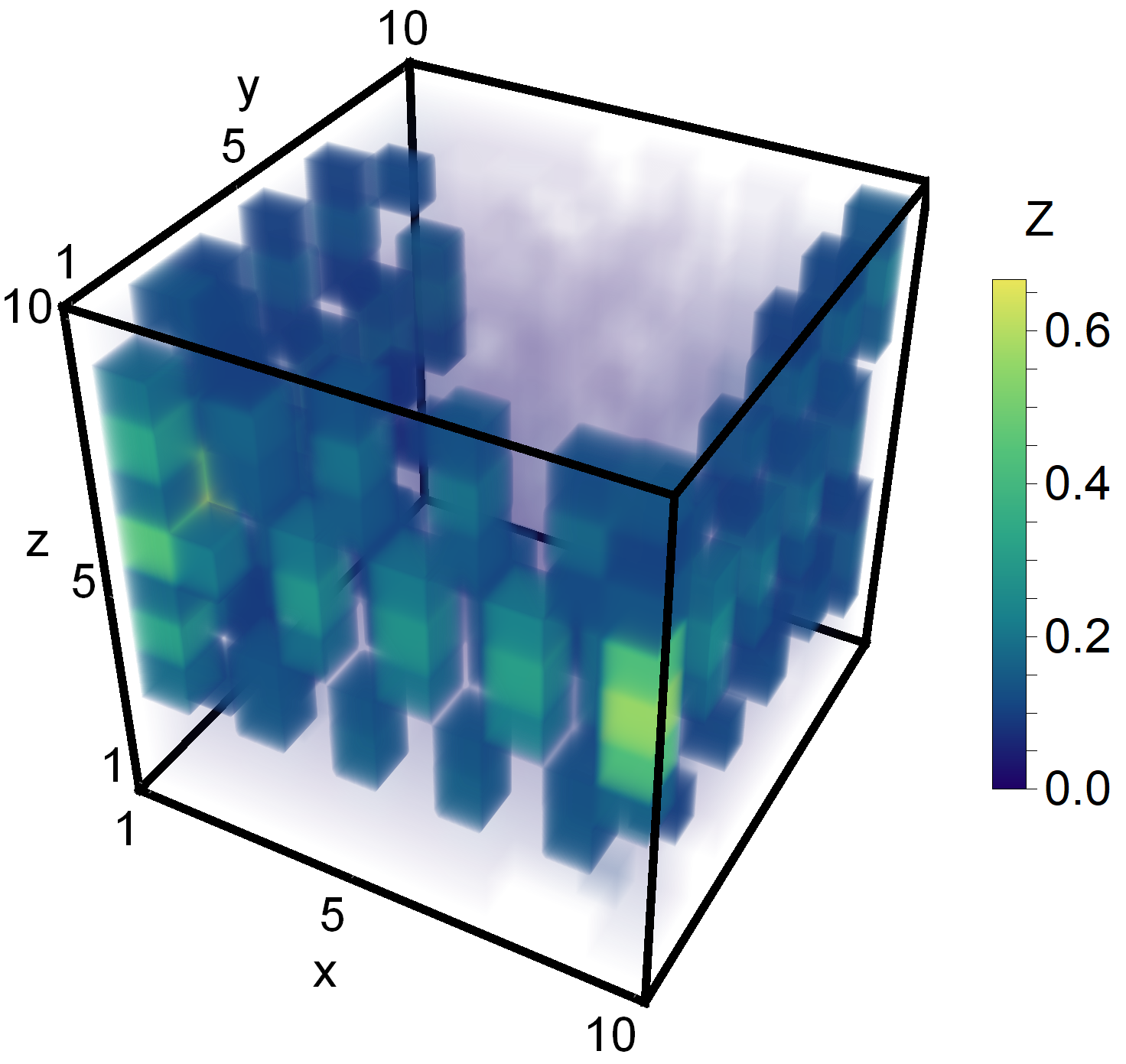}}
\subfigure[]{\includegraphics[width=0.49\linewidth]{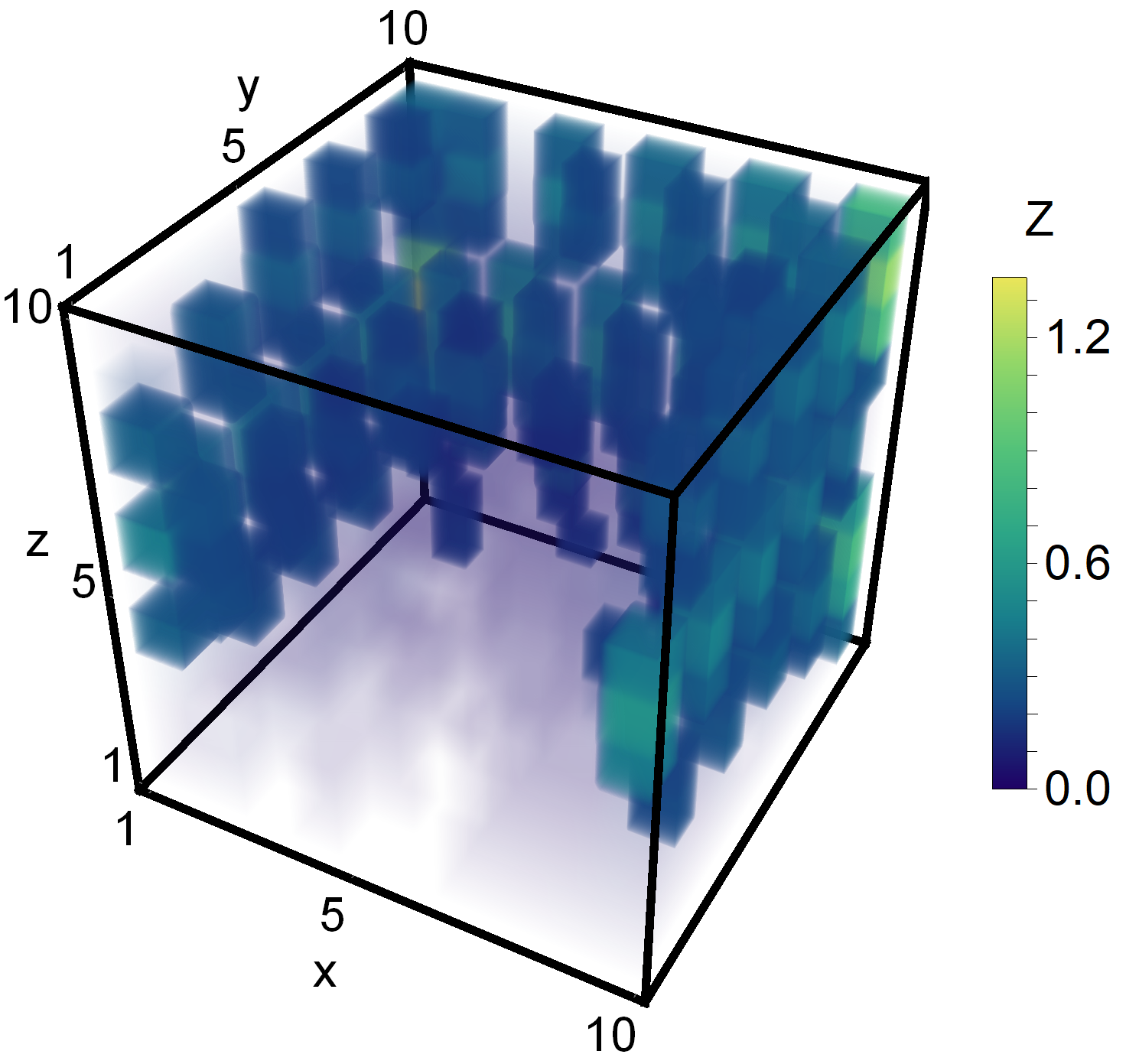}}
\caption{Numerical computation of the on-resonance impedance ($Z$) in a 3D Weyl topolectric circuit with the A (left) and B (right) nodes at (1,1,5) as the fixed input point for $t=t_0=1$, $m=2$ and $t_z=2.5$ [see Eq.~(\ref{eq:stackedCIWSM})]. Respectively, the output point is fixed on the B (left) and A (right) nodes, which we sweep over the entire system. The strong localization of the on-resonance impedance on the $xz$ and $yz$ planes captures the signature of the Fermi arc surface states.
}~\label{fig:Weylsemimetal}
\end{figure}

The circuit construction of the WSM follows the spirit of a stacked SSH chain, shown in Fig.~\ref{fig:stackedSSH}(a). To engineer a WSM we couple layers of Chern circuits, schematically shown in Fig.~\ref{Fig:Chern-lat}, by capacitor connections with capacitance $t_z$. The  construction of the Weyl topolectric circuit is not shown here explicitly. The numerical computation of the on-resonance impedance ($Z$) shows a strong localization on the $xz$ and $yz$ surfaces, see Fig.~\ref{fig:Weylsemimetal}. Selective surface localization therefore bears the signature of the Fermi arc surface states, in turn confirming the realization of a 3D WSM in a topolectric circuit.

\subsection{Stacked QSHI: 3D Dirac semimetal}~\label{subsec:3DDSM}

\begin{figure}[t!]
\includegraphics[width=0.95\linewidth]{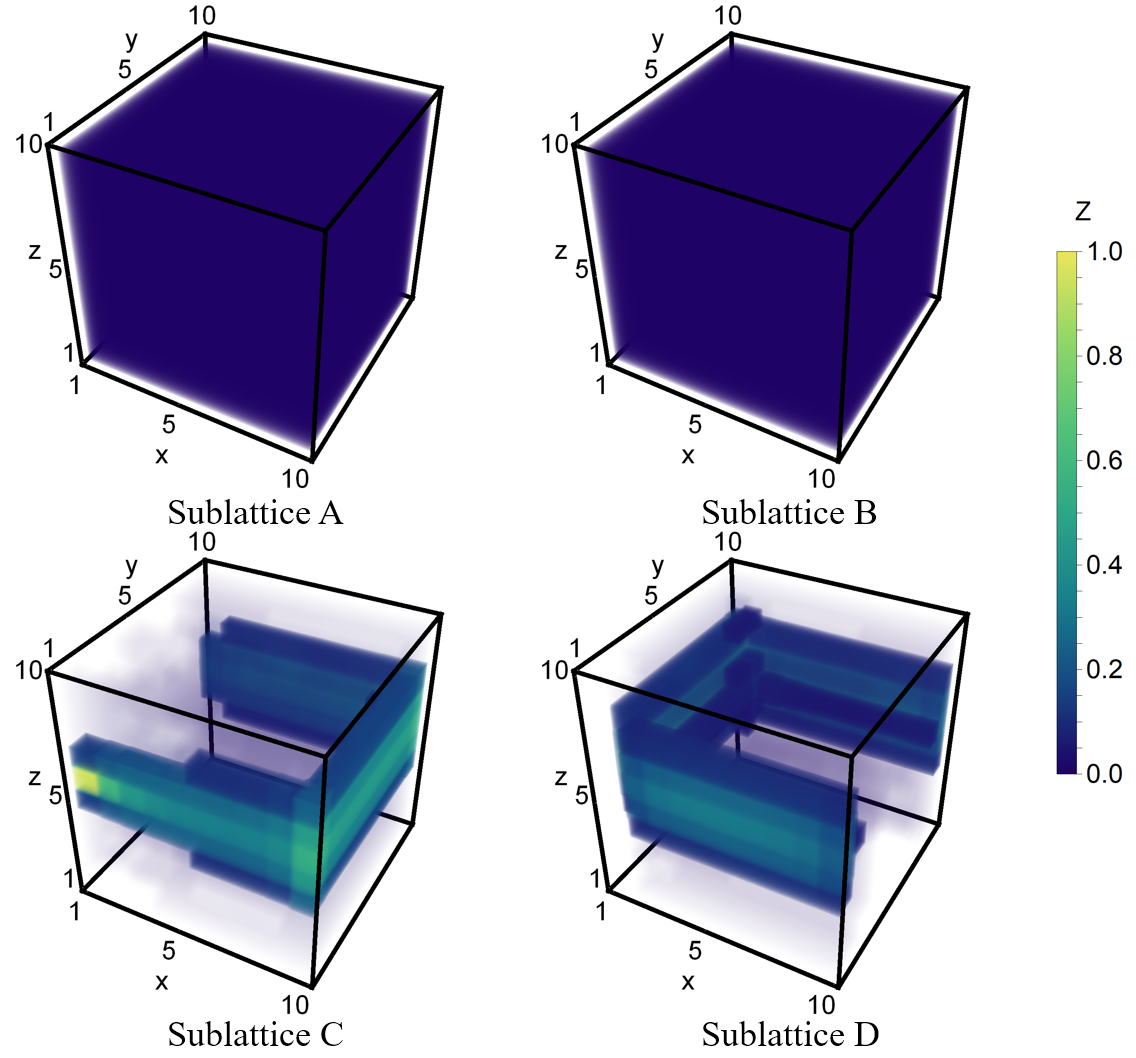}
\caption{Numerical computation of the on-resonance impedance ($Z$) in a 3D Dirac topolectric circuit with the B node at (1,1,5) as the fixed input point for $t=t_0=1$, $m=2$ and $t_z=2$ [see Eq.~(\ref{eq:stackedQSHIDSM})]. The output point is then selected on four nodes or sublattices separately (mentioned explicitly in each subfigure), which we sweep over the entire system. The sublattice selective strong localization of the on-resonance impedance on the C and D sublattices, residing on the $xz$ and $yz$ planes signifies the Fermi arc surface states of a 3D DSM in a topolectric circuit.
}~\label{fig:Diracsemimetal}
\end{figure}

Following the spirit of constructing a 3D WSM from the last section, one can also realize a 3D DSM by stacking 2D layers of quantum spin Hall insulators (QSHIs) in the $z$ direction in a translationally invariant manner. The corresponding Hamiltonian in the momentum space then takes the form
\begin{align}~\label{eq:stackedQSHIDSM}
H^{\rm 3D}_{\rm DSM} &= t \left[ \sin(k_x) \; \Gamma_1 + \sin(k_y) \; \Gamma_2 \right] \nonumber \\
&+ \left[ t_z \cos(k_z) +m - t_0 \sum_{j=x,y} \cos(k_j) \right] \Gamma_3.
\end{align}
The $\Gamma$ matrices follow the representation from Eq.~(\ref{eq:gammarepresentationQSHI}), and $t_z$ denotes the hopping between the nearest-neighbor layers of 2D QSHI in the $z$ direction. For $t=t_0=1$, $m=2$ and $t_z=2$, the system supports a pair of Dirac points separated along the $k_z$ direction and located at ${\bf k}^\star=(0,0\pm \pi/2)$. The resulting Fermi arc surface states in a 3D DSM possess two-fold Kramers degeneracy and connect the Dirac points in the momentum space, similar to the Fermi arc state in a 3D WSM. As $\{ H^{\rm 3D}_{\rm DSM}, \Gamma_5 \}=0$ and $\Gamma_5=\sigma_3 \otimes \tau_0$, the Fermi arc states are sublattice polarized. Specifically, they are localized either on the A and B sublattices or on the C and and D sublattices, similar to the edge modes of the underlying 2D QSHI layers, shown in Fig.~\ref{fig:QSHEsummary}.

A circuit realization of 3D DSM is similar to the other cases  we discussed so far. Specifically, in a Dirac topolectric circuit the layers of 2D QSHI are connected by the capacitors  with capacitance $t_z$. Numerical computation of the on-resonance impedance then reveals sublattice polarization and strong surface localization on the $xz$ and $yz$ planes, see Fig.~\ref{fig:Diracsemimetal}. These observations confirm the existence of the Fermi arc surface states in a 3D Dirac topolectric circuit.

\begin{figure}[t!]
\includegraphics[width=0.95\linewidth]{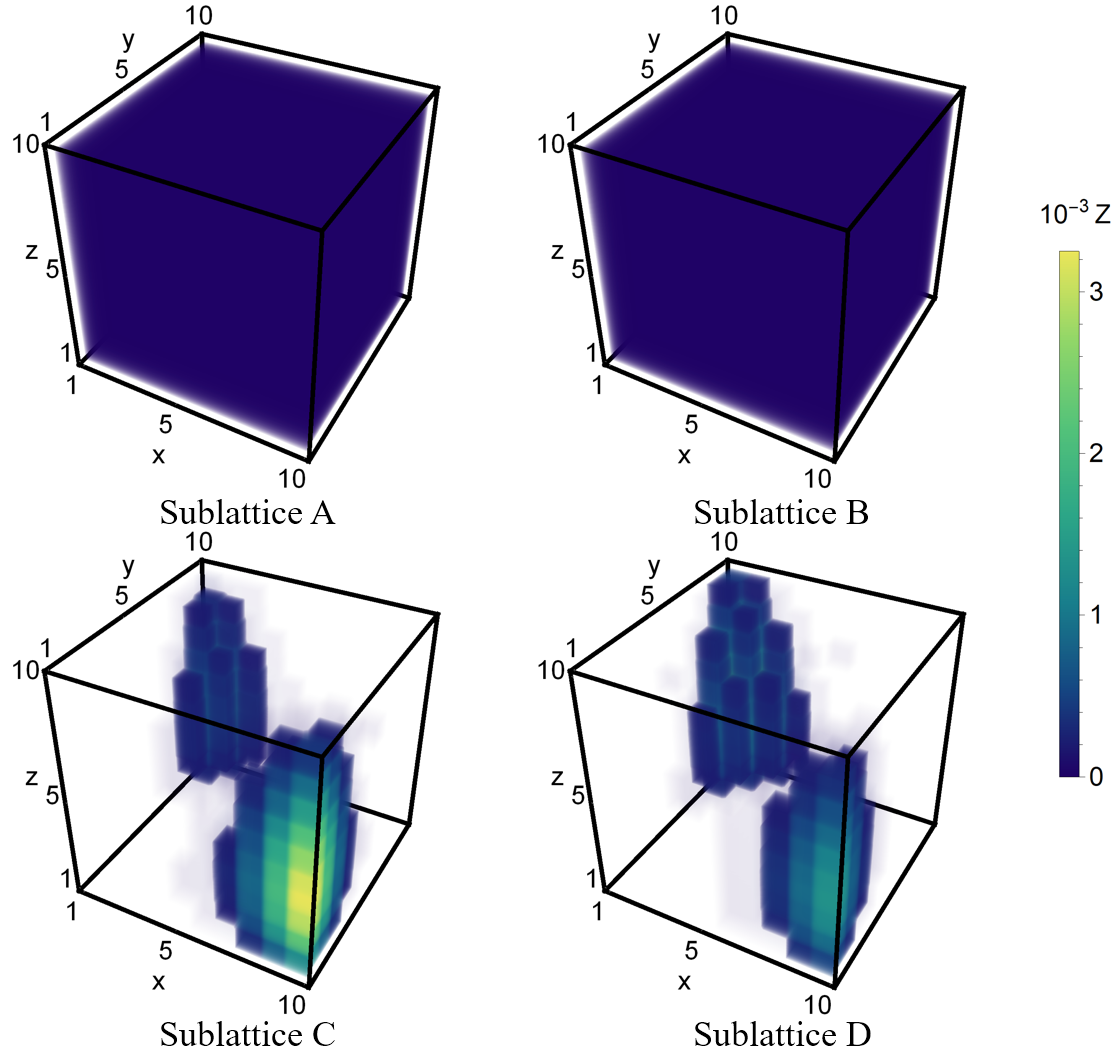}
\caption{Numerical computation of the on-resonance impedance ($Z$) in a 3D second-order Dirac topolectric circuit with the A node at (1,1,5) as the fixed input point for $t=t_0=1$, $m=t_z=2$ and $\Delta=0.8$, see Eqs.~(\ref{eq:stackedHOTIHOTDSM}) and (\ref{eq:stackedQSHIDSM}). The output point is then chosen on four individual nodes or sublattices separately (mentioned in each subfigure), which we scan over the whole system. The on-resonance impedance shows sublattice selective sharp hinge localization, which in turn confirms realization of a 3D higher-order (namely, second-order) DSM in a topolectric circuit.
}~\label{fig:HOTDSM}
\end{figure}

\subsection{Stacked 2D HOTI: 3D quadrupolar Dirac semimetal}~\label{subsec:3D-QDSM}

As a penultimate topic, we discuss the topolectric circuit realization of a 3D second-order or quadrupolar DSM~\cite{Liu-Hughes-PRB2018,calugaru-juricic-roy}. In an electronic system, a 3D quadrupolar DSM is obtained by stacking 2D HOTIs with corner modes in the $z$ direction, for example, while preserving the translational symmetry. The corresponding Hamiltonian in the momentum space assumes the following form
\begin{align}~\label{eq:stackedHOTIHOTDSM}
H^{\rm 3D}_{\rm HOTDSM}=H^{\rm 3D}_{\rm DSM} + \Delta \left[ \cos(k_x)-\cos(k_y) \right] \; \Gamma_4,
\end{align}
where $H^{\rm 3D}_{\rm DSM}$ is defined in Eq.~(\ref{eq:stackedQSHIDSM}) and $\Gamma_4=\sigma_1 \otimes \tau_3$ [see Eq.~\eqref{eq:gammarepresentationQSHI}]. Note that the second term in the above equation, proportional to $\Delta$, causes dimensional reduction of the edge modes associated with each layer of QSHI, yielding four corner modes (see Sec.~\ref{subsec:discretesymmetrybreaking}). The underlying insulating layers then correspond to 2D HOTI. Stacking of such layers of 2D HOTIs produces a pair of higher-order Dirac nodes located at ${\bf k}^\star=(0,0,\pm \pi/2)$ for $t=t_0=1$, $m=t_z=2$ and arbitrary value of $\Delta$. The corner states connecting these two Dirac nodes produce 1D \emph{hinge} modes.

\begin{figure}[t!]
\includegraphics[width=0.95\linewidth]{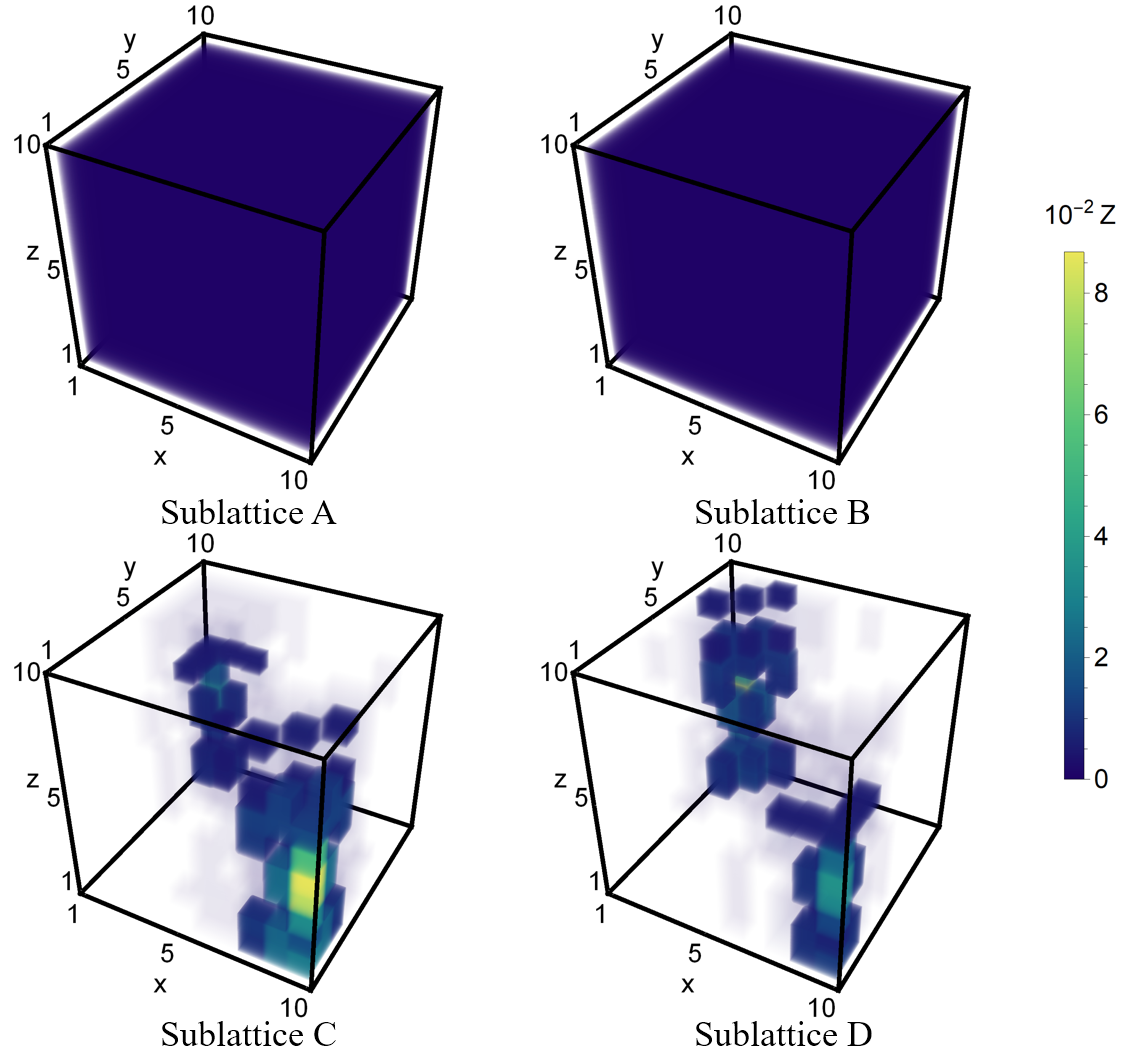}
\caption{Numerical computation of the on-resonance impedance ($Z$) in a 3D second-order Weyl topolectric circuit with the A node at (1,1,5) as the fixed input point for $t=t_0=1$, $m=t_z=2$, $\Delta=1.0$ and $\Delta_1=\Delta_2=2.0$, see Eq.~(\ref{Eq:HOTWeyl}). The output point is then chosen on four individual nodes or sublattices separately (mentioned in each subfigure), which we scan over the whole system. The on-resonance impedance shows sublattice selective sharp hinge localization, which in turn confirms the realization of a 3D second-order WSM in a topolectric circuit.
}~\label{fig:HOTWSM}
\end{figure}

A circuit realization of this model follows the general stacking protocol we discussed so far. Specifically, we couple 2D layers of HOTI circuits, obtained by combing the components shown in Figs.~\ref{fig:QSHEsummary}(a) and \ref{fig:2DQSHI-HOTI}(a), by capacitor connections of capacitance $t_z$. The numerical measurement of the on-resonance impedance then shows sublattice selective sharp hinge localization for the above mentioned parameter values, which we display in Fig.~\ref{fig:HOTDSM} for a specific choice of $\Delta=0.8$. These results demonstrate the existence of 1D hinge modes in a second-order Dirac topolectric circuit.

\subsection{Stacked 2D GHOTI: 3D HOT Weyl semimetal}~\label{subsec:3D-HOTWSM}

Finally, we demonstrate the topolectric circuit realization of a 3D HOT Weyl semimetal. To this end we consider layers of 2D GHOTI [see Sec.~\ref{subsec:antiunitary}] coupled in the $z$ direction along which the translational symmetry is preserved. The resulting Hamiltonian in the momentum space takes the form
\begin{eqnarray}~\label{Eq:HOTWeyl}
H^{\rm 3D}_{\rm HOTWSM} &=& H^{\rm 3D}_{\rm DSM} + \Delta \left[ \cos (k_x)-\cos (k_y) \right] \Gamma_4 \nonumber \\
&+& \Delta_1 \left( i \Gamma_1 \Gamma_2 \right) + \Delta_2 \left( i \Gamma_3 \Gamma_4 \right),
\end{eqnarray}
where the $\Gamma$ matrices are defined in Eq.~(\ref{eq:gammarepresentationQSHI}) and $H^{\rm 3D}_{\rm DSM}$ is shown in Eq.~(\ref{eq:stackedQSHIDSM}). Notice that additional terms proportional to $\Delta_1$ and $\Delta_2$ lift the Kramers degeneracy from the valence and conduction bands~\cite{roy-aniunitaryHOTI}, yielding 2D GHOTI in the absence of the interlayer coupling. When such layers of GHOTI are coupled by translationally invariant interlayer hopping ($t_z$), specifically for $t=t_0=\Delta=1.0$ and $m=t_z=\Delta_1=\Delta_2=2.0$ Kramers \emph{nondegenerate} bands touch each other at ${\bf k}^\star=(0,0,\pi/2)$, yielding a pair of higher-order Weyl nodes. Then each insulating layer of GHOTI between these Weyl nodes hosts four corner modes. By virtue of the translational symmetry,  the collection of such corner modes between these two points then constitutes the 1D hinge modes, yielding a second-order WSM.

In an electric circuit this model is realized by combining the circuit components, shown in Figs.~\ref{fig:QSHEsummary}(a), ~\ref{fig:2DQSHI-HOTI}(a) and ~\ref{fig:2DGHOTI}(a), where each node is supplemented by four subnodes, and subsequently coupling each layer of 2D circuit by capacitor connections of capacitance $t_z$. When the capacitor values are chosen to be the ones mentioned above for the lattice model, we realize a higher-order Weyl topolectric circuit. The numerical measurements of on-resonance impedance, shown in Fig.~\ref{fig:HOTWSM}, display sublattice selective sharp hinge localization, which in turn confirms the circuit realization of a second-order WSM.

\section{Summary and Discussion}~\label{Sec:Conclusions}

To summarize, here we present an alternative derivation of the construction of an arbitrary hopping element, stemming from an underlying lattice tight-binding model, in LC electric circuits, where the electrical nodes play the role of lattice sites. Subsequently, we apply this general protocol to engineer a plethora of topological lattice models in topolectric circuits. In particular, we identity first-order SSH model (Sec.~\ref{subsec:SSH}), Chern (Sec.~\ref{subsec:CI}) and quantum spin Hall (Sec.~\ref{subsec:QSHI}) insulators from the highly endpoint and edge localized on-resonance impedance ($Z$) respectively in $d=1$ and $d=2$. We also propose simple circuit realizations of 2D and 3D higher-order topological insulators (Secs.~\ref{subsec:BBH2D} and ~\ref{subsec:BBH3D}), supporting corner localized and sublattice polarized on-resonance impedance. In addition, we also demonstrate a concrete route to break discrete rotational symmetry and implement Wilson-Dirac mass in topolectric circuits (Sec.~\ref{subsec:discretesymmetrybreaking}). Such a construction allows us to convert a 2D first-order quantum spin Hall insulator (with edge modes) into a higher-order topological insulator (with corner modes). Finally, we construct a generalized second-order topological insulator for which the corner impedance is protected by an antiunitary operator (Sec.~\ref{subsec:antiunitary}). Furthermore, we also show explicit construction of the hierarchy of higher-order topological insulators in three-dimensions, and realization of first-, second- and third-order topological insulators that respectively support surface, hinge and corner impedance as discrete rotational symmetries are systematically broken in a topolectric circuit (Sec.~\ref{Sec:3Dhierarchy}).

The simplicity of our circuit constructions is based on the representation theory of the Clifford algebra. In particular, throughout we exploit the fact that the Clifford algebra of $2^N$-dimensional Hermitian matrices is closed by $(2N+1)$ mutually anticommuting Hermitian matrices, among which $N+1$ ($N$) are purely real (imaginary)~\cite{clifford1, clifford2}. As the existence of topological boundary modes relies on the anticommuting nature of the involved matrices (not on their explicit representations), we choose (whenever possible) matrices multiplying the sine (cosine and constant) functions to be purely imaginary (real), such that the hopping elements in the real space are completely \emph{real}. Exceptions from this scenario are rather sparse, see Secs.~\ref{subsec:CI} and ~\ref{subsec:antiunitary}, for example. One can then implement a lattice topological model on an electric circuit by supplementing each node (representing a lattice site) with only two subnodes, between which the phases of current and voltage differ by a factor of $\exp(i\pi)=-1$, see Sec.~\ref{subsubsec:2nodes}. We also highlight a generalization of this construction involving four (Sec.~\ref{subsubsec:4nodes}) as well as $n$ (Sec.~\ref{subsubsec:nnodes}) subnodes.

Subsequently, we present electric circuit realizations of various gapless topological phases, such as 2D and 3D Dirac semimetals, in Secs.~\ref{subsec:DSM-NLSM} and \ref{subsec:3DDSM}, by respectively stacking 1D SSH and 2D QSHI circuits, while preserving the translational symmetry in the stacking direction. In addition, we also show a concrete realization of Weyl topolectric circuits by stacking 2D Chern insulators (Sec.~\ref{subsec:3DWSM}). These topolectric semimetals are then identified from the on-resonance impedance that mimics the Fermi arc states in the real space. On the other hand, a nodal-line topolectric circuit is identified from on-resonance impedance, localized on the top and bottom surfaces, bearing the signature of drumhead surface states in the real space (Sec.~\ref{subsec:DSM-NLSM}). Finally, we also show realization of higher-order Dirac (Weyl) topolectric circuit, featuring hinge localized impedance, in  Sec.~\ref{subsec:3D-QDSM} (Sec.~\ref{subsec:3D-HOTWSM}).

By focusing on the specific example of the Chern circuit (Sec.~\ref{subsec:CI}), we show that the measurement of on-resonance impedance can be instrumental in mapping the global phase diagram of topological lattice models in topelectric circuits. To this end, we use that the on-resonance impedance is finite only inside the topological phases, while it vanishes in the trivial phase as well as at the topological quantum critical point between two topologically distinct phases (Fig.~\ref{fig:Chernphasediagram}). Therefore, our findings can be experimentally consequential for understanding various paradigmatic toy models of topological phases.

Finally, we comment on a subtle issue regarding the time-reversal symmetry in topolectric circuit. Note that we propose circuit realizations of various topological models that in electronic systems are protected by the time-reversal symmetry (${\mathcal T}$), satisfying ${\mathcal T}^2=-1$ (such as the quantum spin Hall insulator). However, topolectric circuits are constituted by capacitors and inductors, and all these elements are \emph{real}. Therefore, in topolectric circuits ${\mathcal T}^2=+1$. Nevertheless, due to high precision tunability of the circuit elements we believe that these models can still be engineered in topolectric circuits and their topological modes can be observed through on-resonance boundary impedance. This should be so at least when sufficient care is taken to minimize circuit disorder in the setup, given that all the topological phases we discuss here are robust against sufficiently weak randomness.

In the future, our setup should be instrumental to systematically investigate the role of disorder in topolectric circuits. Disorder can be implemented in this setup by randomly and independently varying the grounding elements [the capacitor ($C_a$) or inductor ($L_a$) in Fig.~\ref{Fig:schematiccircuit}, for example] at each node of the circuit, such that the resonance frequency ($\omega_R$) displays a random spatial variation $\delta \omega_R({\bf x})$, with the spatial average $\langle \delta \omega_R({\bf x}) \rangle=0$. Such disordered topolectric circuits can mimic a variety of fundamentally important phenomena in dirty topological systems, among which possibly the most interesting are the topological Anderson insulator in electrical circuits~\cite{zqzhang-prb2019}, gradual melting of the Fermi arc~\cite{slager-juricic-roy:Fermiarc} and hinge~\cite{szabo-roy:disorderHOTDSM} impedance, respectively, in 3D Weyl and higher-order Dirac topolectric circuits. Furthermore, the jurisdiction of topolectric circuits can be further extended by engineering lattice defects to probe topological phases within this setup. Even though lattice defects have been realized in other topological metamaterials, such as photonic~\cite{noh-natphot2018,li-natcomm2018} and phononic~\cite{grinberg-arxiv2019,peterson-arxiv2020,liu-arxiv2020} crystals, their realizations in topoloelectric crystals remain to be studied.

\acknowledgements

V.J. acknowledges the support of the Swedish Research Council (VR 2019-04735). B.R. was partially supported by the startup grant from Lehigh University. We are thankful to Ronny Thomale for useful correspondence.

\appendix

\section{One-shift generator of hopping phase factors}~\label{app:general-proof}

The one-shift matrix is of the form (for a fixed $n$, omitted here for the notational clarity)
\begin{equation}
C_{ij,(1)}=C\left[\delta_{i,j+1}(1-\delta_{i,1})+\delta_{i,1}\delta_{j,n}\right].
\end{equation}
The corresponding effective hopping element then reads
\begin{equation}
t_{(1)}={\rm Tr}[\hat{P}_v \hat{C}_{(1)}\hat{P}_v],
\end{equation}
where $\hat{P}_v$ is the projector onto the subspace generated by the unit vector $v$ with the components being the $n$th roots of unity, namely $v_k=\exp \left[\frac{2i\pi(1-k)}{n} \right]$, where $k=1,...,n$. More explicitly,
\begin{eqnarray}
t_{(1)} &=& C\frac{1}{n^2}\sum_{i,k,l}
v_i v_k^*[\delta_{k,l+1}(1-\delta_{k,1})+\delta_{k,1}\delta_{l,n}]v_l v_i^* \nonumber \\
&=& \exp \left[\frac{2i\pi}{n} \right] \; C.
\end{eqnarray}
An analogous calculation for the $s$-shift matrix, which is a product of $s$ one-shift matrices, given by
\begin{equation}
C_{{ij},(s)}=C\left[\delta_{i,j+s}(1-\delta_{i,s})+\delta_{i,s}\delta_{j,n}\right],
\end{equation}
yields the effective hopping $t_{(s)}=C \exp \left[\frac{2i\pi s}{n} \right]$. Therefore, one-shift is a generator of hopping phase factors.



\end{document}